          \documentclass{cmslatex}
\usepackage[paperwidth=7in, paperheight=10in, margin=.875in]{geometry}
\usepackage{amsmath,graphicx,amssymb,subfigure}
\usepackage{amsfonts}
\usepackage{epstopdf}
\renewcommand{\theequation}{\arabic{section}.\arabic{equation}}

\begin{document}
%          \title{Full title\thanks{}}
\title{A continuum model for distributions of  dislocations incorporating short-range interactions\thanks{This work was partially supported by the Hong Kong Research Grants Council General
Research Fund 606313 and HKUST
Postdoctoral Fellowship Matching Fund. The work of Y.C.Z was partially supported by Natural Science Foundation of China (NSFC)  under the contract no. 11772076. The work of S.Y.D was supported by Natural Science Foundation of China (NSFC)  no. 11701433.}}

\author{
    Xiaohua Niu\footnotemark[2] \and
    Yichao Zhu\footnotemark[4]\ \footnotemark[5] \and
    Shuyang Dai\footnotemark[6] \and
    Yang Xiang\footnotemark[2]
}

\renewcommand{\thefootnote}{\fnsymbol{footnote}}

%\footnotetext[1]{Department of Mathematics, Hong Kong University of Science and Technology.
%(tluoaa@ust.hk, maxiang@ust.hk). The work of the second author was
%partially supported by HKUST School-Based initiatives SBI14SC10. The first
%author thanks the hospitality of the
%Department of Mathematics, Purdue University during Spring 2015 where part
%of this work was conducted.}
%\footnotetext[2]{Department of Mathematics, Purdue University. (yip@math.purdue.edu)}
%

\footnotetext[2]{Department of Mathematics, The Hong Kong University of Science and Technology, Clear Water Bay, Kowloon, Hong Kong, email (xniu@connect.ust.hk, maxiang@ust.hk).}
\footnotetext[4]{State Key Laboratory of Structural Analysis for Industrial Equipment, Department of Engineering Mechanics, Dalian University of Technology, No.2 Linggong Road, Ganjingzi District, Dalian, Liaoning, China, 116024, email (yichaozhu@dlut.edu.cn).}
\footnotetext[5]{International Research Center for Computational Mechanics, Dalian University of Technology.}
\footnotetext[6]{School of Mathematics and Statistics, Wuhan University, Wuhan, Hubei, China, 430072, email (shuyang\_dai@whu.edu.cn).}
%
%          %For each author, make a block with the following four macros:

 %         \author{full name\thanks{ address,  email()}
%         {Put the URL for your home page here if you have
%          one}\and{full name(another author)\thanks{ address,  email() }}}

          %Use \thanks statements for acknowedgements of grants and
          %support. They will appear below all the authors' addresses, so be
          %specific about which author is thanking whom:

          %\thanks{}

          % Use the standard latex environments for theorems, etc. Here is one
          % possible method of declaring them: It numbers all results by the
          % section, and uses a common numbering system for the different
          % environmentts.

          %\date{Received date / Revised version date}

         \pagestyle{myheadings}
         \markboth{Continuum Model for Dislocations Incorporating Short-range Interactions}
         {Xiaohua Niu, Yichao Zhu, Shuyang Dai, Yang Xiang}
         \maketitle

          \begin{abstract}
          Dislocations are the main carriers of the permanent deformation of
crystals. For simulations of engineering applications, continuum models where material
microstructures are represented by continuous density distributions of dislocations
are preferred. It is challenging to capture in the continuum model
the short-range dislocation interactions,  which vanish after the standard averaging procedure from discrete dislocation models. In this study, we consider
systems of parallel straight dislocation walls and develop continuum descriptions
for the short-range interactions of dislocations by using asymptotic analysis.
The obtained continuum short-range interaction formulas are incorporated
in the continuum model for dislocation dynamics based on a pair of
dislocation density potential functions that represent continuous distributions of dislocations.
 This derived continuum model is able to describe the anisotropic dislocation interaction and motion. Mathematically, these short-range interaction terms ensure strong stability property of the continuum model that is possessed by the discrete dislocation dynamics model.
 The derived continuum model is validated by
comparisons with the discrete dislocation dynamical simulation results.
          \end{abstract}



\section{Introduction}

The plastic deformation of crystalline materials is primarily carried out by the motion of a large number of atomistic line defects, i.e. dislocations. Based on the accumulated knowledge about the behavior of individual dislocations \cite{Hirth}, discrete dislocation dynamics (DDD) models \cite{Kubin1992,Ghoniem2000,Xiang2003,Cai2007,Mordehai,Gu2015} have been well developed for the study of crystal plasticity in a wide range of mechanical problems. For engineer applications, however, DDD models are limited to samples of small size (order of microns or below), because of their high computational costs. Hence continuum models, where material microstructures are represented by continuous density distributions of dislocations resulting from the local homogenization of the discrete dislocation networks, are practically preferred \cite{Nye1953,Kroener1963,NelsonToner1981,Mura1987,Groma1997,ElAzab2000,Acharya2001,Arsenlis2002,Groma2003,
Hochrainer2007,Xiang2009_JMPS,ZhuXH2010,Zhu2014_IJP,GeersJMPS2014,Hochrainer2014,Geers_JMPS2015,Zhu_continuum3D,Acharya2015,Schulz2015,
Finel2016,Ngan_JMPS2016,Monavari2016,ZhuScripta2016,dipole1D,ZhuNiuXiang2016,Ngan2017}.

In order to incorporate the orientation-dependent dislocation densities and the anisotropic dislocation interaction and motion in the continuum model, we have employed a pair of dislocation density potential functions (DDPFs) to describe the dislocation distribution \cite{Xiang2009_JMPS,ZhuXH2010,ZhuXiangCMS2012,ZX2014,Zhu_continuum3D}. In this representation, the intersections of the contour lines (of integer multiples of the length of the Burgers vector) of the two DDPFs $\phi$ and $\psi$ are the locations of the dislocations, see Sec.~\ref{sec:ddpf} for the model in two-dimensions (where dislocations are infinite straight lines). Essentially, the DDPF $\psi$ characterizes the local distribution of the active slip planes and the DDPF $\phi$ restricted on a slip plane describes the local dislocation distribution within that plane. As a result, the derived continuum dislocation dynamics model takes the form of a PDE system of two DDPFs $\phi$ and $\psi$, instead of equations of the single variable of scalar dislocation density in the existing two-dimensional continuum models in the literature reviewed above for geometrically necessary dislocations.
While previous continuum model based on DDPFs focused on dislocation glide within slip planes \cite{Xiang2009_JMPS,ZhuXH2010,Zhu2014_IJP,Zhu_continuum3D},
the continuum dislocation dynamics equations derived in this paper incorporate both dislocation motions of glide and climb. The continuum dislocation model based on DDPFs
also provides a mathematical framework for rigorous analysis of the properties of the interaction and dynamics of dislocations and further incorporation of other important dislocation mechanisms at the continuum level (such as the Frank-Read sources \cite{Zhu2014_IJP} and dynamics of dislocation dipoles \cite{dipole1D,ZhuNiuXiang2016}).

In dislocation-density-based continuum models that are derived from the DDD model, the leading order dislocation interaction is given by an integral over the dislocations in the entire system and is referred to as the long-range dislocation interaction. The correction terms that improve a continuum model as an approximation to the DDD model often take the form of higher order derivatives of dislocation densities that depend only on the local arrangement of dislocations, and are referred to as the short-range dislocation interaction terms.
 In the existing dislocation-density-based models of plasticity,
 although the long-range dislocation-dislocation interactions are well-captured by direct averaging, the short-range interactions have to be  incorporated with special treatments.
This is because the mutual interaction force between two dislocations can grow as strong as the order of $1/r$, where $r$ is the dislocation spacing, which leads to a strong dependence of dislocation dynamics on the local discrete arrangement of dislocations and  further influences the plastic behavior of materials. However, when a discrete dislocation network is treated by a dislocation continuum,  such short-range interactions are averaged to zero.
Therefore, the development of continuum modelling of dislocations highly relies on effective ways to capture the short-range interactions on a coarse-grained scale.

 For two-dimensional dislocation configurations where all dislocations are infinitely straight and mutually parallel, Groma et al. \cite{Groma2003} developed a continuum formulation for the short-range  dislocation interaction  based on a statistical approach, and such statistical method was further extended by Dickel et al. \cite{Dickel2014} to identify the role played by dislocation dipoles in crystal plasticity. However, it has been argued by comparisons with discrete dislocation dynamics simulations (Roy et al. \cite{Roy2008}) that the short-range dislocation interaction formulas obtained based on statistic approaches do not necessarily apply to deterministic distributions of dislocations.

A class of representative two-dimensional dislocation configurations widely studied in the literature are distributions (pile-ups) of dislocation walls consisting of straight and mutually parallel dislocations (e.g. \cite{Roy2008,Schuouwenaars2010,Voskoboinikov2007JMPS,Cameron2011,Geers2013,Zhu_2Ddipoles2014,Schulz2014,Schulz2015}). In this scenario, dislocation-dislocation interactions take place in both directions that are in and normal to the dislocation slip planes.
To the best knowledge of the authors', most available analytical results employing dislocation densities were obtained for regular dislocation wall structures, where dislocations within each wall are vertically aligned and uniformly spaced in the direction normal to the dislocation slip planes. For example, in their comparisons with results of discrete dislocation model, Roy et al. \cite{Roy2008} also used semi-continuum analysis (in which discreteness normal to the slip planes are maintained) for pile-up of dislocation walls.
With the matched asymptotic techniques, Voskoboinikov and coworkers \cite{Voskoboinikov2007JMPS} calculated the discrete positions of a simple dislocation structure formed by one horizontal row of straight dislocations near a dislocation lock, where the dislocation density becomes singular under a continuum setting.
Hall \cite{Cameron2011} generalized the approach in Ref.~\cite{Voskoboinikov2007JMPS} to determine the discrete positions of the dislocation walls of infinite length near the grain boundaries. For regular dislocation walls,  Geers and coworkers \cite{Geers2013} identified five regimes for the interaction energy by a single parameter depending on the driving force, the horizonal and the vertical spacing between neighbouring dislocations, and they also studied the continuum limit of the equilibrium state in each regime as the number of the regular walls tends to infinity. The analysis in Ref.~\cite{Geers2013} also suggests that a single field variable describing the dislocation density is not sufficient for the discrete-to-continuum transition for the configuration with dislocation regular walls. Zhu and Chapman \cite{Zhu_2Ddipoles2014} examined the equilibria of periodically arranged dipole walls, and a natural transition between dipolar configurations was found controlled by the dipole height to width ratio.
By investigating the local behavior of the mean-field stress exerted by a row of dislocations, Schulz {\it et al.} added  to the continuum system a dislocation density gradient term depending on the mesh size in their finite element calculations \cite{Schulz2014}. Schmitt {\it et al.} derived continuum internal stress formula for dislocation glide by homogenization of dislocation microstructures under the assumption that the geometrically necessary dislocations form regular walls \cite{Schulz2015}.  Their obtained formula  is similar in its form to the short-range dislocation interaction term obtained by Groma et al. \cite{Groma2003} using statistical approach.

In this paper, we first systematically examine the perturbed regular edge dislocation wall structures and derive continuum short-range interaction formulas  from discrete dislocation dynamics model by asymptotic analysis.
The derived accurate short-range interaction formulation for such
representative deterministic dislocation distributions, together with the available results in the literature reviewed above,  is able to give more complete understanding of the nature
 of the short-range dislocation interactions for parallel dislocations with the same Burgers vector in the continuum model.
 In particular,
our continuum short-range formulation is expressed by higher order derivatives of the dislocation distribution; although it is similar to the continuum formula derived using other approaches \cite{Groma2003,Schulz2015}, the exact expressions are different. Moreover, by using two field variables (two DDPFs), our continuum formulation incorporates the anisotropy of the short-range dislocation interactions in directions along or normal to the dislocation slip planes, in addition to the anisotropic dislocation motions of glide and climb.
Such anisotropy is not included in the available continuum short-range interaction formulas  \cite{Groma2003,Schulz2015}, and although it was examined in Ref.~\cite{Geers2013} by regular dislocation walls, no continuum formulation is available in the existing literature to account for such dislocation interaction anisotropy for general cases.

 We then incorporate these continuum short-range interaction contributions in our continuum PDE model. These terms are local in the sense that they depend on the first and second partial derivatives of the DDPFs instead of their integrals. The full continuum force (including both the long-range and short-range continuum forces) provides a good approximation to the discrete dislocation dynamics model.
   Mathematically, these new terms in the continuum model serve as stabilizing terms that maintain the same stability properties as the discrete dislocation dynamics model. Moreover, since these short-range interaction terms are in the form of second order partial derivatives of the DDPFs $\phi$ and $\psi$, they also serve as regularization terms to the continuum long-range force terms that are in the form of integrals of first partial derivatives of $\phi$ and $\psi$.

The rest of this paper is organized as follows.
In Sec.~\ref{sec:ddd}, we reviewed the discrete dislocation dynamics model, from which the continuum formulation of short-range interactions will be derived. In Sec.~\ref{sec:ddpf}, we present the continuum framework for dislocation walls based on  the representation of dislocation density potential functions, where the force on dislocations consists only of the long-range Peach-Koehler force.
In Sec.~\ref{sec:long-rang}, we show that without short-range interactions, the continuum long-range Peach-Koehler force  is inconsistent with the Peach-Koehler force in the discrete dislocation model for many common dislocation distributions.
In Sec.~\ref{sec:continuum-short},  we  derive continuum expressions for the dislocation short-range interactions from the discrete dislocation dynamics model. We focus on  the dislocation configurations identified in Sec.~\ref{sec:long-rang} where
the continuum long-range force fails to provide stabilizing effect compared with the discrete model.
In Sec.~\ref{eqn:contiuum-ddpf}, we present the DDPF-based continuum dislocation dynamics model that incorporates both the long-range and the short range continuum forces.
 In Sec.~\ref{sec:stability}, we show the new continuum model is indeed able to stabilize the perturbed dislocation structures as the discrete dislocation model does.
In Sec.~\ref{sec:numerical}, numerical simulations are performed to validate the continuum model.

\section{Discrete dislocation  dynamics model}
\label{sec:ddd}

\begin{figure}[htbp]
\centering
\includegraphics[width=80mm]{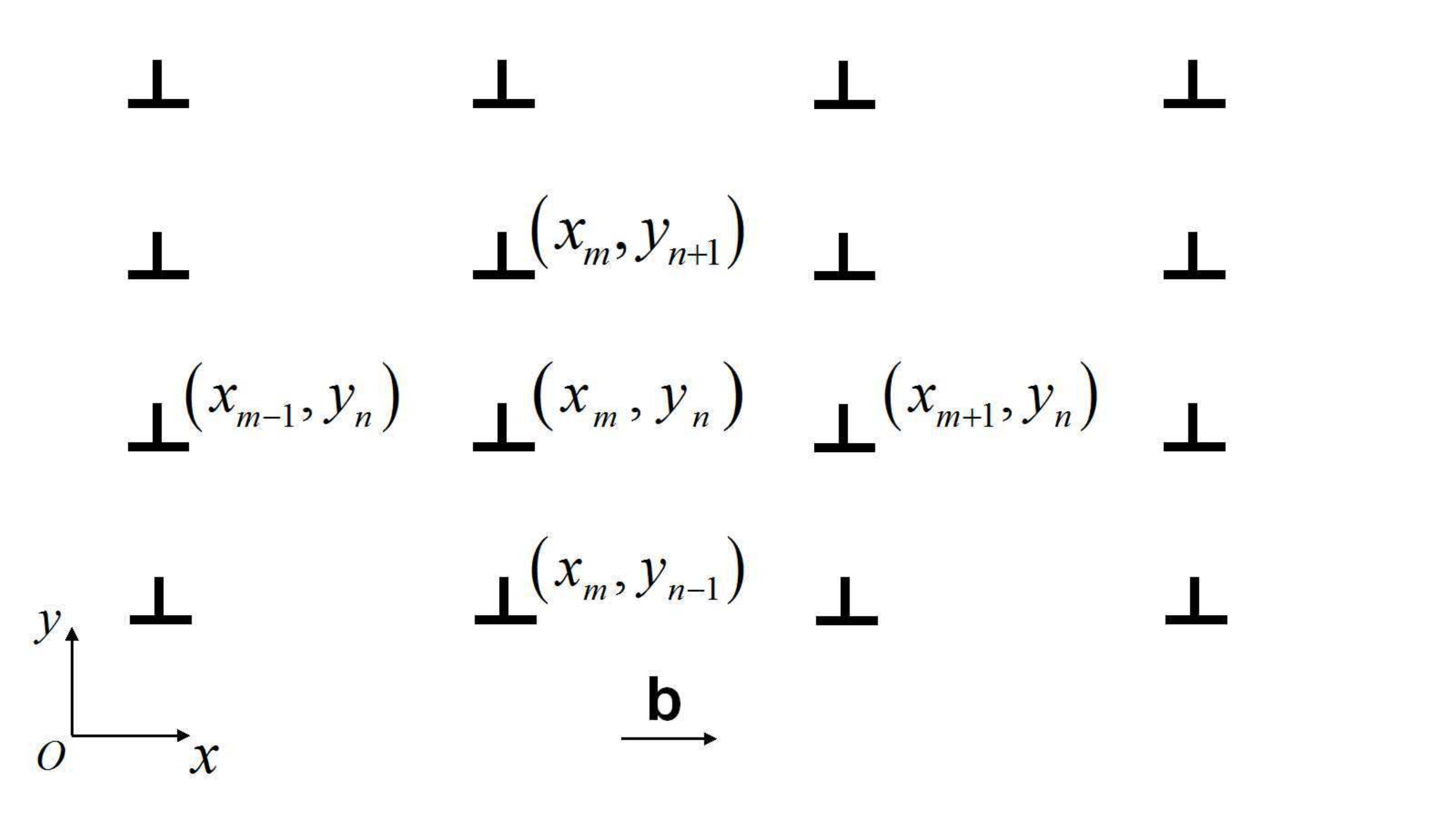}
\caption{A system of parallel straight edge dislocations.}
\label{fig:a}
\end{figure}

In this section, we briefly reviewed the discrete dislocation dynamics model, from which the continuum formulation of short-range interactions will be derived.
We consider a system of parallel straight edge dislocations, see Fig. \ref{fig:a}. In this case, the dislocation dynamics  can be reduced to a two-dimensional spatial problem, in which   the dislocations  are  points in the plane orthogonal to the direction of the dislocation lines, that is, parallel to the $z$-axis.  The Burgers vector $\boldsymbol{b}$  is along the $x$-axis.     The locations of dislocations are denoted by the points $\{(x_m,y_n)\}$ for integer $m$ and $n$.

The Peach-Koehler force $f$ on a dislocation is a configurational force associated with the change of free energy $\mathrm{d}W$ due to a displacement $\mathrm{d}l$ of the dislocation: $\mathrm{d}W=-f\mathrm{d}l$.  The  Peach-Koehler force per unit length on the dislocation is related to the stress field by \cite{Hirth}
\begin{equation}\label{eq:PK}
\textbf{f}=(\boldsymbol{\sigma}\cdot\textbf{b})\times\boldsymbol{\tau}=
\left(\begin{array}{ccc}
\sigma_{\scriptscriptstyle xx} & \sigma_{\scriptscriptstyle xy} & \sigma_{\scriptscriptstyle xz}\\
\sigma_{\scriptscriptstyle yx} & \sigma_{\scriptscriptstyle yy} & \sigma_{\scriptscriptstyle yz}\\
\sigma_{\scriptscriptstyle zx} & \sigma_{\scriptscriptstyle zy} & \sigma_{\scriptscriptstyle zz}\\
\end{array}\right)
\left(\begin{array}{ccc}
    b\\0\\0\\
\end{array}\right)
\times\left(\begin{array}{ccc}
    0\\0\\1\\
\end{array}\right)
=\left(\begin{array}{ccc}
    b \sigma_{\scriptscriptstyle xy}\\-b\sigma_{\scriptscriptstyle xx}\\0\\
\end{array}\right),
\end{equation}
 where $\textbf{b}$ is the Burger's vector with the magnitude $b$,  $\boldsymbol{\tau}=(0,0,1)$ is the dislocation line direction, and  $\boldsymbol{\sigma}$ is the stress tensor. The component of the Peach-Koehler force in the $x$ direction is parallel to the plane containing both the Burgers vector  and the dislocation line direction (which is the slip plane), and is the glide force. The component of the Peach-Koehler force in the $y$ direction is normal to the direction of the Burgers vector $\boldsymbol{b}$ and the dislocation line direction, and is the climb force. From Eq.(~\ref{eq:PK}), we have the glide force  $f_{\text{g}}=b \sigma_{\scriptscriptstyle xy}$ and the climb force $f_{\text{c}}=-b\sigma_{\scriptscriptstyle xx}$.

  Using isotropic linear elasticity theory, an edge dislocation located at the point $(0, 0)$ generates the following stress field \cite{Hirth}
\begin{equation}\sigma_{\scriptscriptstyle xy}(x,y)=\sigma_{\scriptscriptstyle yx}(x,y)=\frac{\mu b}{2\pi(1-\nu)}\frac{x(x^2-y^2)}{(x^2+y^2)^2}\triangleq G_1(x, y),\label{eqn:g1}\end{equation}
\begin{equation}\sigma_{\scriptscriptstyle xx}(x,y)=\frac{-\mu b}{2\pi(1-\nu)}\frac{y(3x^2+y^2)}{(x^2+y^2)^2}\triangleq G_2(x, y),\label{eqn:g2}\end{equation}
where $\mu$ is the shear modulus and $\nu$ is the Poisson ratio, and other stress components vanish.

Therefore, for a dislocation located at $(x_{m_0}, y_{n_0})$, the glide force on it generated by another dislocation  located at $(x_{m},y_{n})$ is
$\frac{\mu b^2}{2\pi(1-\nu)}\frac{(x_{m_0}-x_m)((x_{m_0}-x_m)^2-(y_{n_0}-y_n)^2)}{[(x_{m_0}-x_m)^2+(y_{n_0}-y_n)^2]^2}
$.
By superposition, the total glide force acting on the dislocation located at $(x_{m_0}, y_{n_0})$ is
\begin{equation}\label{eq:glide-dd}f_{\text{g}}^{\text{dd}} (x_{m_0},y_{n_0})= \frac{\mu b^2}{2\pi(1-\nu)}\sum_{(m,n)\neq(m_0,n_0)}\frac{(x_{m_0}-x_{m})[(x_{m_0}-x_{m})^2-(y_{n_0}-y_{n})^2]}
{[(x_{m_0}-x_{m})^2+(y_{n_0}-y_{n})^2]^2}.
\end{equation}
Similarly, the total climb  force acting on the dislocation located at $(x_{m_0}, y_{n_0})$ is
\begin{equation}\label{eq:climb-dd}f_{\text{c}}^{\text{dd}} (x_{m_0},y_{n_0})= \frac{\mu b^2}{2\pi(1-\nu)}\sum_{(m,n)\neq(m_0,n_0)}\frac{(y_{m_0}-y_{m})[3(x_{m_0}-x_{m})^2+(y_{n_0}-y_{n})^2]}
{[(x_{m_0}-x_{m})^2+(y_{n_0}-y_{n})^2]^2}.
\end{equation}

With applied stress, the total glide and climb forces  acting on the dislocation located at $(x_{m_0}, y_{n_0})$  can be written as
\begin{eqnarray}
f_{\text{g}} (x_{m_0},y_{n_0})=f_{\text{g}}^{\text{dd}} (x_{m_0},y_{n_0})+b \sigma_{\scriptscriptstyle xy}^0,\label{eq:glide-dd-app}\\
f_{\text{c}} (x_{m_0},y_{n_0}) =f_{\text{c}}^{\text{dd}} (x_{m_0},y_{n_0})-b \sigma_{\scriptscriptstyle xx}^0,\label{eq:climb-dd-app}
\end{eqnarray}
where $\sigma_{\scriptscriptstyle xx}^0$ and $\sigma_{\scriptscriptstyle xy}^0$ are the components of the applied stress tensor.

In discrete dislocation dynamics, the local dislocation velocity $\textbf{v}$ is given by the following mobility law in terms of the Peach-Koehler force \cite{Kubin1992,Ghoniem2000,Xiang2003,Cai2007} as
$\textbf{v}=\textbf{M}\cdot\textbf{f}$,
where $\textbf{M}$ is the mobility tensor and  $\textbf{f}$ is  the Peach-Koehler force. Following \cite{Xiang2003}, the mobility tensor can be written as
$\textbf{M}=m_{\text{g}}(\textbf{I}-\textbf{n}\otimes\textbf{n})+m_{\text{c}}\textbf{n}\otimes\textbf{n}$,
where  $m_{\text{g}}$ is the mobility constant for dislocation glide, $m_{\text{c}}$ is the mobility constant for dislocation climb, $\textbf{I}$  is the identity matrix, and $\textbf{n}$ is the normal direction of the slip plane. For the edge dislocation array being considered, $\textbf{n}=(0,1,0)^T$, and the dislocation velocity is given by
 \begin{equation}\label{eq:velocity}
\textbf{v}
=\left(\begin{array}{ccc}
    v_{\text{g}} \\ v_{\text{c}}\\0\\
\end{array}\right)
=\left(\begin{array}{ccc}
    m_{\text{g}} f_{\text{g}} \\m_{\text{c}} f_{\text{c}}\\0\\
\end{array}\right).
 \end{equation}
 where the continuum Peach-Koehler force is $\textbf{f}=(f_{\text{g}}, f_{\text{c}},0)^T$.

Note that when the line direction of all the dislocation lines is changed to $\boldsymbol{\tau}=(0,0,-1)$, the Peach-Koehler force components in Eqs.~(\ref{eq:glide-dd}) and (\ref{eq:climb-dd}) do not change because both the dislocation line direction and the stress change their signs. In this case, the total glide and climb forces with applied stress in Eqs.~(\ref{eq:glide-dd-app}) and (\ref{eq:climb-dd-app}) change to
$f_{\text{g}} (x_{m_0},y_{n_0})=f_{\text{g}}^{\text{dd}} (x_{m_0},y_{n_0})-b \sigma_{\scriptscriptstyle xy}^0$ and
$f_{\text{c}} (x_{m_0},y_{n_0}) =f_{\text{c}}^{\text{dd}} (x_{m_0},y_{n_0})+b \sigma_{\scriptscriptstyle xx}^0$.

 \section{Continuum formulation for dynamics of dislocation ensembles using dislocation density potential functions}
 \label{sec:ddpf}

We consider the system of parallel straight edge dislocations as shown in Fig.~\ref{fig:a}.  The number of the dislocations in the vertical direction or  horizontal direction is large and can be considered as infinity.
As all the existing continuum dislocation dynamics models reviewed in the introduction, our continuum model is able to describe smoothly varying dislocation structures and holds in an averaged sense for general dislocation structures by homogenizing the discrete dislocations within some representative volumes centered at each point \cite{Zhu_continuum3D}.

\begin{figure}[htbp]
\centering
  \includegraphics[width=80mm]{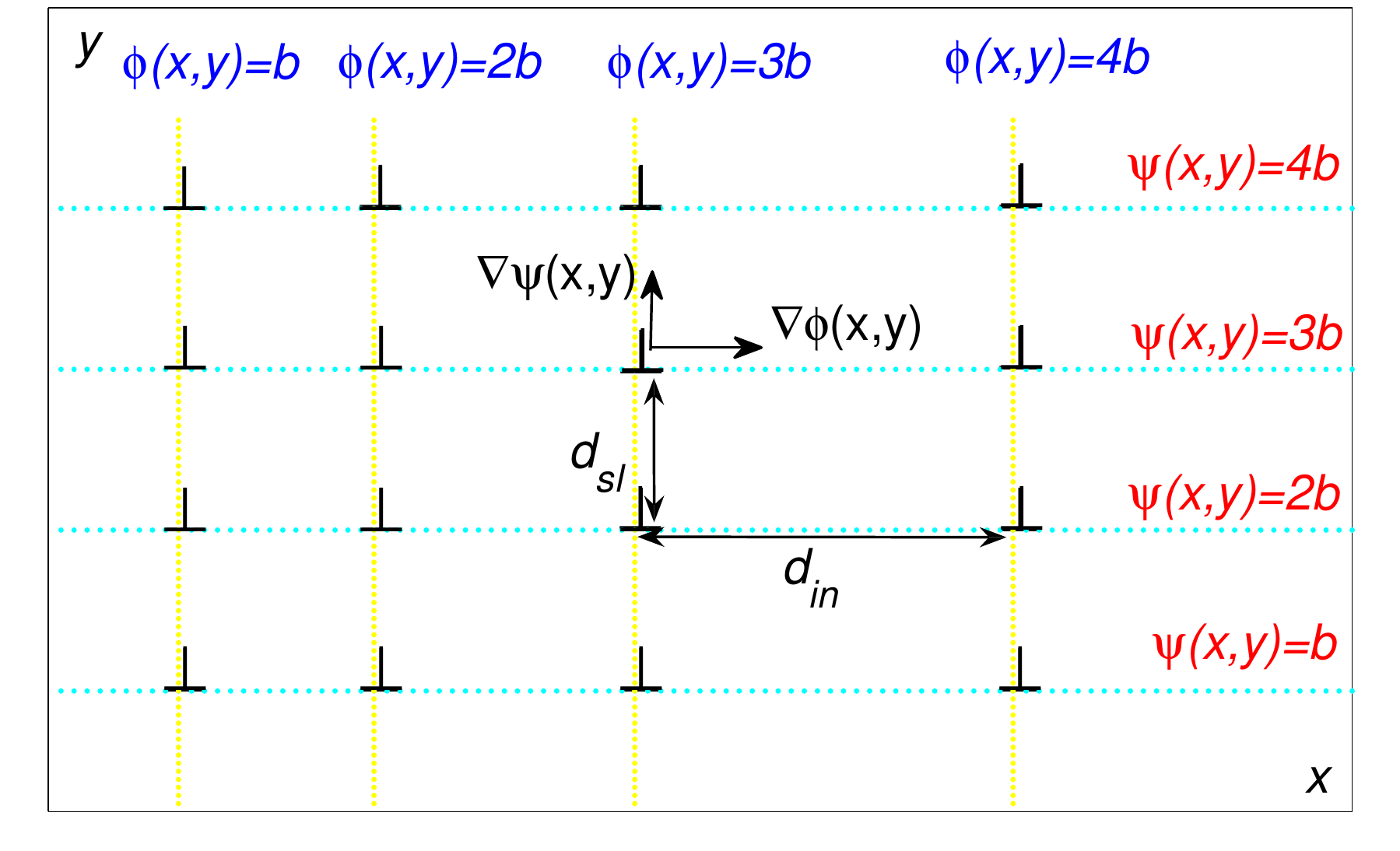}\\
  \caption{Representation of dislocation ensembles by the dislocation density potential functions (DDPFs). Given a smoothly varying dislocation structure, the contour line of one DDPF $\psi$ coincide with the slip planes, while the dislocation lines within a slip plane are given by the contour lines of another DDPF $\phi$ restricted on that plane. The local average active slip plane spacing $d_{\rm sl}$ and the local dislocation spacing within a slip plane $d_{\rm in}$ are given by Eqs.~(\ref{eqn:dsl}) and (\ref{eqn:din}), respectively. Note that in general $\nabla\phi$ is not necessarily normal to $\nabla\psi$.}\label{fig:case1-density}
\end{figure}

To represent the resulting dislocation continuum, we employ a pair of dislocation density potential functions (DDPFs) \cite{Zhu_continuum3D} $\phi(x,y)$ and $\psi(x,y)$, such that, for this two-dimensional problem,
the intersection of the contour lines
\begin{equation}
\phi(x,y)=ib \ \ {\rm and}\ \  \psi(x,y)=jb,
\end{equation}
$i,j=0,\pm1,\pm2,\cdots$, are the dislocation lines, see Fig.~\ref{fig:case1-density}. Given a smoothly varying dislocation structure, the local slip planes are represented by the contour lines of the DDPF $\psi$, while the dislocation lines within a slip plane are described locally by the contour lines of another DDPF $\phi$ restricted on that plane.

With this continuum representation of dislocation distributions,
the local dislocation line direction is determined from the DDPFs by
\begin{equation}
\label{eq:tau}
\boldsymbol{\tau}=\frac{\nabla \phi\times\nabla \psi}{\|\nabla \phi\times\nabla \psi\|}.
\end{equation}
The local normal direction of the dislocation slip plane is in the direction of $\nabla\psi$, and the local average active slip plane spacing is given by
\begin{equation}
\label{eqn:dsl}
d_{\rm sl}=\frac{b}{\|\nabla \psi\|}.
 \end{equation}
  Using the fact that the local dislocation line direction is in the direction of $\nabla\phi\times\nabla\psi$, it can be calculated that the local dislocation spacing within a slip plane is
 \begin{equation}
 \label{eqn:din}
d_{\rm in}=\frac{b\|\nabla \psi\|}{\|\nabla\phi\times\nabla\psi\|}.
\end{equation}
 (In fact, $d_{\rm in}=
b/ \mbox{length of}\ \nabla\phi \ \mbox{in the slip plane}$.)

For the two-dimensional problem considered in this paper,   the Nye dislocation density tensor is reduced to a scalar dislocation density function $\rho(x, y)$, which is the number of dislocations per unit area \cite{Nye1953,Kroener1963,Xiang2006}. (In fact, here the Nye dislocation density tensor $\pmb \alpha=\rho(x,y)\mathbf b\otimes \pmb \tau$, where $\mathbf b=(b,0,0)$ and  $\pmb \tau=(0,0,1)$ or $(0,0,-1)$.)
Here we define $\rho(x,y)$ to be the signed dislocation density: which is positive when the dislocations are in the $+z$ direction and negative when they are in the $-z$ direction.
Since the local dislocation number density in the DDPF framework is
$ \frac{1}{d_{\rm in}d_{\rm sl}}=\frac{\|\nabla \phi\times\nabla \psi\|}{b^2}$,
the signed dislocation density can then be written as
\begin{equation}\label{eqn:density}
    \rho(x,y)=\frac{1}{b^2}(\nabla \phi\times\nabla \psi\cdot\textbf{k}),
\end{equation}
where $\textbf{k}$ is the unit vector in the $+z$ direction.

For example,  consider the case when the distribution of dislocations is uniform in $y$ direction (normal to the slip plane) and nonuniform in $x$ direction (within the slip plane). The DDPFs that describe this dislocation distribution is  $\phi(x,y)=\phi(x)$ and  $\psi(x,y)=\frac{by}{D}$, where $D$ is the uniform active slip plane spacing. The dislocation density in this case is $\rho(x,y)=\frac{\phi'(x)}{bD}$.

For dislocation dynamics problems, the DDPFs $\phi$ and $\psi$ also depend on time $t$ and their evolution implicitly describes the dynamics of dislocations at the continuum level, which is
\begin{equation}\label{eq:evn-eqs1}
\left\{
\begin{array}{l}
\phi_t+\textbf{v}\cdot\nabla \phi=0,\\
\psi_t+\textbf{v}\cdot\nabla \psi=0,
\end{array}
\right.
\end{equation}
where $\textbf{v}=(v_{\text{g}}, v_{\text{c}})^T$ is the local dislocation velocity at the continuum level and is calculated  from the continuum Peach-Koehler force $\textbf{f}=(f_{\text{g}}, f_{\text{c}})^T$ following the mobility law in Eq.~\eqref{eq:velocity} in the two dimensional form. Here the continuum glide force $f_{\text{g}}$ and the continuum climb force $f_{\text{c}}$ are
\begin{eqnarray}
f_{\text{g}}=f_{\text{g}}^{\text{dc}}+(\pmb\tau\cdot {\mathbf k})b\sigma_{xy}^0,\label{eqn:fglide-tot}\\
f_{\text{c}}=f_{\text{c}}^{\text{dc}}-(\pmb\tau\cdot {\mathbf k})b\sigma_{xx}^0,
\end{eqnarray}
where $f_{\text{g}}^{\text{dc}}$ and $f_{\text{c}}^{\text{dc}}$ are the continuum glide and climb forces due to the stress field of dislocations, and the second term in each equation is the force due to the applied stress.

 The  leading order continuum Peach-Koehler force due to the long-range dislocation interaction is given  below  in terms of the DDPFs $\phi$ and $\psi$, using the dislocation density $\rho$ in Eq.~(\ref{eqn:density}):
 \begin{eqnarray}\label{eq:con-glide}\nonumber
f_{\text{g}}^{\text{dc,0}} (x,y)&=&\frac{\mu b^2}{2\pi(1-\nu)}\int^{+\infty}_{-\infty}\int^{+\infty}_{-\infty}\frac{(x-x_1)[(x-x_1)^2-(y-y_1)^2]}
{[(x-x_1)^2+(y-y_1)^2]^2}\rho(x_1,y_1)dx_1dy_1,\\
\end{eqnarray}
\begin{eqnarray}\label{eq:con-climb}\nonumber
f_{\text{c}}^{\text{dc,0}} (x,y)&=&\frac{\mu b^2}{2\pi(1-\nu)}\int^{+\infty}_{-\infty}\int^{+\infty}_{-\infty}\frac{(y-y_1)[3(x-x_1)^2+(y-y_1)^2]}
{[(x-x_1)^2+(y-y_1)^2]^2}\rho(x_1,y_1)dx_1dy_1.\\
\end{eqnarray}
These continuum long-range forces are obtained by straightforward averaging from the discrete dislocation dynamics model in Eqs.~(\ref{eq:glide-dd}) and (\ref{eq:climb-dd}) \cite{Nye1953,Kroener1963,Mura1987}.

While previous continuum model based on DDPFs focused on dislocation glide within slip planes \cite{Xiang2009_JMPS,ZhuXH2010,Zhu_continuum3D},
the continuum dislocation dynamics equations in Eq.~(\ref{eq:evn-eqs1}) incorporate both dislocation motions of glide and climb. Compared with the level set discrete dislocation dynamics method \cite{Xiang2003} in which only the intersection of the {\it zero} level sets of two level set functions is meaningful, the continuum dislocation dynamics equations of the two DDPFs hold {\it everywhere} in the simulation domain, i.e. the intersections of {\it all} the level set pairs of the two DDPFs are meaningful here. As all the existing continuum dislocation dynamics models reviewed in the introduction, our continuum model is able to describe smoothly varying dislocation structures and holds in an averaged sense for general dislocation structures by homogenizing the discrete dislocations within some representative volumes centered at each point \cite{Zhu_continuum3D}.

As to be discussed  in Sec.~\ref{sec:long-rang}, it is essential to include in continuum Peach-Koehler force the contributions due to short-range dislocation interactions, whose accurate expressions will be derived in the next few sections.

\section{Inconsistency between the continuum long-range force and the discrete dislocation model}
\label{sec:long-rang}

 We observe that the continuum Peach-Koehler forces based on the long-range dislocation interaction in Eqs.~\eqref{eq:con-glide} and \eqref{eq:con-climb} are not always consistent with the forces from the discrete dislocation dynamics model, especially when the long-range dislocation interaction vanishes. For example, when the distribution of dislocations is uniform in the $y$ direction, the dislocation density only depends on the spatial variable $x$, i.e.  $\rho(x,y)=\rho(x)$.
Substituting this density into Eq.~(\ref{eq:con-glide}) and using $\int^{+\infty}_{-\infty}\frac{(x^2-y^2)}{(x^2+y^2)^2}dy=0$, we have
\begin{eqnarray}\nonumber
f_{\text{g}}^{\text{dc,0}} (x,y)&=&\frac{\mu b^2}{2\pi(1-\nu)}\int^{+\infty}_{-\infty}\int^{+\infty}_{-\infty}\frac{(x-x_1)[(x-x_1)^2-(y-y_1)^2]}{[(x-x_1)^2+(y-y_1)^2]^2}\rho(x_1)dx_1dy_1\\ \nonumber
 &=& \frac{\mu b^2}{2\pi(1-\nu)}\int^{+\infty}_{-\infty}(x-x_1)\rho(x_1)dx_1\int^{+\infty}_{-\infty}
 \frac{[(x-x_1)^2-(y-y_1)^2]}{[(x-x_1)^2+(y-y_1)^2]^2}dy_1\vspace{1ex}\\
 &=&0.
\end{eqnarray}
Thus the continuum glide force in  Eq.~(\ref{eq:con-glide}) vanishes for this case.

We then calculate the glide force for this case using the discrete dislocation dynamics model.
 Since the distribution of dislocations is uniform in the $y$ direction, the locations of dislocations can be written as $\{(x_m,y_{n_0}+jD)|m,j=0,\pm 1,\pm2,\cdots\}$, where $D$ is the uniform inter-dislocation spacing in the $y$ direction.
  On the dislocation located at $(x_{m_0},y_{n_0})$, the
  glide force calculated from the discrete dislocation dynamics formula in Eq.~(\ref{eq:glide-dd}) is
\begin{eqnarray}\label{eq:f-g-y}\nonumber
f_{\text{g}}^{\text{dd}} (x_{m_0},y_{n_0})
&=&\frac{\mu b^2}{2\pi(1-\nu)} \sum_{m} \sum_{j=-\infty}^{+\infty}{\frac{(x_{m_0}-x_{m})[(x_{m_0}-x_{m})^2-(jD)^2] }{[(x_{m_0}-x_{m})^2+(jD)^2]^2}}\\
&=&\frac{\mu \pi b}{(1-\nu)D^2}\sum_{m\neq m_0} \frac{x_{m_0}-x_{m}}{\cosh(2\pi \frac{x_{m_0}-x_{m}}{D})-1}.
\end{eqnarray}
 This glide force in general is nonzero. This disagreement shows  that in the continuum model, in addition to the leading order contribution from the long-range dislocation interaction, it is essential to incorporate short-range dislocation interactions at higher orders in the coarse-graining process from the discrete dislocation dynamics model.

In this paper, we will derive continuum formulas for these short-range dislocation interactions. We first identify all the cases in which the glide or climb force due to the long-range dislocation interaction vanishes.
The long-range forces are easily calculated in the Fourier space, in which the force formulas in Eq.~(\ref{eq:con-glide}) and (\ref{eq:con-climb}) become
  \begin{equation}\label{eq:Fourier-glide-con}
\hat{f}_{\text{g}}^{\text{dc,0}} (k_1, k_2)=4\pi^2b \hat{G_1}(k_1, k_2)\hat{\rho}(k_1, k_2)=-\frac{2\mu b^2}{1-\nu}\frac{\mathrm{i}k_1k_2^2}{(k_1^2+k_2^2)^2}\hat{\rho}(k_1, k_2),
\end{equation}
\begin{equation}\label{eq:Fourier-climb-con}
\hat{f}_{\text{c}}^{\text{dc,0}} (k_1, k_2)= 4\/\pi^2b\hat{G_2}(k_1,k_2)\hat{\rho}(k_1, k_2)=-\frac{2\mu b^2}{1-\nu}\frac{\mathrm{i}k_2^3}{(k_1^2+k_2^2)^2}\hat{\rho}(k_1,k_2),
\end{equation}
where $\hat{f}$ is the Fourier coefficient of $f$ of the component $e^{\mathrm{i}(k_1x+k_2y)}$, $\mathrm{i}$ is the imaginary unit  and $k_1, k_2$ are the wave numbers. Recall that the functions $G_1(x,y)$ and $G_2(x,y)$ are defined in Eqs.~\eqref{eqn:g1} and \eqref{eqn:g2}.

 {\bf (i) The long-range glide force vanishes, i.e. $f_{\text{g}}^{\text{dc,0}}=0$.}   This is equivalent to $\hat{f}_{\text{g}}^{\text{dc,0}} (k_1, k_2)=0$ for any $k_1$ and $k_2$.
  Following  Eq.~(\ref{eq:Fourier-glide-con}), if  $\hat{f}_{\text{g}}^{\text{dc,0}} (k_1, k_2)=0$, at least one of the following three conditions holds for any fixed $k_1, k_2$:   $k_1 = 0$ but $ k_2\neq0$; $ k_2 = 0$ but $ k_1\neq0$; or $\hat{\rho}(k_1,k_2)=0$  if  $k_1, k_2\neq 0$. Thus all the solutions of $f_{\text{g}}^{\text{dc,0}}=0$   are given by
 \begin{eqnarray}\nonumber
 \rho(x, y)&=&\sum_{k_1}\sum_{k_2}\hat{\rho}(k_1, k_2)e^{\mathrm{i}(k_1x+k_2y)}\\ \nonumber
 &=&\sum_{k_1\neq 0}\hat{\rho}(k_1, 0)e^{\mathrm{i}k_1x}+\sum_{k_2\neq 0}\hat{\rho}(0, k_2)e^{\mathrm{i}k_2y}\\
 &=&\rho_1(x)+\rho_2(y),\label{eqn:vanish1}
 \end{eqnarray}
 where $\rho_1(x)$ and $\rho_2(y)$ are some functions.

{\bf (ii) The long-range climb force vanishes, i.e.  $f_{\text{c}}^{\text{dc,0}}=0$}. This is equivalent to $\hat{f}_{\text{c}}^{\text{dc,0}} (k_1, k_2)=0$ for any $k_1$ and $k_2$.  Following Eq.~(\ref{eq:Fourier-climb-con}), if  $\hat{f}_{\text{c}}^{\text{dc,0}} (k_1, k_2)=0$, at least one of the following two conditions holds for any fixed $k_1, k_2$: $ k_2 = 0$ but $k_1\neq0$; or $\hat{\rho}(k_1,k_2)=0$ if $k_2\neq 0$. Thus all the solutions of $f_{\text{c}}^{\text{dc,0}}=0$   are given by
  \begin{eqnarray}\nonumber
 \rho(x, y)&=&\sum_{k_1}\sum_{k_2}\hat{\rho}(k_1, k_2)e^{\mathrm{i}(k_1x+k_2y)}\\ \nonumber
 &=&\sum_{k_1\neq 0}\hat{\rho}(k_1, 0)e^{\mathrm{i}k_1x} \\
 &=&\rho_3(x) ,\label{eqn:vanish2}
 \end{eqnarray}
 where $\rho_3(x)$ is some function.

In summary, the long-range glide force in the continuum model vanishes if and only if the dislocation density has the form  $\rho(x,y)=\rho_1(x)+\rho_2(y)$, and the long-range climb force in the continuum model vanishes if and only if the dislocation density has the form  $\rho(x,y)=\rho_3(x)$ (which means that
 the dislocation distribution is uniform in the $y$ direction). However, the forces calculated from the discrete dislocation dynamics model are not necessarily zero, see the example in Eq.~(\ref{eq:f-g-y}).
 In these cases, it is essential to keep the next order forces that represent the short-range dislocation interaction due to the discreteness of dislocation distributions, in the coarse-graining process from the discrete dislocation dynamics model.
 In the next section, we examine these cases and derive continuum force expressions  to capture such short-range interactions of dislocations.

\section{Continuum force formulation due to short-range dislocation interactions}\label{sec:continuum-short}

In this section, we  derive continuum expressions for the dislocation short-range interactions from the discrete dislocation dynamics model.
We focus on
the dislocation configurations identified in Sec.~\ref{sec:long-rang} where
the continuum long-range force fails to provide stabilizing effect compared with the discrete model. These dislocation distributions  are uniform either within the slip planes (in the $x$ direction) or in the direction normal to the slip planes (in the $y$ direction), i.e.,
 \begin{equation}\label{eqn:linear_rho}
 \rho=\rho(x) \ {\rm or}\  \rho(y).
 \end{equation}
We consider the dislocation configurations that are not far from a unform distribution (i.e. in the linear regime of the deviations). The perturbations are small in the sense of the maximum norm. We neglect the force due to applied stress in this section.

\begin{figure}[htbp]  \centering
\subfigure[Case 1]{\label{fig:case1}\includegraphics[width=2.5in]{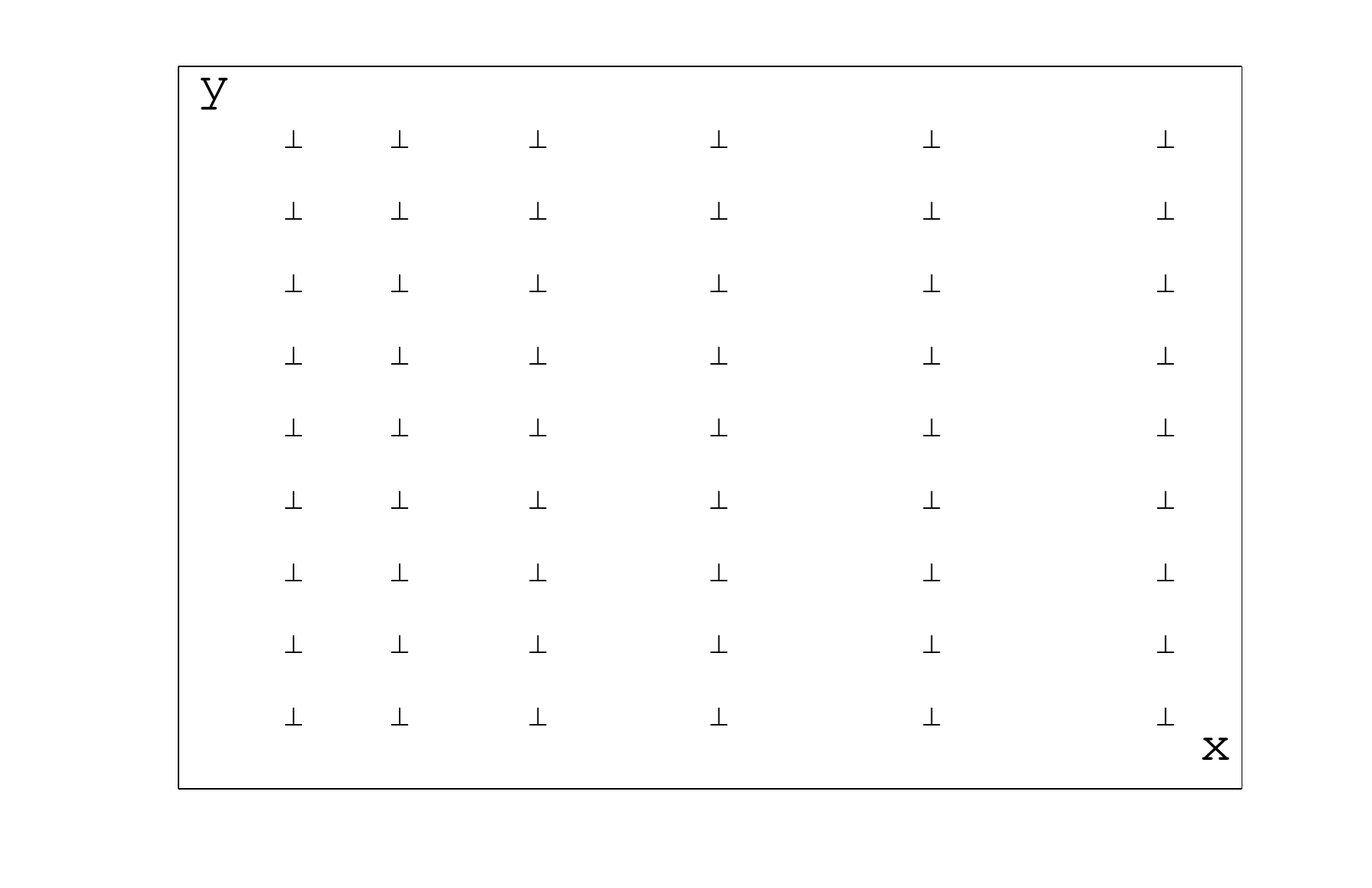}}
\subfigure[Case 2]{\label{fig:case2}\includegraphics[width=2.5in]{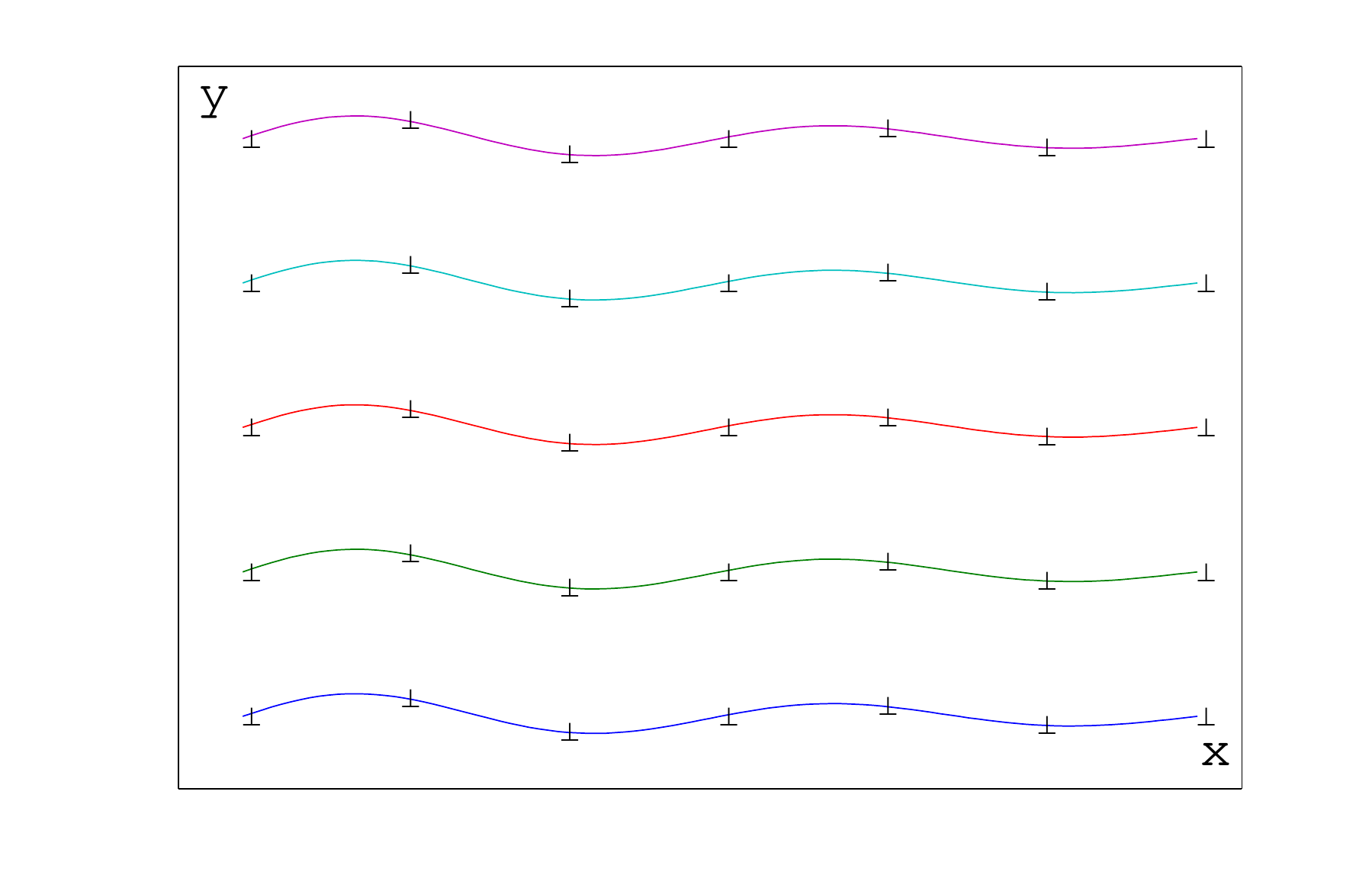}}
\subfigure[Case 3]{\label{fig:case3}\includegraphics[width=2.5in]{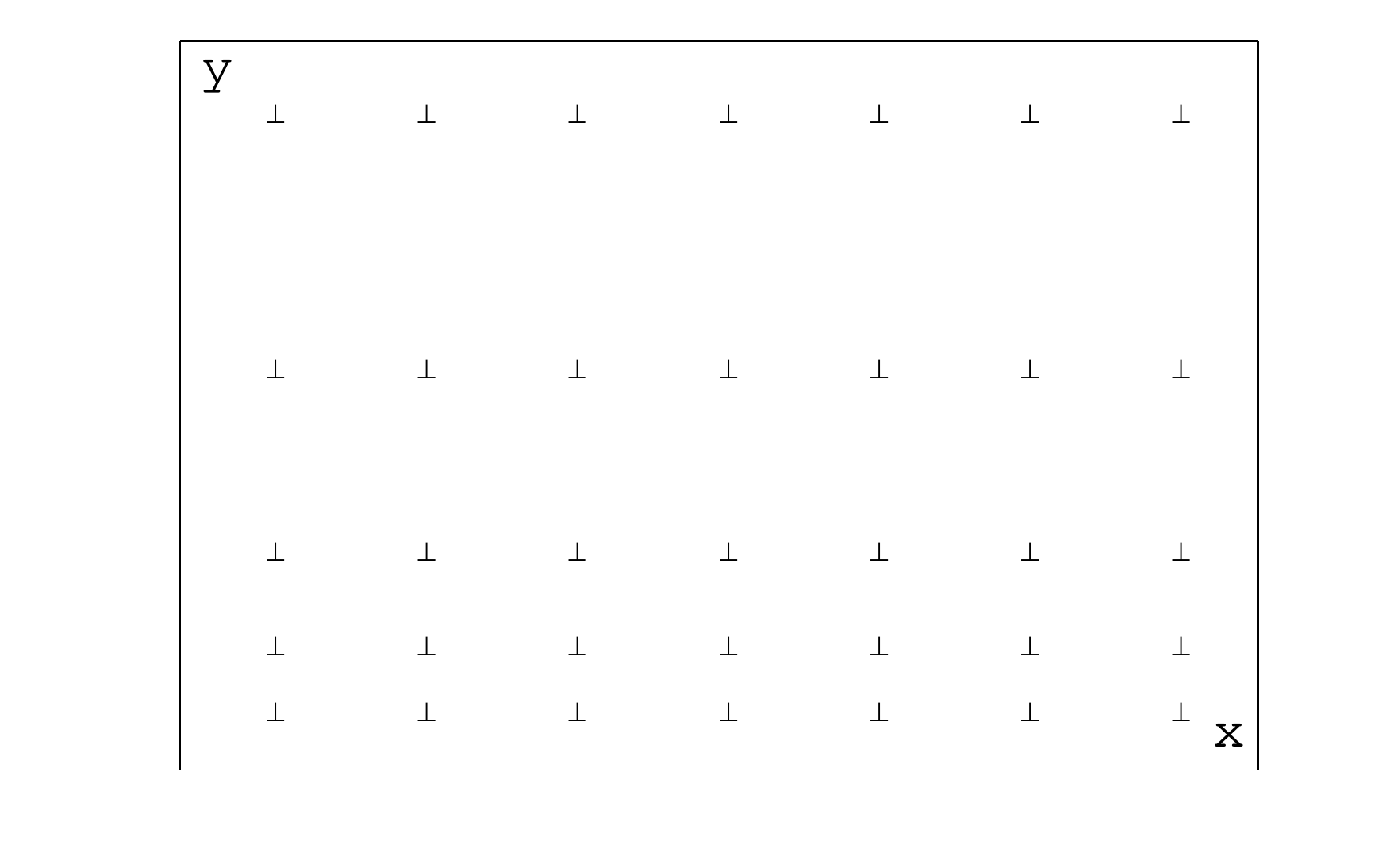}}
\subfigure[Case 4]{\label{fig:case4}\includegraphics[width=2.5in]{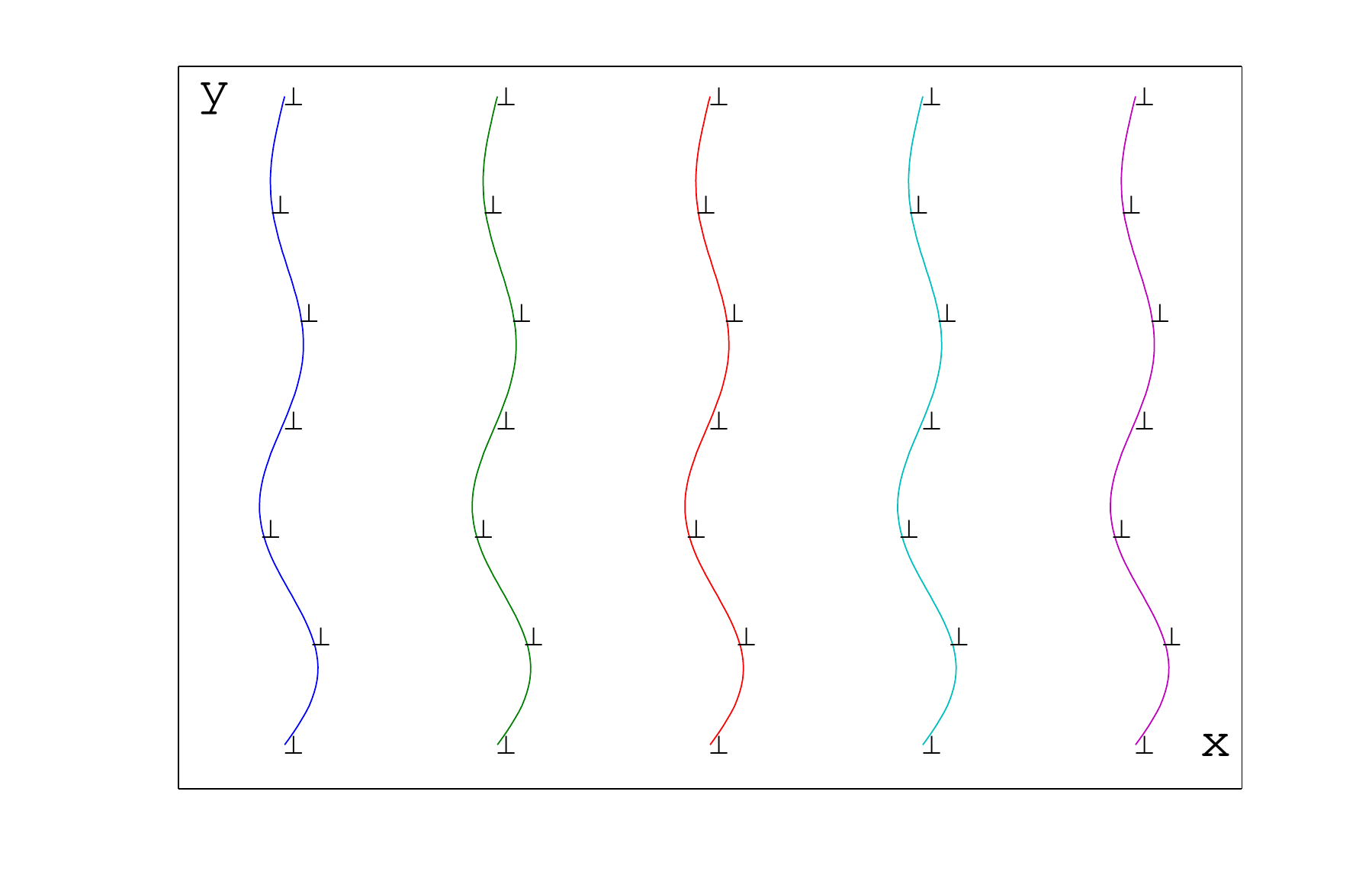}}
 \caption{Four cases of dislocation distributions with vanishing glide or climb force due to the long-range dislocation interaction.  Case 1: $\phi=\phi(x)$, $\psi=by/D$. Case 2: $\phi=bx/B$, $\psi=by/D+\tilde{\psi}(x)$.
  Case 3:  $\phi=bx/B$, $\psi=\psi(y)$. Case 4: $\phi=bx/B+\tilde{\phi}(y)$, $\psi=by/D$. See the text for the description of each case.}  \label{fig:case1-4} %% label for entire figure
 \end{figure}

 Using the representation of DDPFs described in Sec.~\ref{sec:ddpf}, such a perturbed uniform dislocation wall can be described by
 \begin{equation}
 \phi=\frac{b}{B}x+\tilde{\phi}, \ \ \ \ \psi=\frac{b}{D}y+\tilde{\psi},
 \end{equation}
where $B$ is the inter-dislocation spacing in a slip plane and $D$ is the inter-slip plane spacing in the uniform dislocation wall. From the formula of $\rho$ in Eq.~(\ref{eqn:density}), it is easy to show that Eq.~(\ref{eqn:linear_rho}) holds under the following necessary condition in the linear regime that the perturbations in a DDPF $\phi$ or $\psi$ are either functions of $x$ or $y$, i.e.
 \begin{equation}
 \left\{
 \begin{array}{l}
 \tilde{\phi}=\tilde{\phi}(x) \ {\rm or}\ \tilde{\phi}(y)\vspace{1ex}\\
 \tilde{\psi}=\tilde{\psi}(x) \ {\rm or}\ \tilde{\psi}(y).
 \end{array}
 \right.
 \end{equation}
 These dislocation configurations can be summarized into four cases as shown in  Fig.~\ref{fig:case1-4}.

 In Case 1, the dislocation distribution is uniform in the direction normal to the slip planes, but nonuniform in a slip plane. This dislocation structure can be described using DDPFs $\phi$ and $\psi$ as $\phi=\phi(x)$, $\psi=by/D$.
 In Case 2, each row of dislocations has a small perturbation in the direction normal to the slip planes,
   and the perturbations are uniform in the direction normal to the slip planes. This dislocation structure is given by $\phi=bx/B$, $\psi=by/D+\tilde{\psi}(x)$, where $\tilde{\psi}(x)$ is some function.
    In Case 3, the dislocation distribution is uniform in any slip plane, but nonuniform in the direction normal to the  slip planes.  This dislocation structure is given by $\phi=bx/B$, $\psi=\psi(y)$.
   Finally, in Case 4, each column of dislocations has a small perturbation, and the perturbations are uniform for all the columns of dislocations.  This dislocation structure is given by $\phi=bx/B+\tilde{\phi}(y)$, $\psi=by/D$.

We then derive for each of these four cases a continuum formula of the short-range dislocation interaction force from the discrete dislocation dynamics model reviewed in Sec.~\ref{sec:ddd}. In this discrete to continuum process, we employ asymptotic analysis under the assumption  that $L>>B,D$ where $L$ is the length unit of the continuum model. This means that there are a large number of dislocations contained in a unit area of the domain of the continuum model.
Note that in this limit process, $b/B$ and $b/D$ are fixed finite (small) numbers, and $B$ and $D$ are greater than a few multiples of the Burgers vector length $b$ such that the core regions of different dislocations are not overlapped.

 Note that although we use linear assumption, the obtained continuum model still holds for configurations significantly deviated from the uninform distributions. See the numerical examples in Sec.~\ref{sec:numerical}.

\subsection{Case 1}

 The structure of dislocations in this case is shown schematically in Fig.~\ref{fig:case1}, which
 is uniform in the direction normal to the slip planes (in the $y$ direction), but nonuniform in a slip plane (in the $x$ direction).  This dislocation structure is described by
 \begin{equation}\label{eqn:case1}
 \phi=\phi(x)=\frac{b}{B}x+\tilde{\phi}(x), \ \ \psi=\psi(y)=\frac{b}{D}y,
 \end{equation}
 where $\tilde{\phi}(x)$ is some small perturbation such that $\tilde{\phi}(x)<<b$ and $\phi'(x)>0$.
Using Eq.~(\ref{eqn:density}), the dislocation density $\rho=\rho(x)=\frac{1}{D}(\frac{1}{B}+\frac{\tilde{\phi}'(x)}{b})$, and accordingly, the continuum Peach-Koehler force due to the long-range dislocation interaction vanishes as shown in Sec.~\ref{sec:long-rang}.  We will derive  a continuum formula of the short-range dislocation interaction force from the discrete dislocation dynamics model.

 We first consider the glide force. In this case, the discrete dislocation dynamics model in Eq.~(\ref{eq:glide-dd}) gives the following expression for the glide force on the dislocation located at $(x_m,y_n=nD)$:
\begin{eqnarray}\label{eq:case1-dis}
f_{\text{g}}^{\rm dd}(x_m,y_n)&=&\frac{\mu b^2}{2\pi(1-\nu)}\sum_{j\neq m} \sum_{k=-\infty}^{+\infty}
{\frac{(x_m-x_j)[(x_m-x_j)^2-(kD)^2]}{[(x_m-x_j)^2+(kD)^2]^2}}\nonumber\\
&=&\frac{\pi \mu b^2}{(1-\nu)D^2}\sum_{j\neq m}\frac{x_m-x_j}{\cosh 2\pi \frac{x_m-x_j}{D}-1}\nonumber\\
&=&\frac{\pi \mu b^2}{(1-\nu)D^2}\sum_{j=1}^{+\infty}\left(\frac{x_m-x_{m+j}}{\cosh 2\pi \frac{x_m-x_{m+j}}{D}-1}+\frac{x_m-x_{m-j}}{\cosh 2\pi \frac{x_m-x_{m-j}}{D}-1}\right).
\end{eqnarray}

 We will derive a continuum expression from Eq.~\eqref{eq:case1-dis} in the limit of the length unit of the continuum model $L>>B$, $D$ and $b$.   The continuum expression will be based on the DDPF $\phi(x)$ in Eq.~(\ref{eqn:case1}) such that $\phi(x_m)=mb$, $m=0,\pm 1, \pm 2, \cdots$.
 We then have
\begin{equation}\label{eq:case1-x-phi-02}
x_m-x_{m+j}=-jB+\frac{B}{b}[\tilde{\phi}(x_{m+j})- \tilde{\phi}(x_{m})].
\end{equation}
 Using the assumption $\tilde{\phi}<<b$, we can make the following Taylor expansion at $x_m-x_{m+j}=-jB$:
\begin{equation}\label{eq:case1-x-phi-03}
{\textstyle \frac{x_m-x_{m+j}}{\cosh 2\pi \frac{x_m-x_{m+j}}{D}-1}=\frac{-jB}{\cosh 2\pi \frac{jB}{D}-1}+\frac{B}{b}\cdot\frac{\cosh 2\pi \frac{jB}{D}-1- 2\pi \frac{jB}{D}\sinh2\pi \frac{jB}{D}}{(\cosh 2\pi \frac{jB}{D}-1)^2}[\tilde{\phi}(x_{m+j})- \tilde{\phi}(x_{m})]+\cdots.} \end{equation}
We can then approximation the glide force in Eq.~(\ref{eq:case1-dis}) by
\begin{equation}\label{eq:case1-dis-02}
{\textstyle f_{\text{g}}^{\rm dd}(x_m,y_n)
\approx \frac{\pi \mu b^2}{(1-\nu)D^2}\sum_{j=1}^{+\infty} \frac{B}{b}\frac{\cosh 2\pi \frac{jB}{D}-1- 2\pi \frac{jB}{D}\sinh2\pi \frac{jB}{D}}{(\cosh 2\pi \frac{jB}{D}-1)^2}[\tilde{\phi}(x_{m-j})+\tilde{\phi}(x_{m+j})- 2\tilde{\phi}(x_{m})].}
\end{equation}

 Following Eq.~(\ref{eq:case1-x-phi-02}), we have
\begin{equation}
\tilde{\phi}(x_{m-j})+\tilde{\phi}(x_{m+j})- 2\tilde{\phi}(x_{m})
=\frac{b}{B}(2x_m-x_{m+j}-x_{m-j})
=-\frac{b}{B}(jb)^2x_{\phi\phi}
=\frac{b^3}{B}\frac{\phi_{xx}}{\phi^3_x}j^2.
\end{equation}
Note that since we have assumed  $\phi'(x)>0$, $x$ can also be considered as a function of $\phi$. Thus Eq.~(\ref{eq:case1-dis-02}) can be approximated by
\begin{eqnarray}\label{eq:case1-dis-03}
f_{\text{g}}^{\rm dd}(x_m,y_n) \nonumber
&\approx &\frac{\pi \mu b^4}{(1-\nu)D^2}\frac{\phi_{xx}}{\phi^3_x}\sum_{j=1}^{+\infty}  \frac{\cosh 2\pi \frac{jB}{D}-1- 2\pi \frac{jB}{D}\sinh2\pi \frac{jB}{D}}{(\cosh 2\pi \frac{jB}{D}-1)^2}j^2 \\
&=&\frac{\pi \mu bB}{1-\nu}\phi_{xx}\sum_{j=1}^{+\infty}  \frac{[\cosh 2\pi \frac{jB}{D}-1- 2\pi \frac{jB}{D}\sinh2\pi \frac{jB}{D}](\frac{jB}{D})^2}{(\cosh 2\pi \frac{jB}{D}-1)^2},\nonumber\\
&=&-\frac{\pi \mu bD}{1-\nu}g_1\left(\frac{B}{D}\right)\phi_{xx},
\end{eqnarray}
where the function $g_1(s)$ is defined as
\begin{equation}
\label{eq:fg1}
g_1(s)=\sum_{j=1}^{+\infty}  \frac{[2\pi js\sinh(2\pi js)-\cosh(2\pi js)+1](js)^2s}{[\cosh(2\pi js)-1]^2}.
\end{equation}

In the continuum model, it would be more convenient to have a simple formula for the coefficient instead of the summation in Eq.~\eqref{eq:fg1}. Obtaining analytical formula for such a summation is difficult. In the following,  we will derive an approximate formula for it.

 First, when $B/D$ is small, the summation in Eq.~(\ref{eq:fg1}) can be considered as an approximation to some integral with $\Delta x=B/D$ as follows
 \begin{eqnarray}
 \label{eq:fg1_app}
g_1\left(\frac{B}{D}\right)&=&\frac{1}{2}\sum_{j\neq 0} \frac{[2\pi \frac{jB}{D}\sinh2\pi \frac{jB}{D}-\cosh 2\pi \frac{jB}{D}+1](\frac{jB}{D})^2}{(\cosh 2\pi \frac{jB}{D}-1)^2}\cdot\frac{B}{D}\nonumber\\
&\approx&\frac{1}{2}\left[\int^{+\infty}_{-\infty} \frac{(2\pi x\sinh2\pi x-\cosh 2\pi x+1)x^2}{(\cosh 2\pi x-1)^2}dx\right.\nonumber\\
&&-\left.\lim_{x\rightarrow 0}\frac{(2\pi x\sinh2\pi x-\cosh 2\pi x+1)x^2}{(\cosh 2\pi x-1)^2}\cdot \frac{B}{D}\right]\nonumber\\
&=&\frac{1}{2}\left(\int^{+\infty}_{-\infty} \frac{2\pi x^3\sinh2\pi x}{(\cosh 2\pi x-1)^2}dx
-\int^{+\infty}_{-\infty} \frac{x^2}{\cosh 2\pi x-1}dx
-\frac{1}{2\pi^2}\frac{B}{D}\right)\nonumber\\
&=&\frac{1}{2}\left(\int^{+\infty}_{-\infty} \frac{2x^2}{\cosh 2\pi x-1}dx
-\frac{1}{2\pi^2}\frac{B}{D}\right)\nonumber\\
&=&\frac{1}{6\pi}-\frac{1}{4\pi^2}\frac{B}{D}.
\end{eqnarray}
Note that in these calculations, the approximation from the summation in the first line to the integral in the second line is based on the trapezoidal rule and the fact that the integrand decays exponentially as $x\rightarrow\pm\infty$.
Thus by Eqs.~\eqref{eq:case1-dis-03}--\eqref{eq:fg1_app},   we have the following continuum approximation of the glide force on the dislocation
\begin{equation}\label{eq:case1-result3}
f_{\text{g}}^{\rm dc}=-\frac{\mu b^2}{6(1-\nu)|\psi_y|}\left( 1-\frac{3}{2\pi}\frac{|\psi_y|}{|\phi_x|}\right)\phi_{xx}.
\end{equation}
Here we have used $\frac{b}{B}\approx|\phi_x|$ and $\frac{b}{D}=|\psi_y|$ by Eq.~(\ref{eqn:case1}).

\begin{figure}[htbp]
\centering
\includegraphics[width=4.5in]{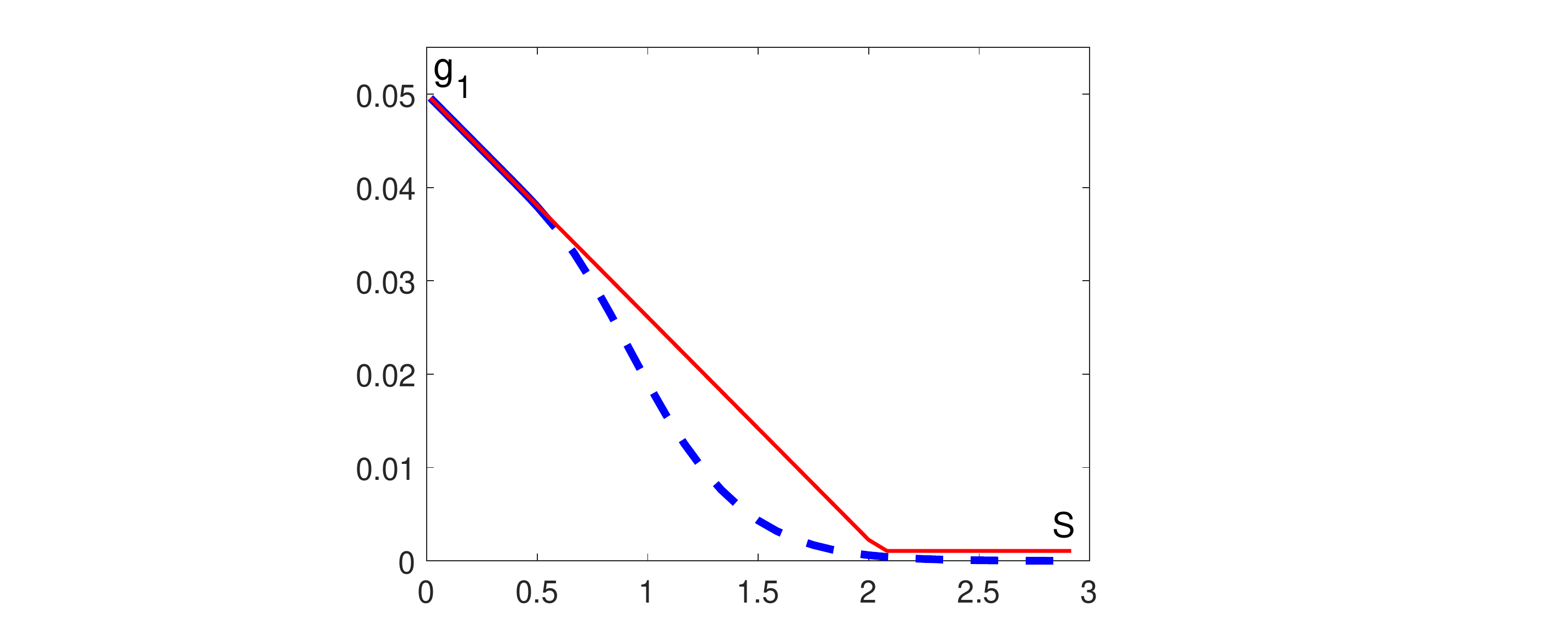}
\caption{Comparison of the approximation of the function $g_1(s)$ in Eq.~\eqref{eq:truncationg1} (the red piecewise linear curve) and its exact formula in Eq.~\eqref{eq:fg1} (the blue dash curve, calculated numerically) for  different values of $s$, where $\varepsilon=0.02$.}
\label{fig:case1-01}
\end{figure}

Note that the above approximation holds when $B/D$ is small. When $B/D$ is large, all the terms in the summation in $g_1$ are exponentially small controlled by $e^{-\frac{B}{D}}$, and accordingly $g_1$ is exponentially small. On the other hand, there is an important property that  $g_1> 0$ always holds. Thus when $B/D$ is large, we use $\varepsilon/(6\pi)$ to approximate $g_1$, where $\varepsilon$ is some small positive constant.
That is,
\begin{equation}\label{eq:truncationg1}
g_1(s)\approx\left\{
  \begin{array}{ll}
    \frac{1}{6\pi}-\frac{s}{4\pi^2}, & \hbox{if $1-\frac{3}{2\pi}s>\varepsilon$}; \vspace{1ex}\\
    \frac{\varepsilon}{6\pi}, & \hbox{otherwise for}\ s\geq 0.
  \end{array}
\right.
\end{equation}
Fig.~\ref{fig:case1-01} shows good match between the results from the approximation of the function $g_1(s)$ and its exact formula in Eq.~\eqref{eq:fg1} for different values of $s$.

Using the approximations in the two regimes obtained above, we have  the following continuum approximation of the glide force on the dislocation for all values of $B/D$:
\begin{equation}\label{eq:case1-result3new}
f_{\text{g}}^{\rm dc}=-\frac{\mu b^2}{6(1-\nu)|\psi_y|}\left[ 1-\frac{3}{2\pi}\frac{|\psi_y|}{|\phi_x|}\right]_{\varepsilon +}\phi_{xx},
\end{equation}
where the notation $[\cdot]_{\varepsilon +}$ is defined as
\begin{equation}\label{eq:truncation}
[h]_{\varepsilon +}=\left\{
  \begin{array}{ll}
    h, & \hbox{if $h>\varepsilon$;} \\
    \varepsilon, & \hbox{if $h\leq\varepsilon$.}
  \end{array}
\right.
\end{equation}
We would like to remark that in addition to its accuracy,  the form of the continuum force formula in Eq.~\eqref{eq:case1-result3new} is also essential to maintain the strict stability of the evolution equations, see  Eq.~\eqref{eq:case1-evn3}.

 Note that when the line direction of these dislocations changes to $\boldsymbol{\tau}=( 0,0,-1)$, we may have
 $\phi_x<0$, and this case can be included by modifying the continuum glide force in Eq.~(\ref{eq:case1-result3new}) as
 \begin{equation}\label{eq:case1-result5}
f_{\text{g}}^{\rm dc}=-{\rm sgn}(\phi_x)\frac{\mu b^2}{6(1-\nu)|\psi_y|}\left[ 1-\frac{3}{2\pi}\frac{|\psi_y|}{|\phi_x|}\right]_{\varepsilon +}\phi_{xx},
\end{equation}
where the function ${\rm sgn}(s)$ gives the sign of $s$. This continuum expression does not depend on the sign of $\psi_y$.

Next we derive continuum expression of the climb force for this case. On the dislocation at $(x_m,y_n)$, the climb force  from the discrete dislocation dynamics model in Eq.~(\ref{eq:climb-dd}) is
\begin{equation}\label{eq:case1-dis-climb}
f_{\text{c}}^{\rm dd}(x_m,y_n)=\frac{\mu b^2}{2\pi(1-\nu)}\sum_{j\neq m} \sum_{k=-\infty}^{+\infty}
{\frac{(0-kD)(3(x_m-x_j))^2+(0-kD)^2)}{[(x_m-x_j)^2+(0-kD)^2]^2}}
=0.
\end{equation}
Thus the continuum expression of the climb force in this case is
\begin{equation}\label{eq:case1-result4}
f_{\text{c}}^{\rm dc}\equiv 0.
\end{equation}

Substituting the continuum expressions of $f_{\text{g}}^{\rm dc}$ and $f_{\text{c}}^{\rm dc}$ in Eqs.~(\ref{eq:case1-result5}) and  (\ref{eq:case1-result4}) into the evolution equation of $\phi$ in (\ref{eq:evn-eqs1}), with the mobility law in Eq.~\eqref{eq:velocity},
the final form of the evolution equation for Case 1 is
%\begin{eqnarray}\label{eq:case1-evn3}
%\phi_t+ m_{\text{g}}b\sigma_{xy}^0|\phi_x|- \frac{m_{\text{g}}\mu b^2}{6(1-\nu)}\frac{ |\phi_x|}{|\psi_y| }\left[ 1 -\frac{3}{2\pi}\frac{|\psi_y|}{|\phi_x|}\right]_{\varepsilon+}\phi_{xx}=0.
%\end{eqnarray}
\begin{eqnarray}\label{eq:case1-evn3}
\phi_t- \frac{m_{\text{g}}\mu b^2}{6(1-\nu)}\frac{ |\phi_x|}{|\psi_y| }\left[ 1 -\frac{3}{2\pi}\frac{|\psi_y|}{|\phi_x|}\right]_{\varepsilon+}\phi_{xx}=0.
\end{eqnarray}

\subsection{Case 2}

The structure of dislocations in this case is shown schematically in Fig.~\ref{fig:case2}.
Each row of dislocations has a small perturbation in the direction normal to the slip planes (in the $y$ direction),
   and the perturbations are uniform in the $y$ direction.
   This dislocation structure is described by
 \begin{equation}\label{eqn:case2}
 \phi=\frac{b}{B}x, \ \ \psi=\frac{b}{D}y+\tilde{\psi}(x),
 \end{equation}
where $\tilde{\psi}(x)$ is some small perturbation with $\tilde{\psi}(x)<<b$ and $Bb/D$.
The continuum Peach-Koehler force due to the long-range dislocation interaction vanishes as shown in Sec.~\ref{sec:long-rang}.

In the discrete model of this case, if we denote the locations of the dislocations on the $\psi=0$ row by
$(x_j=jB, y_j)$ for $j=0,\pm 1, \pm 2, \cdots$, i.e.,
\begin{equation}\label{eq:case2-y}
\frac{b}{D}y_j+\tilde{\psi}(x_j)=0,
 \end{equation}
  the glide force on the dislocation $(x_{m},y_{m})$ using Eq.~(\ref{eq:glide-dd}) is
 \begin{eqnarray}\label{eq:case2-dis-glide}
\nonumber
f_{\text{g}}^{\text{dd}}(x_{m},y_{m}) &=&\frac{\mu b^2}{2\pi(1-\nu)}\sum_{j\neq m}\sum_{k=-\infty}^{+\infty}
{\frac{(x_m-x_j)[(x_m-x_j)^2-(y_m-(y_j+kD))^2]}{[(x_m-x_j)^2+(y_m-(y_j+kD))^2]^2}}\\
 &=&\frac{\pi \mu b^2}{ (1-\nu)D^2}\sum_{j\neq m}\frac{ (x_m-x_j)[\cosh2\pi \frac{x_m-x_j}{D} \cos2\pi \frac{y_m-y_j}{D}-1]}{(\cosh 2\pi \frac{x_m-x_j}{D}-\cos 2\pi \frac{y_m-y_j}{D})^2}\nonumber\\
 &\approx&\frac{\pi \mu b^2}{ (1-\nu)D^2}\nonumber\\
 &&\cdot\sum_{j\neq m}{\textstyle \frac{ (x_m-x_j)\left[\cosh2\pi \frac{x_m-x_j}{D}-1-\left(\cosh2\pi \frac{x_m-x_j}{D}+2\right) \left(1-\cos2\pi \frac{y_m-y_j}{D}\right)
  \right]}{\left(\cosh 2\pi \frac{x_m-x_j}{D}-1\right)^2}}.\nonumber\\
 \end{eqnarray}
Here we have summed up the contributions from each column first. When $j=m$, the glide force on the dislocation $(x_{m},y_{m})$ imposed by the vertical dislocation array containing this dislocation itself is zero. The last approximation is obtained by Taylor expansions
using the fact that $\cosh 2\pi \frac{x_m-x_j}{D}-1>>1- \cos2\pi \frac{y_m-y_j}{D}$ for $j\neq m$,
which is due to $x_j=jB$ and $y_j<<D$ and $B$.  The latter can be derived from the assumption $\tilde{\psi}(x)<<b$ and $Bb/D$ and the definition of $y_j$ in Eq.~\eqref{eq:case2-y}.
%The relative error of this approximation is $O({\displaystyle \max_j}|y_j|/B)^2$.

Next we derive a continuum expression from the summation in Eq.~\eqref{eq:case2-dis-glide} when $B, D << L$,  the length unit of the continuum model. As in Eq.~\eqref{eq:case1-dis} in Case 1, the summation in Eq.~\eqref{eq:case2-dis-glide} can be written in a symmetric way as
 \begin{eqnarray}\label{eq:case2-dis-glide2}
\nonumber
f_{\text{g}}^{\text{dd}}(x_{m},y_{m})
 &\approx&\frac{\pi \mu b^2}{ (1-\nu)D^2}\\
 &&\cdot\sum_{j=1}^{+\infty}\left\{{\textstyle \frac{ (x_m-x_{m+j})\left[\cosh2\pi \frac{x_m-x_{m+j}}{D}-1-\left(\cosh2\pi \frac{x_m-x_{m+j}}{D}+2\right) \left(1-\cos2\pi \frac{y_m-y_{m+j}}{D}\right)
  \right]}{\left(\cosh 2\pi \frac{x_m-x_{m+j}}{D}-1\right)^2}}\right.\nonumber\\
 &&+\left.{\textstyle \frac{ (x_m-x_{m-j})\left[\cosh2\pi \frac{x_m-x_{m-j}}{D}-1-\left(\cosh2\pi \frac{x_m-x_{m-j}}{D}+2\right) \left(1-\cos2\pi \frac{y_m-y_{m-j}}{D}\right)
  \right]}{\left(\cosh 2\pi \frac{x_m-x_{m-j}}{D}-1\right)^2}}\right\}.\nonumber\\
 \end{eqnarray}
Using $x_j=jB$, Eq.~\eqref{eq:case2-y}, and the assumption $y_j<<D$, we can calculate as in Case 1 that
\begin{eqnarray}\label{eq:case2-dis-02}
f_{\text{g}}^{\rm dd}(x_m,y_m) &\approx &\frac{2\mu \pi^3}{ (1-\nu)D^2}\sum_{j=1}^{+\infty}jB\frac{\cosh 2\pi \frac{jB}{D}+2 }{(\cosh 2\pi \frac{jB}{D}-1)^2} \nonumber\\
&&\cdot [\tilde{\psi}(x_{m+j})- \tilde{\psi}(x_{m-j})][\tilde{\psi}(x_{m+j})- 2\tilde{\psi}(x_{m})+\tilde{\psi}(x_{m-j})]\nonumber\\
&\approx &\frac{4\mu\pi^3D^2 }{1-\nu}\tilde{\psi}_{xx}\tilde{\psi}_x \sum_{j=1}^{+\infty}\left(\frac{jB}{D}\right)^4\frac{\cosh 2\pi \frac{jB}{D}+2 }{(\cosh 2\pi \frac{jB}{D}-1)^2}\nonumber\\
&= &O(\tilde{\psi}^2)\nonumber\\
&\approx &0.
\end{eqnarray}
 Note that we only keep linear terms of the small perturbation $\tilde{\psi}$. Thus the continuum expression of the glide force in this case is
\begin{equation}\label{eq:case2-result4}
f_{\text{g}}^{\rm dc}\equiv 0.
\end{equation}

Next we will derive a continuum expression of the climb force in this case. The discrete expression given by Eq.~\eqref{eq:climb-dd} is
\begin{eqnarray}\label{eq:case2-dis-climb}
\nonumber
f_{\text{c}}^{\text{dd}}(x_{m},y_{m}) &=&\frac{\mu b^2}{2\pi(1-\nu)}\sum_{j\neq m}\sum_{k=-\infty}^{+\infty}
\frac{(y_m-(y_j+kD))[3(x_m-x_j)^2+(y_m-(y_j+kD))^2]}{[(x_m-x_j)^2+(y_m-(y_j+kD))^2]^2}\\ \nonumber
&=&\frac{\mu b^2}{ 2(1-\nu)D}\sum_{j\neq m}\frac{ \sin2\pi\frac{y_m-y_j}{D}}{(\cosh2\pi \frac{x_m-x_j}{D} -\cos2\pi \frac{y_m-y_j}{D})^2} \\\nonumber
&~&\cdot\left\{\cosh 2\pi \frac{x_m-x_j}{D}-\cos 2\pi \frac{y_m-y_j}{D}+2\pi \frac{x_m-x_j}{D} \sinh2\pi \frac{x_m-x_j}{D}\right\}\\
&\approx&\frac{\pi\mu b^2}{ (1-\nu)D^2}\sum_{j\neq m}\frac{(y_m-y_j)(\cosh 2\pi \frac{x_m-x_j}{D}-1+2\pi \frac{x_m-x_j}{D} \sinh2\pi \frac{x_m-x_j}{D})}{(\cosh2\pi \frac{x_m-x_j}{D} -1)^2}.\nonumber\\
\end{eqnarray}

Using the same method as before, Eq.~\eqref{eq:case2-dis-climb} can be approximated by
\begin{eqnarray}\label{eq:case2-con-climb}
f_{\text{c}}^{\text{dd}}(x_{m},y_{m})
\approx\frac{\pi \mu bD^2}{(1-\nu)B}\tilde{\psi}_{xx}\sum_{j=1}^{+\infty}  \frac{[\cosh 2\pi \frac{jB}{D}-1+ 2\pi \frac{jB}{D}\sinh2\pi \frac{jB}{D}](\frac{jB}{D})^2}{(\cosh 2\pi \frac{jB}{D}-1)^2}\cdot\frac{B}{D}.\nonumber\\
\end{eqnarray}
 Further using  $|\psi_y|=\frac{b}{D}$,  $|\phi_x|=\frac{b}{B}$, and taking into consider the dislocations in the opposite direction (i.e. $\psi_y<0$), as in Case 1, we have
\begin{eqnarray}\label{eq:case2-con-climb2}
f_{\text{c}}^{\text{dc}}={\rm sgn}(\psi_y)
\frac{\pi \mu b^2|\phi_x|}{(1-\nu)|\psi_y|^2}g_2\left(\frac{|\psi_y|}{|\phi_x|}\right)\psi_{xx},
\end{eqnarray}
where function $g_2$ is defined as $g_2(\frac{B}{D})=\sum_{j=1}^{+\infty}  \frac{[\cosh 2\pi \frac{jB}{D}-1+ 2\pi \frac{jB}{D}\sinh2\pi \frac{jB}{D}](\frac{jB}{D})^2}{(\cosh 2\pi \frac{jB}{D}-1)^2}\cdot\frac{B}{D}$.
 Substituting the obtained $f_{\text{g}}^{\text{dc}}$  and $f_{\text{c}}^{\text{dc}}$ into the evolution equation of $\psi$ in Eq.~\eqref{eq:evn-eqs1}, with the mobility law in Eq.~\eqref{eq:velocity}, the evolution equation of dislocations for this case can be written as
 $\psi_t+
\frac{\pi m_{\text{c}} \mu b^2|\phi_x|}{(1-\nu)|\psi_y|}g_2\left(\frac{|\psi_y|}{|\phi_x|}\right)\psi_{xx}=0$. It is easy to see that $g_2(s)>0$ for $s>0$. This means that the obtained evolution equation is not wellposed. In order to obtain a wellposed equation, we can keep higher order derivative terms in the continuum approximation, which will make the equation very complicated. Alternatively, we simply choose a simple regularization term of second order to ensure the wellposedness of the continuum model, which leads to the following evolution equation for Case 2:
%\begin{eqnarray}\label{eq:case2-evn-3}
%\psi_t-m_{\text{c}}b\sigma_{xx}^0|\psi_y|- \frac{m_{\text{c}}\mu b^2 }{6(1-\nu) }\varepsilon\psi_{xx} =0,
%\end{eqnarray}
\begin{eqnarray}\label{eq:case2-evn-3}
\psi_t- \frac{m_{\text{c}}\mu b^2 }{6(1-\nu) }\varepsilon\psi_{xx} =0,
\end{eqnarray}
where $\varepsilon>0$ is the same small parameter as that in Eq.~\eqref{eq:truncationg1}.

\subsection{Case 3}

The structure of dislocations in this case is shown schematically in  Fig.\ref{fig:case3}, which is uniform in each slip plane (in the $x$ direction), but slip planes of these dislocations are nonuniform (in the $y$ direction). This dislocation structure is described by
 \begin{equation}\label{eqn:case3}
\phi=\frac{b}{B}x, \ \ \psi=\frac{b}{D}y+\tilde{\psi}(y),
 \end{equation}
where $\tilde{\psi}(y)$ is some small perturbation such that $\psi'(y)>0$.

Using Eq.~(\ref{eqn:density}), the dislocation density in this case is
\begin{equation}\label{eqn:case3density}
\rho=\rho(y)=\frac{1}{B}\left(\frac{1}{D}+\frac{\tilde{\psi}'(y)}{b}\right).
\end{equation}
Based on the conclusions in Sec.~\ref{sec:long-rang} (Eqs.~(\ref{eqn:vanish1}) and (\ref{eqn:vanish2})),
 the continuum long-range glide force vanishes, whereas the continuum long-range climb force does not. Therefore, in this case,  the integral expression in Eq.~(\ref{eq:con-climb}) is able to give a nonvanishing leading order continuum approximation for the climb force, and we only need to derive a continuum formula for the short-range glide force.

Using the discrete model in Eq.~(\ref{eq:glide-dd}), the glide force on the dislocation located at
$(x_m=mB,y_n)$ in this case is
\begin{equation}\label{eq:case3-dis-glide}
f^{\text{dd}}_{\text{g}}(x_m,y_n)=
\frac{\mu b^2}{2\pi(1-\nu)}\sum_{k\neq n} \sum_{j=-\infty}^{+\infty}
{\frac{-jB[(jB)^2-(y_n-y_k)^2]}{[(jB)^2+(y_n-y_k)^2]^2}}
=0.
\end{equation}
This means that the glide force in this case indeed vanishes. Therefore, in this case,
\begin{equation}\label{eq:case3-result4}
f_{\text{g}}^{\rm dc}\equiv 0.
\end{equation}

{\bf Remark}: In this case, we have shown that the  integral expression in Eq.~(\ref{eq:con-climb}) is able to give a nonvanishing leading order continuum approximation for the climb force. It is interesting to note that this integral expression with the dislocation density $\rho$ in Eq.~(\ref{eqn:case3density}) in this case can be further simplified to a local expression: $f_{\text{c}}^{\text{dc,0}}=\frac{2\mu b}{(1-\nu)B}\tilde{\psi}(y)$, if the perturbation $\tilde{\psi}$ goes to zero at infinity.

\subsection
{Case 4}

The structure of dislocations in this case is shown schematically in Fig.~\ref{fig:case4}.
Each column of dislocations has a small perturbation in their own slip planes (in the $x$ direction),
   and the perturbations are uniform in the $x$ direction.
   This dislocation structure is described by
 \begin{equation}\label{eqn:case4}
 \phi=\frac{b}{B}x+\tilde{\phi}(y), \ \ \psi=\frac{b}{D}y,
 \end{equation}
where $\tilde{\phi}(y)$ is some small perturbation with $\tilde{\phi}(y)<<b$ and $Db/B$.
The continuum Peach-Koehler force due to the long-range dislocation interaction vanishes as shown in Sec.~\ref{sec:long-rang} because the scalar dislocation density calculated by Eq.~\eqref{eqn:density} is a constant.

In the discrete model of this case, we denote the locations of the dislocations on the $\phi=0$ column by
$(x_k, y_k=kD)$ for $k=0,\pm 1, \pm 2, \cdots$, i.e.,
\begin{equation}\label{eq:case4-x}
\frac{b}{B}x_k+\tilde{\phi}(y_k)=0.
 \end{equation}
  The glide force on the dislocation $(x_{n},y_{n})$ using Eq.~(\ref{eq:glide-dd}) is
\begin{eqnarray} \label{eq:case4-dis-glide-1}\nonumber
f^{\text{dd}}_{\text{g}}(x_{n},y_{n})&=&\frac{\mu b^2}{2\pi(1-\nu)}\sum_{k\neq n}\sum_{j=-\infty}^{+\infty}
{\frac{(x_n-(x_k+jB))[(x_n-(x_k+jB))^2-(y_n-y_k)^2]}{[(x_n-(x_k+jB))^2+(y_n-y_k)^2]^2}}\\ \nonumber
&=&\frac{\mu b^2}{2(1-\nu)B}\sum_{k\neq n}\frac{\sin 2\pi \frac{x_n-x_k}{B}}{(\cosh 2\pi \frac{y_n-y_k}{B}-\cos2\pi \frac{x_n-x_k}{B})^2}\\ \nonumber
&~&\cdot \left( \cosh2\pi \frac{y_n-y_k}{B}-\cos2\pi \frac{x_n-x_k}{B}-2\pi \frac{y_n-y_k}{B}\sinh 2\pi \frac{y_n-y_k}{B}\right)\\
&\approx&\frac{\mu b^2}{2(1-\nu)B}\sum_{k\neq n}\frac{\sin 2\pi \frac{x_n-x_k}{B}
(\cosh2\pi \frac{y_n-y_k}{B}-1-2\pi \frac{y_n-y_k}{B}\sinh 2\pi \frac{y_n-y_k}{B})}{(\cosh 2\pi \frac{y_n-y_k}{B}-1 )^2}.\nonumber\\
\end{eqnarray}
Here we have summed up the contributions from each row first. When $k=n$, the glide force on the dislocation $(x_{n},y_{n})$ imposed by the row of dislocations containing this dislocation itself is zero.
The last approximation is obtained by Taylor expansions
using the fact that $\cosh 2\pi \frac{y_n-y_k}{B}-1>>1- \cos2\pi \frac{x_n-x_k}{B}$ for $k\neq n$,
which is due to $y_k=kD$ and $x_k<<B$ and $D$.  The latter can be derived from the assumption $\tilde{\phi}(y)<<b$ and $Db/B$ and the definition of $x_k$ in Eq.~\eqref{eq:case4-x}. The relative error of this approximation is $O({\displaystyle \max_k}|x_k|/D)^2$.

Following  Eq.~\eqref{eq:case4-x}, we have the Taylor expansion  that
\begin{eqnarray}\label{eq:case4-x-tilde} \nonumber
 x_k-x_n &=&-\frac{B}{b}\tilde{\phi}(y_k)+\frac{B}{b}\tilde{\phi}(y_n)\\
 &=&-\frac{B}{b}\tilde{\phi}_y(y_n)(y_k-y_n)-\frac{B}{2b}\tilde{\phi}_{yy}(y_n)(y_k-y_n)^2+O((y_k-y_n)^3).
\end{eqnarray}
Using Eqs.~\eqref{eq:case4-dis-glide-1} and \eqref{eq:case4-x-tilde} and $y_k=kD$, we have
\begin{eqnarray} \label{eq:case4-dis-glide-2}
f^{\text{dd}}_{\text{g}}(x_{n},y_{n})&\approx&\frac{\mu b^2}{2(1-\nu)B}\sum_{k=1}^{+\infty}\left(\sin 2\pi \frac{x_n-x_{n+k}}{B}+\sin 2\pi \frac{x_n-x_{n-k}}{B}\right)\nonumber\\
&&\cdot\frac{
\cosh2\pi \frac{kD}{B}-1-2\pi \frac{kD}{B}\sinh 2\pi \frac{kD}{B}}{(\cosh 2\pi \frac{kD}{B}-1 )^2}\nonumber\\
&\approx&\frac{\pi\mu b}{(1-\nu)B}\tilde{\phi}''(y_n)\sum_{k=1}^{+\infty}\frac{
(\cosh2\pi \frac{kD}{B}-1-2\pi \frac{kD}{B}\sinh 2\pi \frac{kD}{B})(kD)^2}{(\cosh 2\pi \frac{kD}{B}-1 )^2}.
\end{eqnarray}

Using the definition of the function $g_1$ in Eq.~\eqref{eq:fg1} and the approximation in Eq.~\eqref{eq:truncationg1}, we have the continuum approximation
\begin{equation}\label{eq:case4-glide-result}
f^{\text{dc}}_{\text{g}}=-\frac{\pi\mu b B^2}{(1-\nu)D}\ g_1\left(\frac{D}{B}\right)\tilde{\phi}_{yy}\approx-\frac{\mu b^2 |\psi_y| }{6(1-\nu)|\phi_x|^2}\left[1-\frac{3}{2\pi}\frac{|\phi_x|}{|\psi_y|}\right]_{\varepsilon+}\phi_{yy}.
\end{equation}
Here we have used $\tilde{\phi}_{yy}=\phi_{yy}$.

As in Case 1, when the line direction of these dislocations changes to $\boldsymbol{\tau}=( 0,0,-1)$, we may have
 $\phi_x<0$, and this case can be included by modifying the continuum glide force in Eq.~\eqref{eq:case4-glide-result} as
 \begin{equation}\label{eq:case4-result5}
f^{\text{dc}}_{\text{g}}=-{\rm sgn}(\phi_x)\frac{\mu b^2 |\psi_y| }{6(1-\nu)|\phi_x|^2}\left[1-\frac{3}{2\pi}\frac{|\phi_x|}{|\psi_y|}\right]_{\varepsilon+}\phi_{yy}.
\end{equation}
This continuum expression does not depend on the sign of $\psi_y$.

As in the previous cases, we also calculate the continuum approximation of the climb force in this case from the discrete model in Eq.~(\ref{eq:climb-dd}), and the result is
\begin{eqnarray} \nonumber
f^{\text{dd}}_{\text{c}}(x_{n},y_{n})&=&\frac{\mu b^2}{2\pi(1-\nu)}\sum_{k\neq n}\sum_{j=-\infty}^{+\infty}
{\frac{(y_n-y_k)[3(x_n-(x_k+jB))^2+(y_n-y_k)^2]}{[(x_n-(x_k+jB))^2+(y_n-y_k)^2]^2}}\\ \nonumber
&=&\frac{\mu b^2}{2(1-\nu)B}\sum_{k\neq n}\frac{1}{(\cosh 2\pi \frac{y_n-y_k}{B}-\cos2\pi \frac{x_n-x_k}{B})^2}\\ \nonumber
&&{\textstyle \cdot [-2\pi \frac{y_n-y_k}{B}(\cosh 2\pi \frac{y_n-y_k}{B}\cos2\pi \frac{x_n-x_k}{B}-1)}\\ \nonumber
&&{\textstyle +2 \sinh 2\pi \frac{y_n-y_k}{B}(\cosh 2\pi \frac{y_n-y_k}{B} -\cos2\pi \frac{x_n-x_k}{B})]}\\ \nonumber
&=&O(\tilde{\phi}'(y_n)\tilde{\phi}''(y_n))\\
&\approx&0. \label{eq:case4-dis-climb-2}
\end{eqnarray}
Again, we have used  the fact that $\cosh 2\pi \frac{y_n-y_k}{B}-1>>1- \cos2\pi \frac{x_n-x_k}{B}$ for $k\neq n$, to obtain the expansions.
Therefore, in this case,
\begin{equation}\label{eq:case4-result4}
f_{\text{c}}^{\rm dc}\equiv 0.
\end{equation}

Substituting Eqs.~\eqref{eq:case4-result5} and \eqref{eq:case4-result4} into Eq.~(\ref{eq:evn-eqs1}), we have the following evolution equation for this case:
\begin{equation}\label{eq:case4-evn-3}
\phi_t
- \frac{m_{\text{g}}\mu b^2 }{6(1-\nu)}\frac{ |\psi_y| }{|\phi_x|}\left[1-\frac{3}{2\pi}\frac{|\phi_x|}{|\psi_y|}\right]_{\varepsilon+}\phi_{yy}=0.
\end{equation}

\section{Continuum dislocation dynamics model incorporating short-range interactions}\label{eqn:contiuum-ddpf}

In this section, we present the continuum dislocation dynamics model that incorporates the short range dislocation interactions discussed in the previous section.

\subsection{The continuum dislocation dynamics model based on DDPFs}

We have shown in Sec.~\ref{sec:long-rang} that a continuum model with only the long-range Peach-Koehler force is not always able to capture the behaviors of discrete dislocation dynamics. It will be shown in Sec.~\ref{sec:stability} that such inconsistency leads to insufficiency in the stabilizing effect of the continuum model compared with the discrete dislocation dynamics model.  As a result, in numerical simulations using such a continuum model, there is no effective mechanism to eliminate some numerical oscillations generated during the simulations.

In Sec.~\ref{sec:ddpf}, we have presented
the framework of our DDPF-based continuum dislocation dynamics model, see Eq.~\eqref{eq:evn-eqs1}.
 We incorporate into our continuum model   the continuum  short-range forces obtained in the previous section for the cases where the continuum long-range glide or climb force vanishes.
 With these short-range terms and including the contributions of the applied stress field, the continuum dislocation dynamics equations in Eq.~\eqref{eq:evn-eqs1} become
\begin{equation}\label{eq:evn-general-1}
 \left\{
 \begin{array}{l}
\phi_t
+{\mathbf v}\cdot\nabla\phi
= \frac{m_{\text{g}}\mu b^2}{6(1-\nu)}\frac{ |\phi_x|}{|\psi_y| }\left[ 1 -\frac{3}{2\pi}\frac{|\psi_y|}{|\phi_x|}\right]_{\varepsilon+}\phi_{xx}
+\frac{m_{\text{g}}\mu b^2 }{6(1-\nu)}\frac{ |\psi_y| }{|\phi_x|}\left[1-\frac{3}{2\pi}\frac{|\phi_x|}{|\psi_y|}\right]_{\varepsilon+}\phi_{yy}, \vspace{1ex}\\
\psi_t
+{\mathbf v}\cdot\nabla\psi
=\frac{m_{\text{c}}\mu b^2 }{6(1-\nu) }\varepsilon\psi_{xx},
\end{array}
\right.
\end{equation}
where
\begin{eqnarray}
{\mathbf v}&=&(v_{\text g},v_{\text c}),\vspace{1ex}\label{eqn:6-2}\\
v_{\text g}&=& {\textstyle \frac{m_{\text{g}}}{b}} (\pmb\tau\cdot\mathbf k)\ G_1*(\nabla \phi \times \nabla \psi\cdot \boldsymbol{k})+m_{\text{g}}(\pmb\tau\cdot {\mathbf k})b\sigma_{xy}^0,\vspace{1ex}\nonumber\\
&=&{\textstyle \frac{m_{\text{g}} \mu }{2\pi(1-\nu)}(\pmb\tau\cdot\mathbf k)
\int^{+\infty}_{-\infty}\int^{+\infty}_{-\infty}
\frac{(x-x_1)[(x-x_1)^2-(y-y_1)^2]}{[(x-x_1)^2+(y-y_1)^2]^2}
[\nabla \phi (x_1,y_1)\times \nabla \psi(x_1,y_1)\cdot \boldsymbol{k}] \ dx_1dy_1} \vspace{1ex}\nonumber\\
&~&+m_{\text{g}}(\pmb\tau\cdot {\mathbf k})b\sigma_{xy}^0, \vspace{1ex}\\
v_{\text c}&=&-{\textstyle \frac{m_{\text{c}}}{b}} (\pmb\tau\cdot\mathbf k)\ G_2*(\nabla \phi \times \nabla \psi\cdot \boldsymbol{k})-m_{\text{c}}(\pmb\tau\cdot {\mathbf k})b\sigma_{xx}^0, \vspace{1ex}\nonumber\\
&=&{\textstyle \frac{m_{\text{c}} \mu }{2\pi(1-\nu)}(\pmb\tau\cdot\mathbf k)
\int^{+\infty}_{-\infty}\int^{+\infty}_{-\infty}
\frac{(y-y_1)[3(x-x_1)^2+(y-y_1)^2]}{[(x-x_1)^2+(y-y_1)^2]^2}
[\nabla \phi (x_1,y_1)\times \nabla \psi(x_1,y_1)\cdot \boldsymbol{k}] \ dx_1dy_1} \vspace{1ex}\nonumber\\
&~& -m_{\text{c}}(\pmb\tau\cdot {\mathbf k})b\sigma_{xx}^0,\vspace{1ex}\\
\pmb \tau&=&{\textstyle \frac{\nabla \phi \times \nabla \psi}{\|\nabla \phi \times \nabla \psi\|}},\vspace{1ex}\\
\mathbf k&=&(0,0,1)^T,
\end{eqnarray}
with
\begin{eqnarray}
G_1(x,y)&=&{\textstyle \frac{\mu b}{2\pi(1-\nu)}\frac{(x-x_1)[(x-x_1)^2-(y-y_1)^2]}{[(x-x_1)^2+(y-y_1)^2]^2}},\vspace{1ex}\label{eqn:6-7}\\
G_2(x,y)&=&{\textstyle -\frac{\mu b}{2\pi(1-\nu)}\frac{(y-y_1)[3(x-x_1)^2+(y-y_1)^2]}{[(x-x_1)^2+(y-y_1)^2]^2}}.\label{eqn:6-8}
\end{eqnarray}

Under the assumptions that the length of the Burgers vector $b<<L$, where $L$ is the unit length of the continuum model, and the average dislocation spacing $B\sim D<<L$, it is easy to find that the ratio of the second order partial derivative terms vs the long-range terms $\mathbf v\cdot \nabla \phi$ and $\mathbf v\cdot \nabla \psi$ in Eq.~\eqref{eq:evn-general-1} is $O(b/L)<<1$. Here we have used the fact that $\phi_x=O(b/B)$, $\phi_{xx}=O(b/(BL))$, and similar orders for other partial derivatives of $\phi$ and $\psi$.

Recall that continuum short-range interaction terms provide good approximations to the discrete dislocation model when the continuum long-range force vanishes for some non-trivial perturbed dislocation walls. For a general dislocation distribution described by the continuum model, the full continuum force (including both the long-range and short-range continuum forces) still provides a good approximation to the discrete dislocation dynamics model  under the assumption that a point in the continuum model corresponds to one of these dislocation microstructures of perturbed regular dislocation walls, which is a common technique for the coarse-graining from micro- or meso-scopic models to continuum models.
   Mathematically, these short-range terms in the continuum model serve as stabilizing terms that maintain the same stability properties as the discrete dislocation dynamics model, as will be shown in  Sec.~\ref{sec:stability}.

  Recall also that the main advantage of continuum model based on DDPFs \cite{Xiang2009_JMPS,ZhuXH2010,Zhu2014_IJP,Zhu_continuum3D} is being able to describe the orientation-dependent dislocation densities of curved dislocations. The dislocation glide within its slip plane due to the long-range Peach-Koehler force is regularized by the local curvature term due to line tension effect.
  In the continuum dynamics equations in Eq.~\eqref{eq:evn-general-1} obtained in this paper, the short-range interaction terms are in the form of second partial derivatives of the DDPFs and are able to provide regularization in the cross-section of the dislocations for both glide and climb.
   Combining these two regularization effects, we expect to have a full well-posed continuum dislocation dynamics model based on DDPFs. Moreover, the use of two DDPFs $\phi$ and $\psi$ in the continuum dislocation dynamics model enables the study of the anisotropic behaviors of dislocation ensembles within and out of their slip planes. These will be further explored in the future work.

\subsection{Continuum model for dislocation glide}\label{subsec:glidemodel}
In this subsection, we consider the dynamics of dislocations only by their glide. In this case, we assume the average inter-slip plane distance is $D$ \cite{Zhu_continuum3D}, that is, $\psi(x,y)=\frac{b}{D}y$ is always fixed. Applying our continuum model in Eq.~\eqref{eq:evn-general-1} to this case, i.e., the dislocations only move in the $x$ direction. In this case, Eq.~\eqref{eq:evn-general-1} becomes
\begin{eqnarray}\label{eq:evn-glide}
&&\phi_t
+{\textstyle \frac{m_{\text{g}}}{D}}|\phi_x| G_1*\phi_x+ m_{\text{g}}b\sigma_{xy}^0|\phi_x|\vspace{1ex}\nonumber\\
&&\hspace{0.5in} = {\textstyle\frac{m_{\text{g}}\mu bD}{6(1-\nu)} |\phi_x|\left[ 1 -\frac{3D}{2\pi b|\phi_x|}\right]_{\varepsilon+}\phi_{xx}
+\frac{m_{\text{g}}\mu b^3 }{6(1-\nu)D|\phi_x|}\left[1-\frac{3D|\phi_x|}{2\pi b}\right]_{\varepsilon+}\phi_{yy}},
\end{eqnarray}
where
\begin{eqnarray}
&& G_1*\phi_x(x,y)={\textstyle \frac{ \mu b}{2\pi(1-\nu)}
\int^{+\infty}_{-\infty}\int^{+\infty}_{-\infty}
\frac{(x-x_1)[(x-x_1)^2-(y-y_1)^2]}{[(x-x_1)^2+(y-y_1)^2]^2}
 \phi_x (x_1,y_1) \ dx_1dy_1}.
\end{eqnarray}
%
%It is easy to see that Eq.~\eqref{eq:evn-glide} reduces to Eq.~\eqref{eq:case1-evn3} in Case 1
%when the dislocation distribution is uniform in the $y$ direction, and it reduces to Eq.~\eqref{eq:case4-evn-3} in Case 4
%when the dislocation distribution is uniform
% in the $x$ direction.

In this case, the continuum model in Eq.~\eqref{eq:evn-glide} can be written as:
\begin{eqnarray}\label{eq:evn-glide-phi}
\phi_t+v_{\text{g}}\phi_x=0,
\end{eqnarray}
where the total glide velocity $v_{\text{g}}=m_{\text{g}}f_{\text{g}}$,
 the continuum total glide force $f_{\text{g}}=f_{\text{g}}^{\text{dc}}+{\rm sgn}(\phi_x)b\sigma_{xy}^0$ as given by Eq.~\eqref{eqn:fglide-tot}, and the continuum force due to interactions between dislocations
\begin{eqnarray}\label{eq:evn-glide-conserv}
f_{\text{g}}^{\rm dc}={\textstyle {\rm sgn}(\phi_x)\left\{
\frac{1}{D} G_1*\phi_x
-\frac{m_{\text{g}}\mu bD}{6(1-\nu)} \left[ 1 -\frac{3b}{2\pi D|\phi_x|}\right]_{\varepsilon+}\phi_{xx}
-\frac{m_{\text{g}}\mu b^3 }{6(1-\nu)D\phi_x^2}\left[1-\frac{3b|\phi_x|}{2\pi D}\right]_{\varepsilon+}\phi_{yy}\right\}}\nonumber\\
\end{eqnarray}
including both the long-range interaction force (the first term) and the short-range interaction forces (the remaining two terms) on the dislocations.

When the dislocation distribution is uniform in the $y$ direction, which is Case 1 in Sec.~\ref{sec:continuum-short}, Eq.~\eqref{eq:evn-glide} reduces to
\begin{eqnarray}\label{eq:case1-evn3-simulation}
\phi_t+ m_{\text{g}}b\sigma_{xy}^0|\phi_x|- \frac{m_{\text{g}}\mu b^2D}{6(1-\nu)} |\phi_x|\left[ 1 -\frac{3b}{2\pi D|\phi_x|}\right]_{\varepsilon+}\phi_{xx}=0.
\end{eqnarray}
In this case, the continuum total force in Eq.~\eqref{eq:evn-glide-conserv} reduces to Eq.~\eqref{eq:case1-result5}.

\subsection{Comparison with scalar dislocation density based continuum models}

In this subsection,
we examine the evolution of the signed dislocation density $\rho$ defined
 Eq.~\eqref{eqn:density} in terms of the DDPFs $\phi$ and $\psi$.

We first consider the continuum model of $\phi$ and $\psi$ in the form of Eq.~\eqref{eq:evn-eqs1}. From Eqs.~\eqref{eqn:density} and \eqref{eq:evn-eqs1}, we can calculate that
 \begin{equation}\label{eq:evn-rho}
\rho_t+\nabla\cdot(\rho {\mathbf v})=0,
\end{equation}
where $\textbf{v}=(v_{\text{g}}, v_{\text{c}})^T$ is the dislocation velocity.
In fact,
\begin{eqnarray} \label{eq:evn-rho-0}
\rho_t&=&\frac{1}{b^2}(\phi_x\psi_y-\psi_x\phi_y)_t\nonumber\\ \nonumber
&=&\frac{1}{b^2}(\phi_{xt}\psi_y+\phi_x\psi_{yt}-\psi_{xt}\phi_y-\psi_x\phi_{yt})\\ \nonumber
&=&\frac{1}{b^2}\{(-\boldsymbol{v}\cdot\nabla\phi)_{x}\psi_y+(-\boldsymbol{v}\cdot\nabla\psi)_{y}\phi_x-(-\boldsymbol{v}\cdot\nabla\psi)_{x}\phi_y-(-\boldsymbol{v}\cdot \nabla\phi)_{y}\psi_x\}\\ \nonumber
&=&\frac{1}{b^2}\{(-v_1\phi_x\psi_y+v_1\psi_x\phi_y)_x+(-v_2\phi_x\psi_y+v_2\psi_x\phi_y)_y\} \\
&=&-\nabla\cdot (\rho\boldsymbol{v}).
\end{eqnarray}
In most of the continuum dislocation dynamics models in the literature,  the evolution equation is written in the conservative form in Eq.~\eqref{eq:evn-rho}.
Here we only consider the geometrically necessary dislocations. When only the long-range Peach-Koehler force is considered, the dislocation velocity $\textbf{v}$ is expressed by the mobility law in Eq.~\eqref{eq:velocity} and the long-range force $\textbf{f}=(f_{\text{g}}, f_{\text{c}})^T$ in Eqs.~\eqref{eq:con-glide}  and \eqref{eq:con-climb} in terms of $\rho$. These form a closed evolution equation for the dislocation density $\rho$.

However, the modified continuum dislocation dynamics models incorporated with short-range interaction terms in Eq.~\eqref{eq:evn-general-1} in general is not able to be described fully by the evolution of $\rho$.
The reason is that in our continuum model incorporates the anisotropy of dislocation structure and motion within and out of the slip planes, whereas the single scalar dislocation density $\rho$ is only able to describe isotropy dislocation structure and motion.
When we only consider the glide motion of dislocations as in Sec.~\ref{subsec:glidemodel}, following Eq.~\eqref{eqn:density}, the dislocation density is $\rho=(\nabla \phi \times \nabla \psi\cdot \boldsymbol{k})/b^2 =\frac{1}{bD}\phi_x$, and Eq.~\eqref{eq:evn-glide-phi} can be written as
\begin{eqnarray}\label{eq:evn-glide-conserv-rho0}
\rho_t+(\rho v_{\text{g}})_x=0,
\end{eqnarray}
where $v_{\text{g}}=m_{\text{g}}f_{\text{g}}$,
$f_{\text{g}}=f_{\text{g}}^{\text{dc}}+{\rm sgn}(\rho)b\sigma_{xy}^0$, and
\begin{eqnarray}\label{eq:evn-glide-conserv-rho}
f_{\text{g}}^{\rm dc}={\textstyle {\rm sgn}(\rho)\left\{
b G_1*\rho
-\frac{m_{\text{g}}\mu b^2D^2}{6(1-\nu)}\left[ 1 -\frac{3}{2\pi D^2\rho}\right]_{\varepsilon+}\rho_x
-\frac{m_{\text{g}}\mu b^2 }{6(1-\nu)D^2}\frac{1 }{\rho^2}\left[1-\frac{3D^2\rho}{2\pi}\right]_{\varepsilon+}\phi_{yy}\right\}}.\nonumber\\
\end{eqnarray}
Although Eq.~\eqref{eq:evn-glide-conserv-rho0} is in a conservative form of the dislocation density $\rho$, the continuum total glide force in Eq.~\eqref{eq:evn-glide-conserv-rho} also depends on $\phi_{yy}$, which cannot be simply expressed in terms of $\rho$.
Especially,
for the dislocation structure of Case 4 shown in Fig.~\ref{fig:case4}, the dislocation density $\rho\equiv1/(BD)$, thus the representation by $\rho$ alone is not able to tell the difference between this dislocation structure and a uniform distribution.

 For the dislocation structure of  Case 1 shown in Fig.~\ref{fig:case1} (without the applied stress), our continuum model
 in Eq.~\eqref{eq:case1-evn3}
 can indeed be rewritten as an evolution equations of the dislocation density $\rho$  following  $\rho=\frac{1}{bD}\phi_x$,  which is
 \begin{eqnarray}\label{eq:case1-evn4}
\rho_t- \frac{m_{\text{g}}\mu b^2}{6(1-\nu)}\left(D^2|\rho|\left[ 1 -\frac{3}{2\pi D^2\rho}\right]_{\varepsilon+}\rho_x\right)_x=0.
\end{eqnarray}
In this case, only the local short-range force is nonvanishing, which is
\begin{eqnarray}\label{eq:evn-glide-conserv-rho-case1}
f_{\text{g}}^{\rm dc}=-{\rm sgn}(\rho)
\frac{m_{\text{g}}\mu b^2D^2}{6(1-\nu)} \left[ 1 -\frac{3}{2\pi D^2\rho}\right]_{\varepsilon+}\rho_x.
\end{eqnarray}
In the available continuum formulas in the literature for this case  using different methods \cite{Groma2003,Schulz2015},  their local forces are proportional to $\rho_x/|\rho|$ when only the geometrically necessary dislocations are considered. The corresponding term in our continuum model for this case
in Eqs.~\eqref{eq:case1-evn4} and  \eqref{eq:evn-glide-conserv-rho-case1}  is $\rho_xD^2$,
which means that for this special case, the isotropic dislocation density $\rho$ in the denominator in the models in the literature should be replaced by a more accurate expression $1/D^2$ where $D$ is the average inter-dislocation distance  normal to the slip plane.
 Again we can see that our model using two DDPFs $\phi$ and $\psi$ are able to anisotropy of dislocation structure and motion within and out of the slip planes,  which is not able to be achieved by using the traditional scalar dislocation density $\rho$.

 \section{Stability using  the new continuum model}\label{sec:stability}

In this section,  we examine the stability of the uniform dislocation distributions using the derived continuum model in Eq.~\eqref{eq:evn-general-1}. Consider a uniform distribution of dislocations represented by $\phi_0=\frac{b}{B}x$, $\psi_0=\frac{b}{D}y$. This uniform distribution subject to a small perturbation can be written as
\begin{equation}
\left\{
\begin{array}{l}
\phi=\frac{b}{B}x+\tilde{\phi}(x,y,t), \\
\psi=\frac{b}{D}y+\tilde{\psi}(x,y,t),
\end{array}
\right.
\end{equation}
where  $\tilde{\phi}(x,y,t)$ and $\tilde{\psi}(x,y,t)$ are small perturbation functions.
Using Eq.~\eqref{eqn:density}, the dislocation density for this distribution up to linear order of the small perturbations is
\begin{equation}
\rho(x, y, t)=\frac{(\nabla {\phi} \times \nabla {\psi})\cdot\boldsymbol{k} }{b^2}\approx\frac{1}{BD}+\frac{1}{bD}\tilde{\phi}_x+\frac{1}{bB}\tilde{\psi}_y.
\end{equation}

 Substituting the above $\phi$ and $\psi$ into the continuum model in Eq.~\eqref{eq:evn-general-1} with Eqs.~\eqref{eqn:6-2}--\eqref{eqn:6-8}, the linearized evolution equations of  $\tilde{\phi}(x,y,t), \tilde{\psi}(x,y,t)$, written in the Fourier space, is
 \begin{eqnarray}\nonumber
\hat{\tilde{\phi}}_t
&=&-\frac{2m_{\text{g}}\mu   b^2 }{1-\nu}\left\{\frac{1}{BD}\frac{k_1^2k_2^2}{(k_1^2+k_2^2)^2}
+\frac{D}{12B }\left[ 1 -\frac{3}{2\pi}\frac{B}{D}\right]_{\varepsilon+}k_1^2
+\frac{ B }{12D}\left[1-\frac{3}{2\pi}\frac{D}{B}\right]_{\varepsilon+}k_2^2\right\}
\hat{\tilde{\phi}}\\
&~&-\frac{2m_{\text{g}}\mu   b^2 }{(1-\nu)B^2}\frac{k_1k_2^3}{(k_1^2+k_2^2)^2}\hat{\tilde{\psi}}, \label{eq:evn-fourier-21}\\
 \hat{\tilde{\psi}}_t
&=&-\frac{2m_{\text{c}}\mu   b^2 }{(1-\nu)D^2}\frac{k_1k_2^3}{(k_1^2+k_2^2)^2}\hat{\tilde{\phi}}
-\frac{2m_{\text{c}}\mu   b^2 }{1-\nu}\left[\frac{1}{BD}\frac{k_2^4}{(k_1^2+k_2^2)^2}
+\frac{1 }{12}\varepsilon k_1^2\right]\hat{\tilde{\psi}}, \label{eq:evn-fourier-22}
\end{eqnarray}
where $k_1$ and $k_2$ are frequencies in the $x$ and $y$ directions, respectively. Here we have used
 $\hat{G_1}(k_1,k_2)=-i\frac{\mu b}{2\pi^2(1-\nu)}\frac{k_1k_2^2}{(k_1^2+k_2^2)^2}$ and
 $\hat{G_2}(k_1,k_2)=-i\frac{\mu b}{2\pi^2(1-\nu)}\frac{k_2^3}{(k_1^2+k_2^2)^2}$
 for $G_1(x,y)$ and $G_2(x,y)$ in Eqs.~\eqref{eqn:6-7} and \eqref{eqn:6-8}.

The evolution of  $\hat{\tilde{\phi}}$ and  $\hat{\tilde{\psi}}$  described by Eqs.~\eqref{eq:evn-fourier-21} and \eqref{eq:evn-fourier-22} is determined by the two eigenvalues of the coefficient matrix solved from
the characteristic  polynomial
\begin{equation}\label{eq: characteristic  polynomial}\left|
      \begin{array}{cc}
     \lambda+A+a_1+a_2&R \\
      C& \lambda+S+s_1\\
      \end{array}
    \right|=0,
 \end{equation}
where
 \begin{eqnarray}         \nonumber
&A=\frac{2m_{\text{g}}\mu   b^2 }{(1-\nu)BD}\frac{k_1^2k_2^2}{(k_1^2+k_2^2)^2}, \ \
R=\frac{2m_{\text{g}}\mu   b^2 }{(1-\nu)B^2}\frac{k_1k_2^3}{(k_1^2+k_2^2)^2}, \vspace{1ex}\\ \nonumber
&C=\frac{2m_{\text{c}}\mu   b^2 }{(1-\nu)D^2}\frac{k_1k_2^3}{(k_1^2+k_2^2)^2}, \ \
 S=\frac{2m_{\text{c}}\mu   b^2 }{(1-\nu)BD}\frac{k_2^4}{(k_1^2+k_2^2)^2}, \vspace{1ex}\\   \nonumber
&a_1=\frac{m_{\text{g}}\mu   b^2D }{6(1-\nu)B }\left[ 1 -\frac{3}{2\pi}\frac{B}{D}\right]_{\varepsilon+}k_1^2,   \ \
a_2=\frac{m_{\text{g}}\mu   b^2 B }{6(1-\nu)D}\left[1-\frac{3}{2\pi}\frac{D}{B}\right]_{\varepsilon+}k_2^2,  \vspace{1ex}\\  \nonumber
&s_1=\frac{\varepsilon m_{\text{c}}\mu   b^2 }{6(1-\nu)}k_1^2.
   \end{eqnarray}
   The Fourier coefficients of the small perturbations $\hat{\tilde{\phi}}$ and  $\hat{\tilde{\psi}}$ decay when the two eigenvalues $\lambda_1,\lambda_2<0$.

  Due to $AS=RC$, the  characteristic  polynomial in Eq.~\eqref{eq: characteristic  polynomial} becomes
 \begin{equation}
 \lambda^2+(A+a_1+a_2+S+s_1)\lambda+(a_1+a_2)(S+s_1)+As_1=0.
 \end{equation}
% \begin{equation}
% \Delta=(A+a_1+a_2-S-s_1)^2+4AS.
% \end{equation}
Thus the two eigenvalues are
 \begin{eqnarray}
\lambda_{1,2}&=&{\textstyle \frac{-(A+a_1+a_2+S+s_1)\pm\sqrt{(A+a_1+a_2+S+s_1)^2-4[(a_1+a_2)(S+s_1)+As_1]}}{2}}\label{eqn:eigen1}\\
&=&{\textstyle \frac{-(A+a_1+a_2+S+s_1)\pm\sqrt{(A+a_1+a_2-S-s_1)^2+4AS}}{2}}\label{eqn:eigen2}.
 \end{eqnarray}
Note that $A, S, a_1, a_2, s_1\geq0$. By Eq.~\eqref{eqn:eigen2}, we know that both $\lambda_1$ and $\lambda_2$ are real. It is easy to conclude from Eq.~\eqref{eqn:eigen1} that $\lambda_{1,2}<0$ when $k_1\neq 0$ or $k_2\neq 0$ (because the term $4[(a_1+a_2)(S+s_1)+As_1]>0$ in this case), and $\lambda_1=\lambda_2=0$ when $k_1=k_2=0$. Therefore, when $(k_1,k_2)\neq (0,0)$, $\hat{\tilde{\phi}}$ and  $\hat{\tilde{\psi}}$  always decay and the uniform distribution of dislocations is stable using the derived continuum model in Eq.~\eqref{eq:evn-general-1}.

This stability result provides a basis for wellposedness of the continuum model in Eq.~\eqref{eq:evn-general-1} as well as stability of numerical solutions for it. These topics will be further explored in the future work.
When only the continuum long-range Peach-Koehler force is considered, i.e., the second partial derivative terms on the right-hand side of the PDE system in Eq.~\eqref{eq:evn-general-1} vanish, the linearized equations for the small perturbations $\tilde{\phi}$ and $\tilde{\psi}$ in the Fourier space are
 \begin{eqnarray}
\hat{\tilde{\phi}}_t
&=&-\frac{2m_{\text{g}}\mu   b^2 }{1-\nu}\frac{1}{BD}\frac{k_1^2k_2^2}{(k_1^2+k_2^2)^2}
\hat{\tilde{\phi}}-\frac{2m_{\text{g}}\mu   b^2 }{(1-\nu)B^2}\frac{k_1k_2^3}{(k_1^2+k_2^2)^2}\hat{\tilde{\psi}}, \label{eq:evn-fourier-212}\\
 \hat{\tilde{\psi}}_t
&=&-\frac{2m_{\text{c}}\mu   b^2 }{(1-\nu)D^2}\frac{k_1k_2^3}{(k_1^2+k_2^2)^2}\hat{\tilde{\phi}}
-\frac{2m_{\text{c}}\mu   b^2 }{1-\nu}\frac{1}{BD}\frac{k_2^4}{(k_1^2+k_2^2)^2}
\hat{\tilde{\psi}}. \label{eq:evn-fourier-222}
\end{eqnarray}
 Same as the discussion in Sec.~\ref{sec:long-rang}, when $k_1=0$ or $k_2=0$, there is no stabilizing force (which is the glide force) for $\tilde{\phi}$; and when $k_2=0$, there is no stabilizing force (which is the climb force) for $\tilde{\psi}$. In these cases, numerical oscillations in simulations cannot be stabilized without the second order partial derivative terms.

Recall that the second order partial derivative terms in  Eq.~\eqref{eq:evn-general-1} are based on the short-range interactions of dislocations. Those terms coming from the short-range glide forces  (in the $\phi$-equation) agree with the glide forces using the discrete dislocation model for uniform dislocation distributions subject to small perturbations in the glide direction. For the climb force,
a regularization term (in the $\psi$-equation) is added in addition to the stabilizing effect provided by the
long-range climb force.

\section{Numerical simulations}
\label{sec:numerical}

In this section, we perform numerical simulations to validate the derived continuum model. In addition to the nondimensionalization before simulations, we set Poisson ratio $\nu=1/3$.

\subsection{Comparisons of the continuum force with the discrete model}

 We first examine the total glide force in the continuum model including both the long-range and short-range contributions given by Eq.~\eqref{eq:evn-glide-conserv} by comparisons with the discrete dislocation dynamics model.
  Recall that the continuum short-range glide force terms are derived from the discrete dislocation model for uniform dislocation distributions subject to small perturbations in the glide direction.

We first consider the dislocation distributions of Case 1 in Sec.~\ref{sec:continuum-short}, where the dislocation distributions are uniform in the $y$ direction. This problem is reduced to a one-dimensional problem depending only on $x$.

{\bf Example 1}

\begin{figure}[htbp]\centering
\subfigure[]
{\label{fig:case1-eg2-xphi1}\includegraphics[width=2.0in]{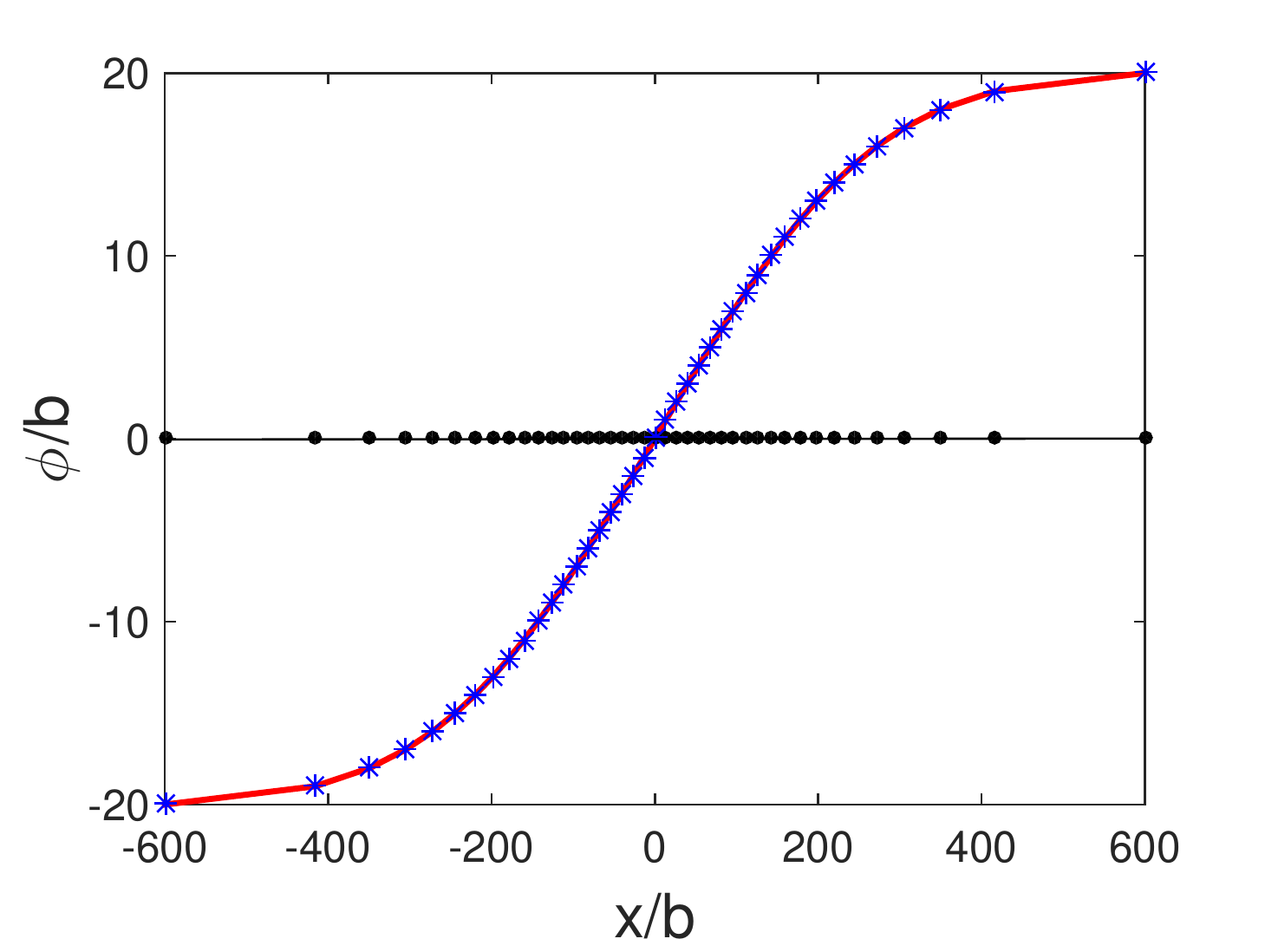}}
\subfigure[]
{\label{fig:case1-eg2-r1}\includegraphics[width=2.0in]{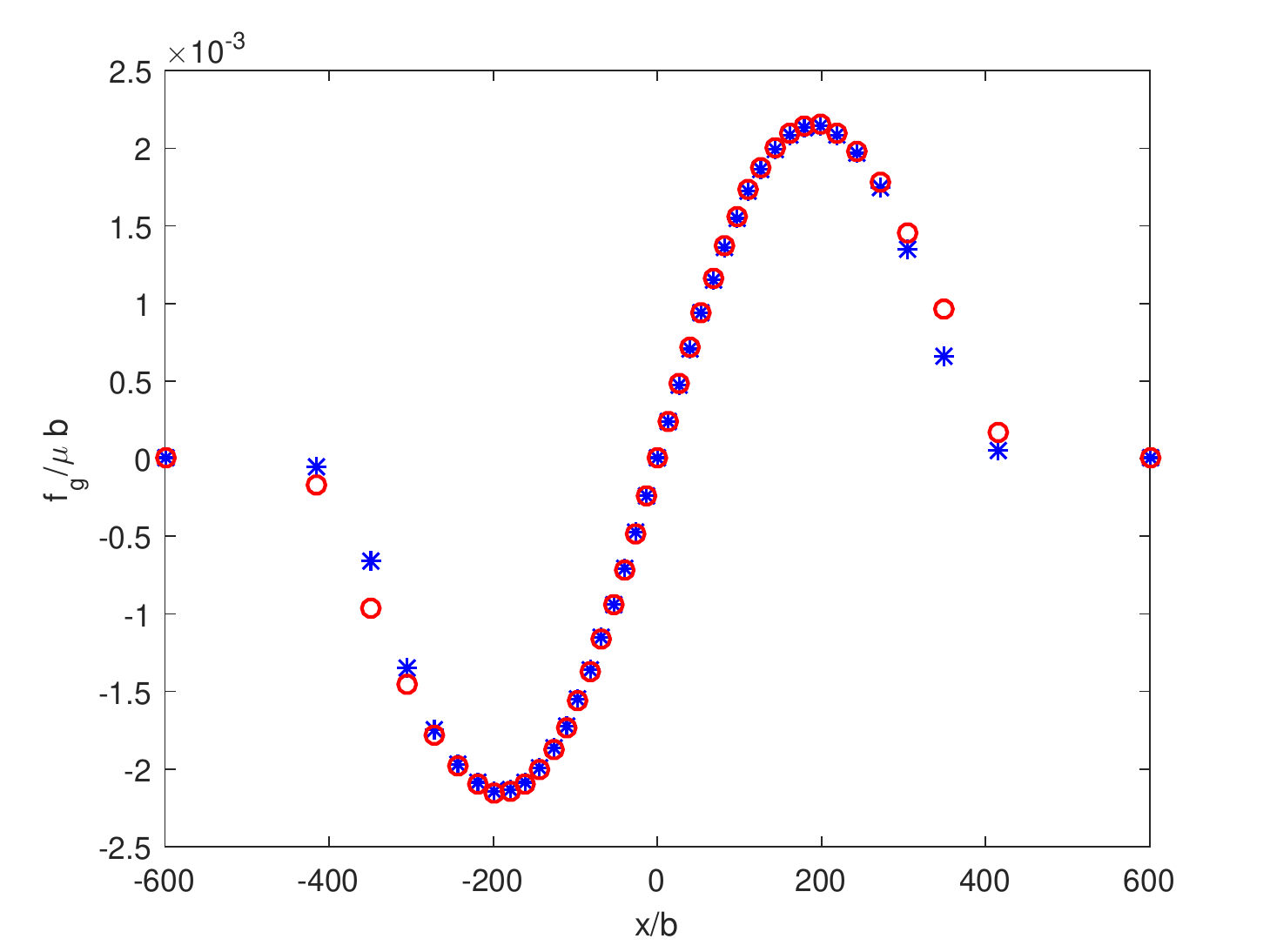}}\\
\subfigure[]
{\label{fig:case1-eg2-xphi2}\includegraphics[width=2.0in]{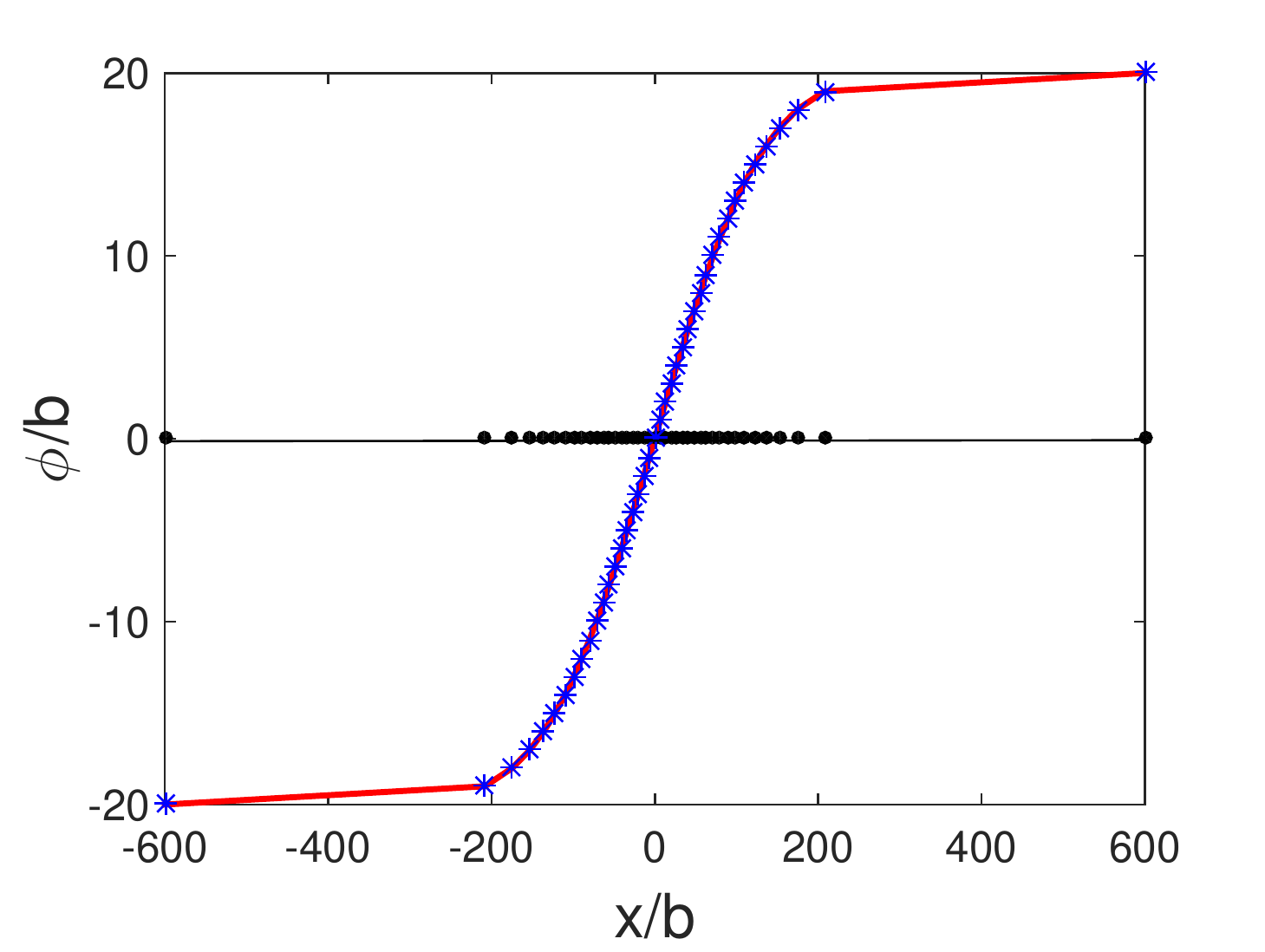}}
\subfigure[]
{\label{fig:case1-eg2-r2}\includegraphics[width=2.0in]{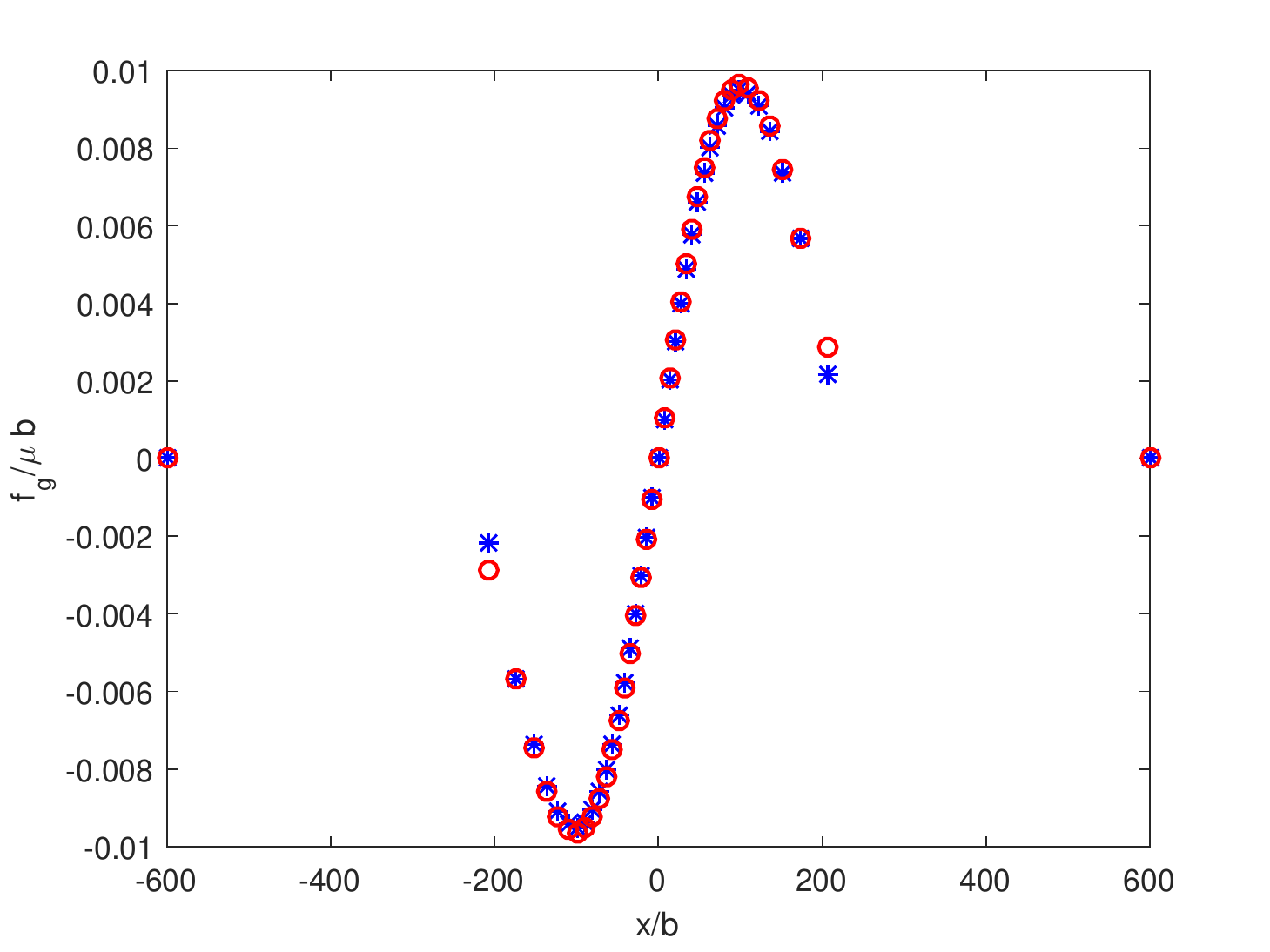}}\\
\subfigure[]
{\label{fig:case1-eg2-xphi3}\includegraphics[width=2.0in]{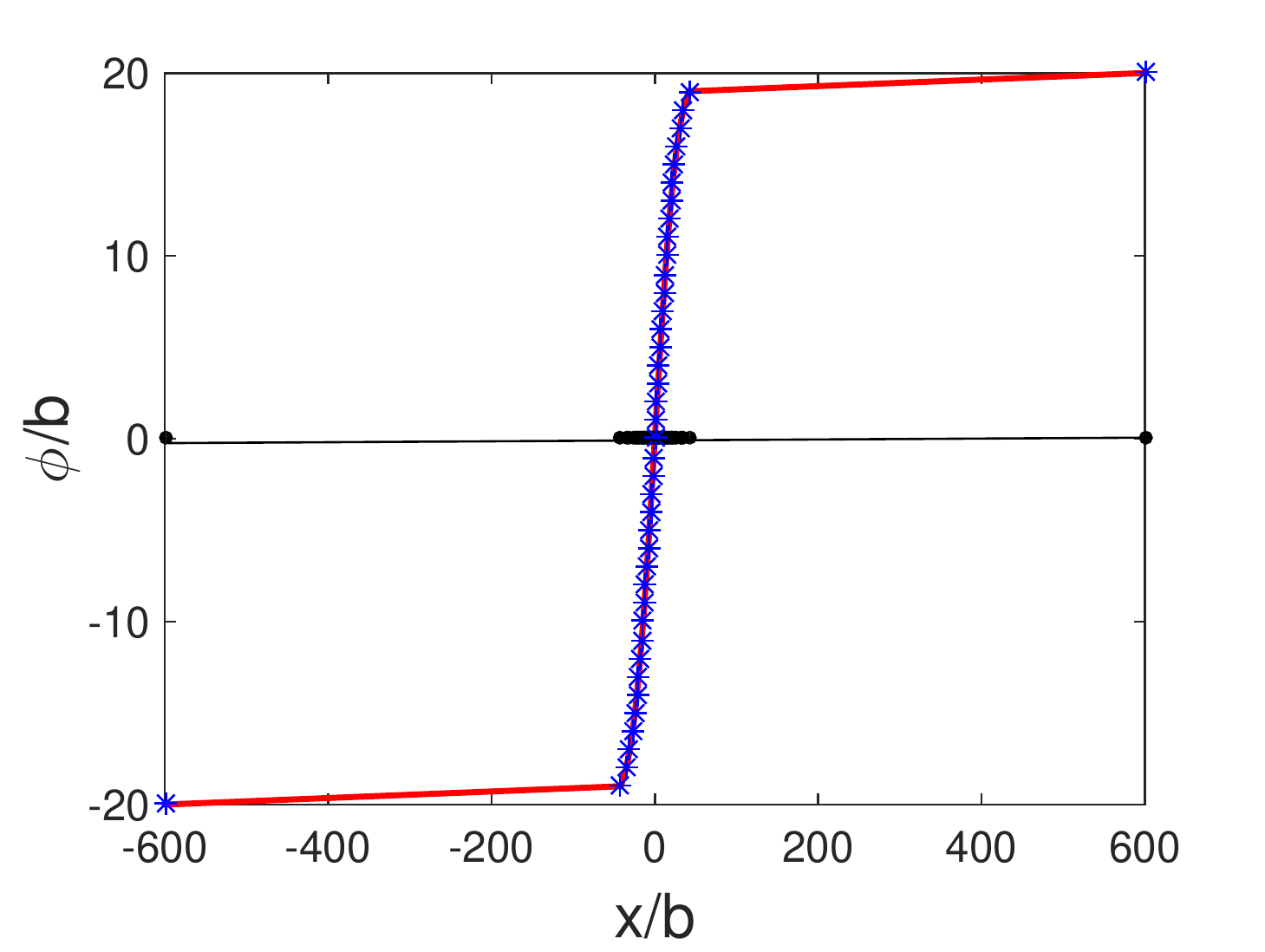}}
\subfigure[]
{\label{fig:case1-eg2-r3}\includegraphics[width=2.0in]{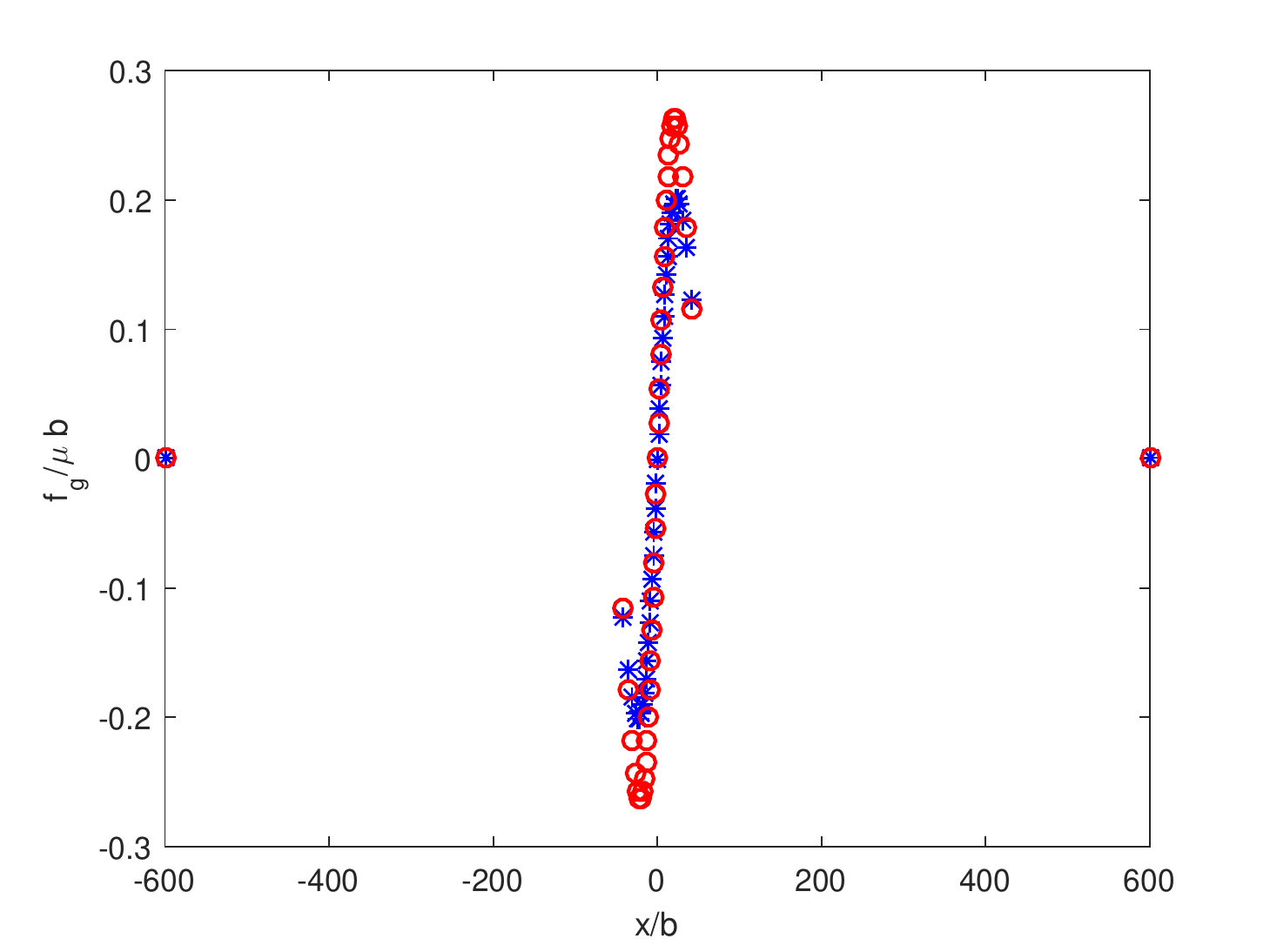}}
\caption{Example 1: Continuum glide force compared with the discrete model for distributions of dislocation walls for different values of  concentration width $w$ (defined in Eq.~\eqref{eqn:width}).  The concentration width $w=10B$ in (a) and (b), $w=5B$ in (c) and (d), and $w=B$ in (e) and (f). Images (a), (c), and (e) show the profile of $\phi(x)$ (red curve) and locations of the dislocation walls for each value of the width $w$. The black dots on the horizontal line indicate the locations of the dislocation walls, and the blue dots show the corresponding values of $\phi$ in the continuum model. Images (b), (d), and (f) show values of the glide force $f_{\rm g}$ on the dislocation walls calculated by using the continuum model (red circles) and by using the discrete dislocation model (blue stars).
} \label{fig:case1-eg2}
\end{figure}

Assume the dislocation distribution is described by
\begin{equation}\label{eqn:width}
\phi(x)=
\begin{cases}
-\frac{Nb}{2} &\mbox{if $x=-\frac{NB}{2}$ }\\
\frac{Nb}{2}{\rm erf}(\frac{x}{w}) &\mbox{if $-\frac{NB}{2}<x<\frac{NB}{2}$ }\\
\frac{Nb}{2} &\mbox{if $x=\frac{NB}{2}$ }
\end{cases}
\end{equation}
where ${\rm erf}(x)=\frac{2}{\sqrt{\pi}}\int_0^x e^{-u^2} du$, and $\psi(y)=\frac{b}{D}y$. Periodic boundary condition is assumed in the $x$ direction. We set $D=50b$, $B=30b$, and $N=40$. The dislocation walls are concentrated within the region in the center with width $w$. We perform simulations for the cases of $w=10B$, $w=5B$, $w=B$. The profiles of the DDPF $\phi$ and the locations of the dislocation walls are shown in Fig.~\ref{fig:case1-eg2} (a), (c), and (e), and the corresponding glide forces calculated by the continuum model in Eq.~\eqref{eq:evn-glide-conserv} (which reduces to Eq.~\eqref{eq:case1-result5} in this case) and by the discrete dislocation model are plotted in Fig.~\ref{fig:case1-eg2} (b), (d), and (f), respectively. It can be seen that the results of the  continuum model agree
 excellently with those of discrete model for smoothly varying (the case of $w=10B$ in Fig.~\ref{fig:case1-eg2} (a),(b)) and even concentrated (the case of $w=5B$ in Fig.~\ref{fig:case1-eg2} (c),(d)) distributions of dislocation walls. For  extremely concentrated distribution of dislocation walls as shown in Fig.~\ref{fig:case1-eg2} (e) with $w=B$, the overall continuum approximation Fig.~\ref{fig:case1-eg2} (f) is still reasonably good.   At the two ends of the concentrated distribution where the dislocation density changes dramatically, our continuum approximation gives the strongest force as in the discrete model, although there are discrepancies in the exact values.  (Recall that the continuum formulations are derived based on smoothly-varying dislocation densities.)

{\bf Example 2}

\begin{figure}[htpb]
\centering
\subfigure[]
{\label{fig:case1a1}\includegraphics[width=1.5in]{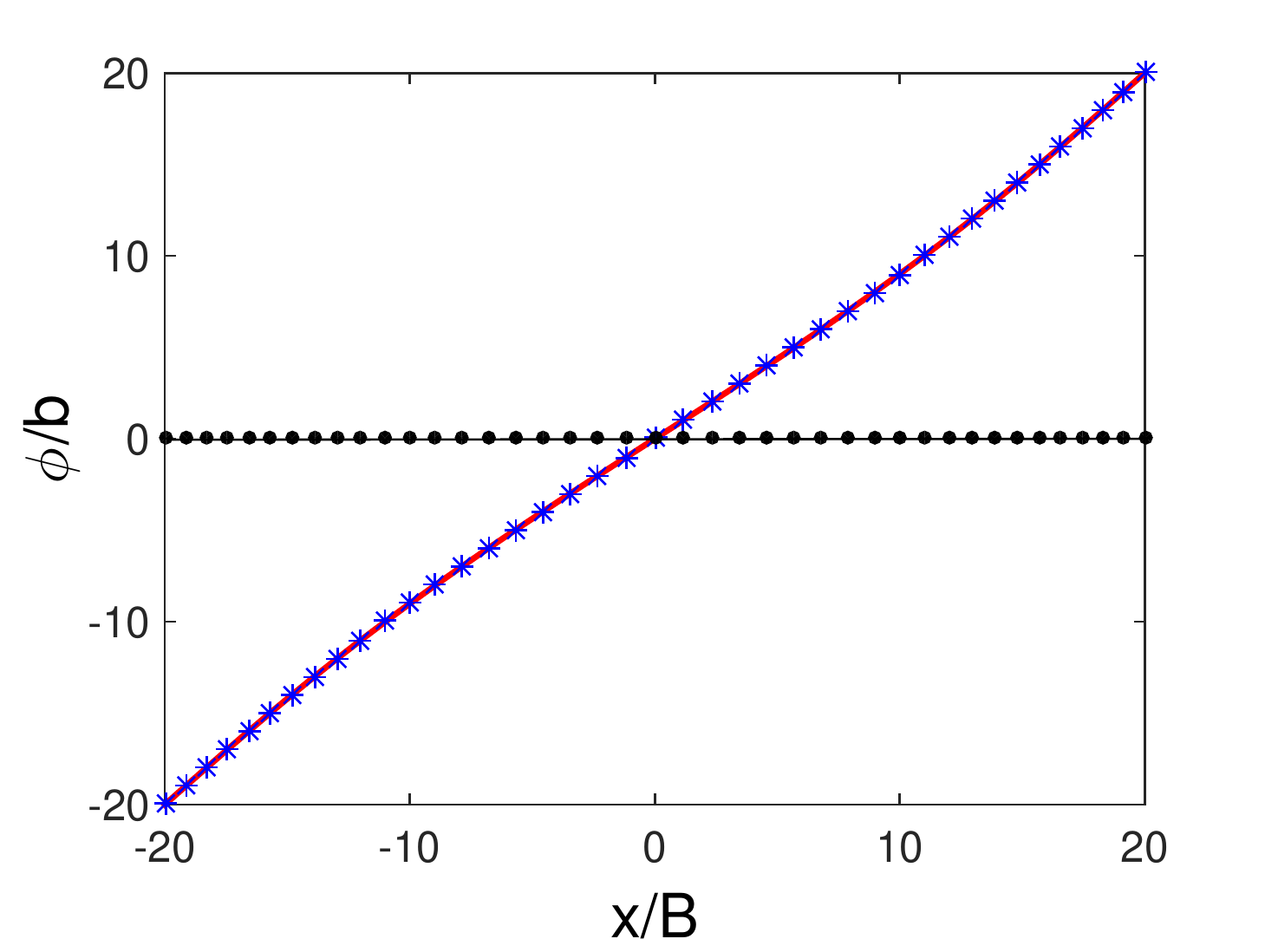}}
\subfigure[]
{\label{fig:case1b1}\includegraphics[width=1.5in]{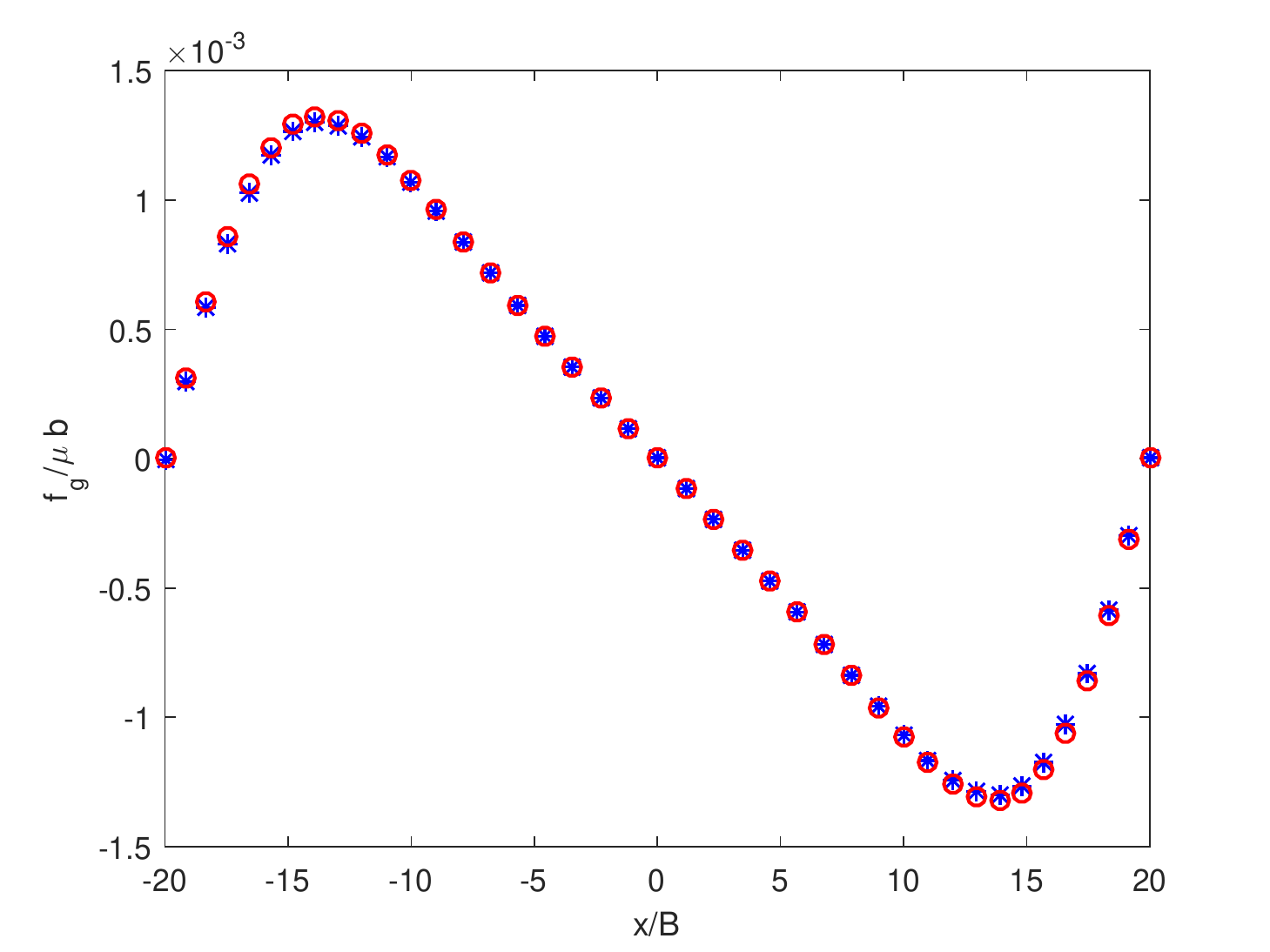}}
\subfigure[]
{\label{fig:case1b2}\includegraphics[width=1.5in]{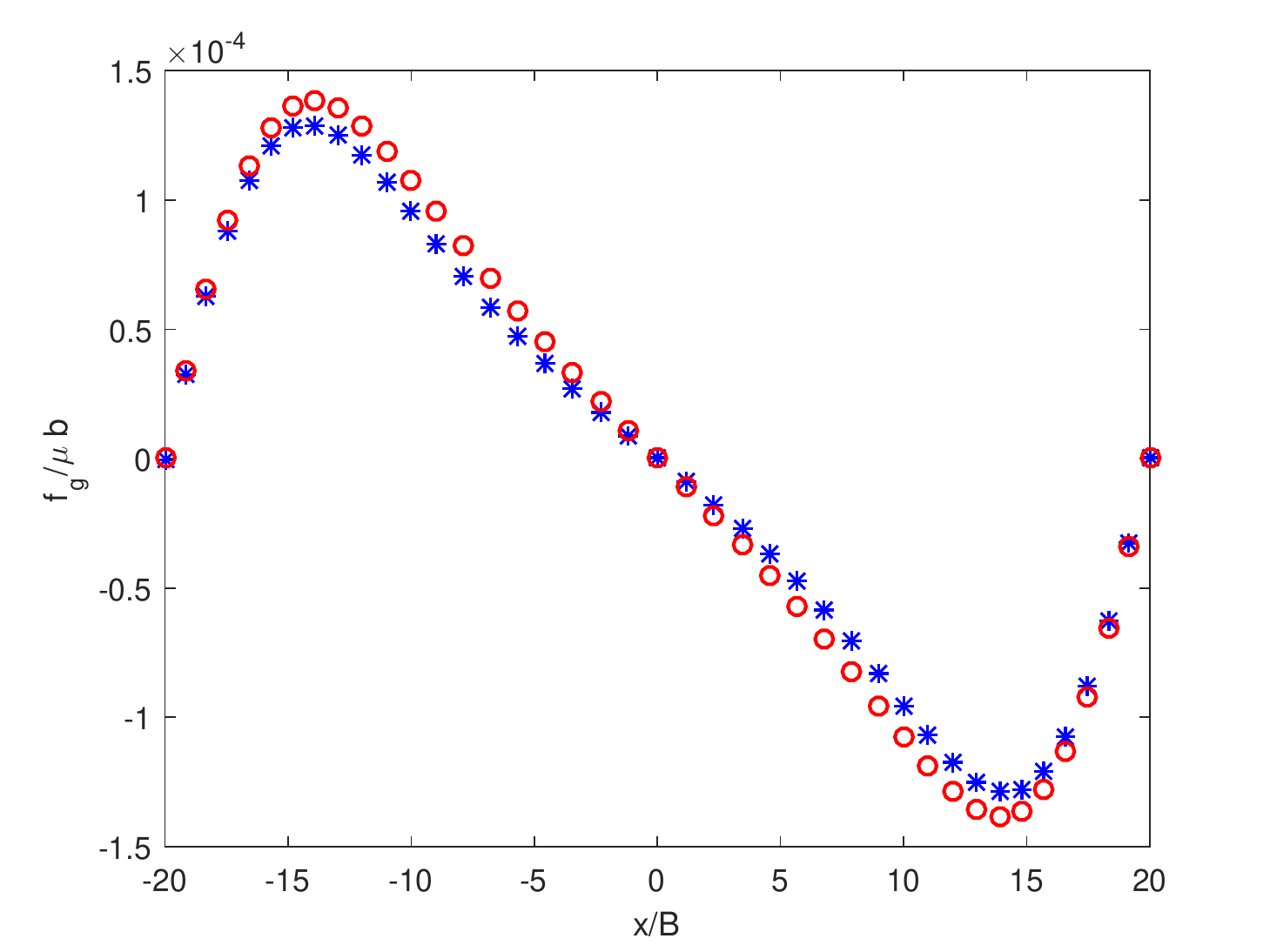}}\\
  \subfigure[]
{\label{fig:case1b3}\includegraphics[width=1.5in]{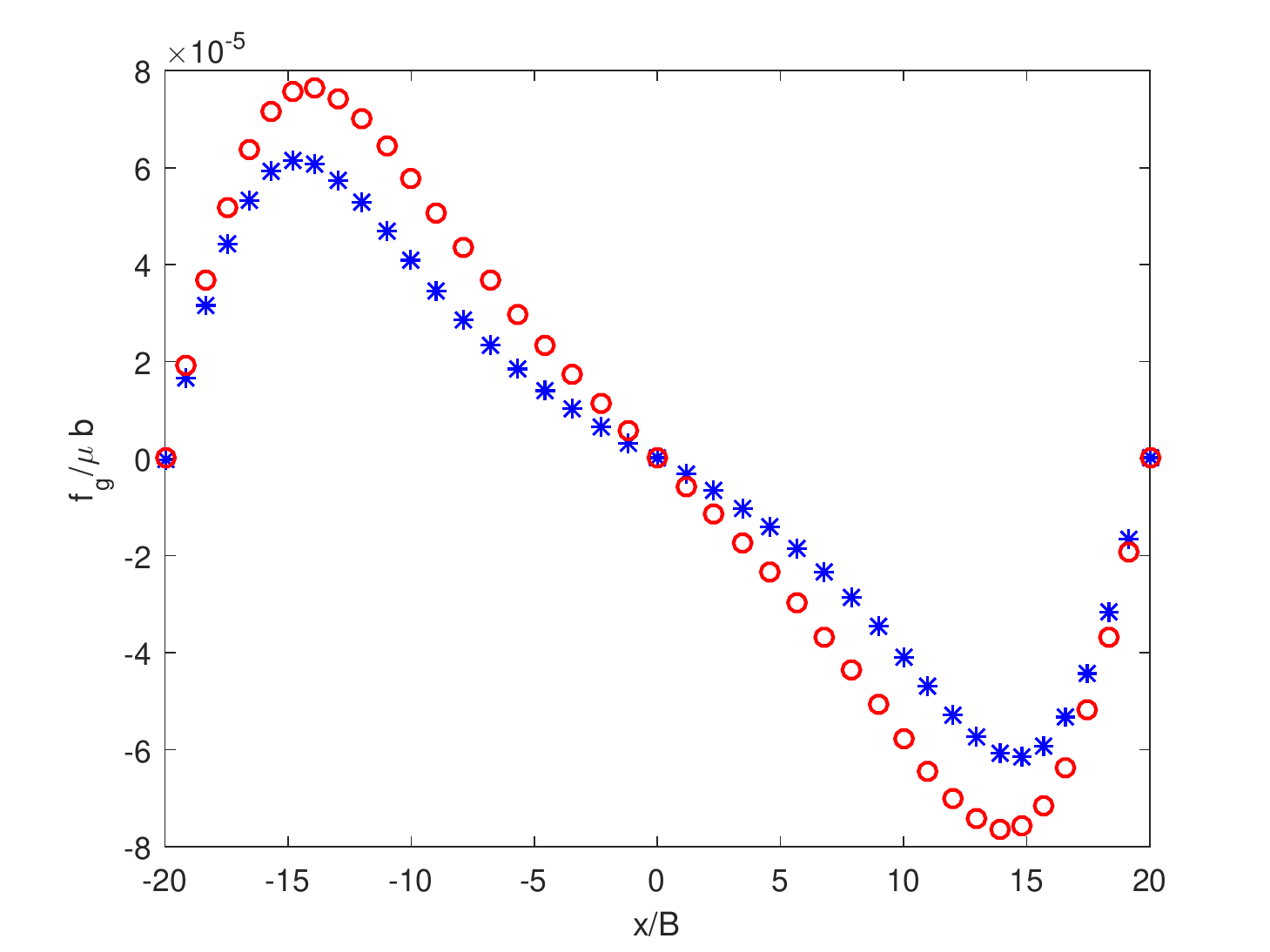}}
\subfigure[]
{\label{fig:case1b4}\includegraphics[width=1.5in]{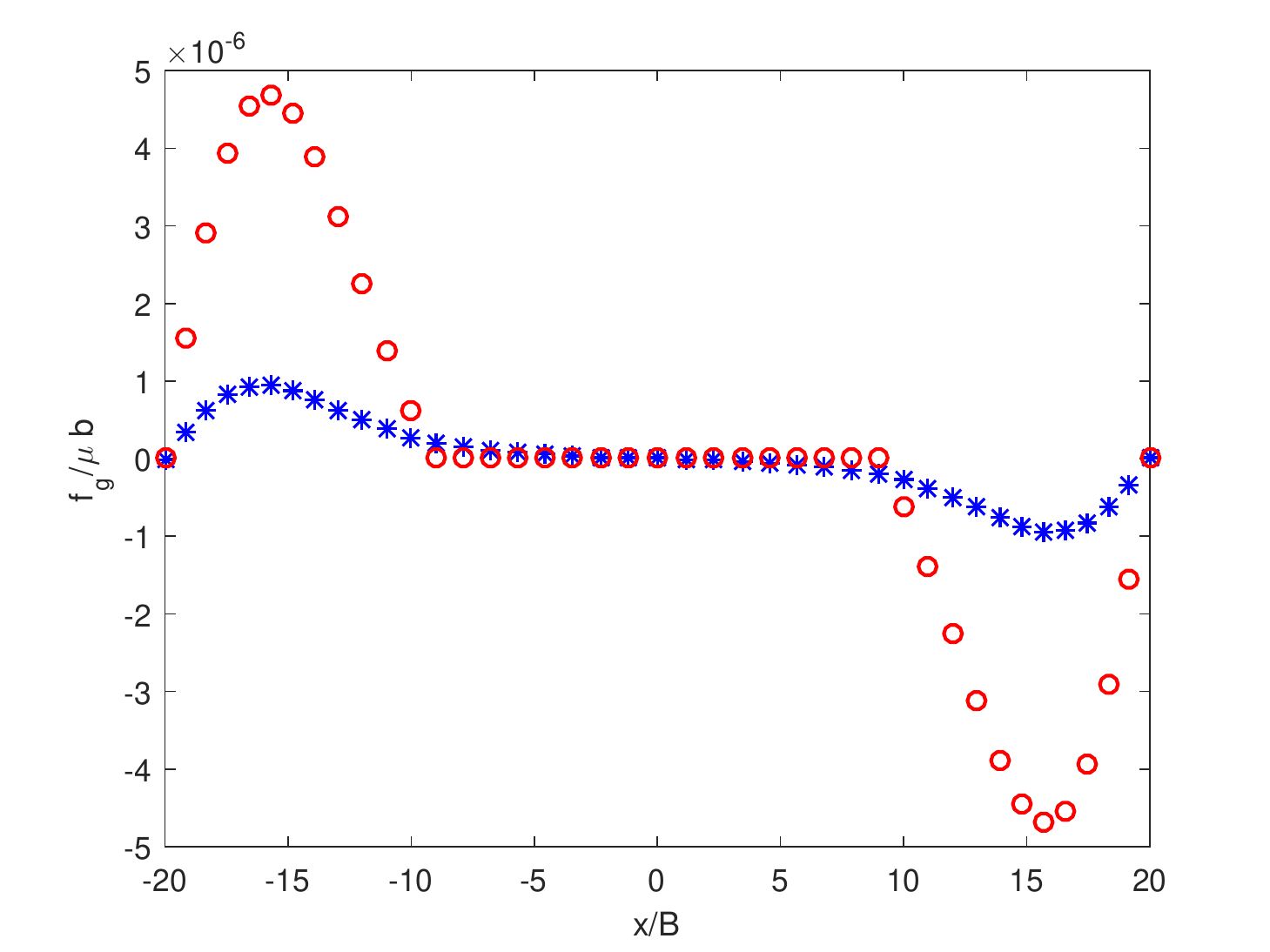}}
\subfigure[]
{\label{fig:case1b5}\includegraphics[width=1.5in]{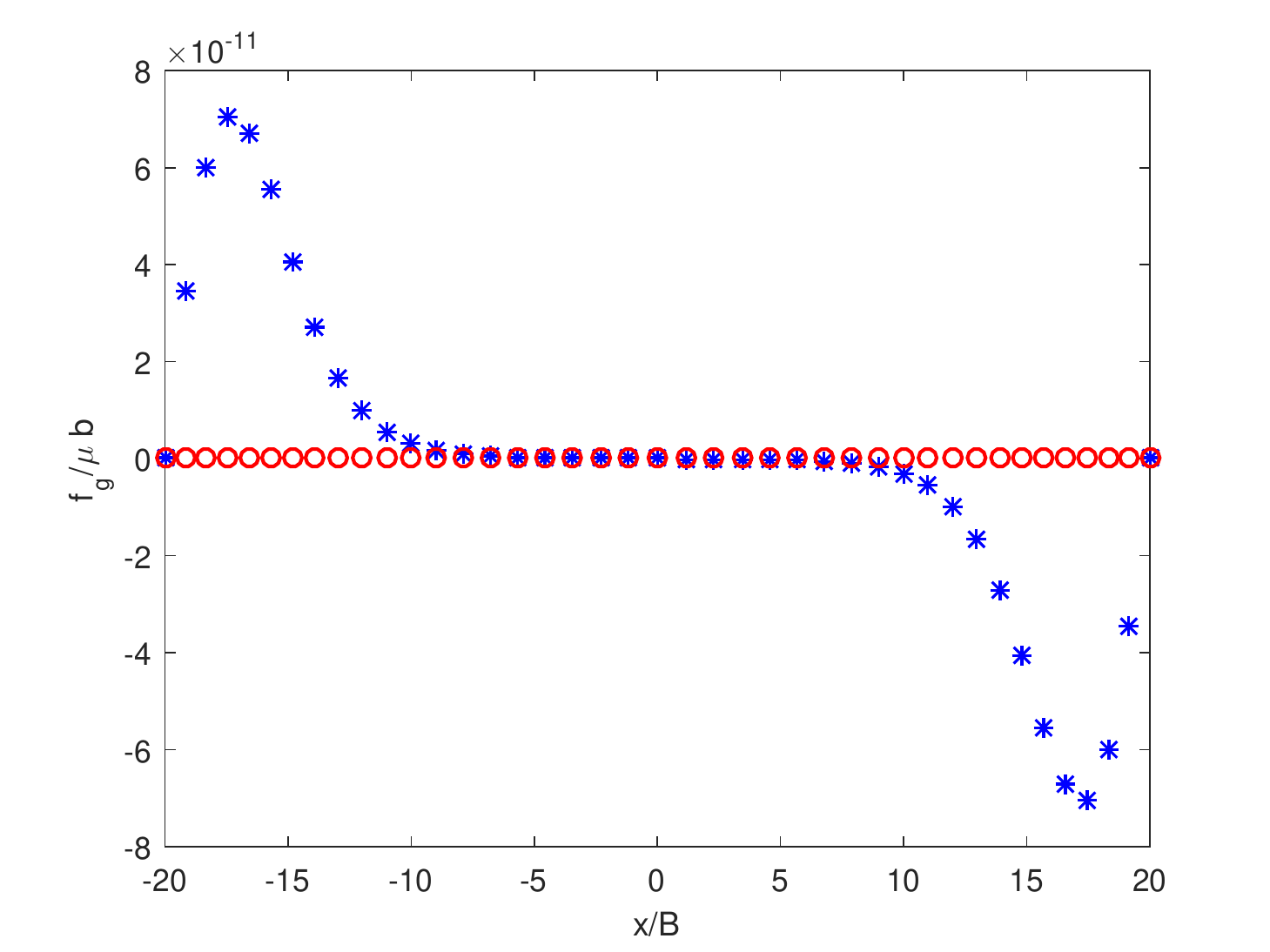}}
  \caption{Example 2: Continuum glide force compared with that of the discrete model for distributions of dislocation walls for different values of the ratio $B/D$ for distributions of dislocation walls with uniform active slip plane spacing (given by Eq.~\eqref{eq:case1-x-phi1}). (a) The profile of the DDPF $\phi$ (red curve) and locations of  the dislocation walls. The black dots on the horizontal line indicate the locations of the dislocation walls, and the blue dots show the corresponding values of $\phi$ in the continuum model. Images (b)-(f) show the continuum glide force (red circles) compared with the force calculated from the discrete dislocation model (blue stars) for the cases of (b) $B=15b$, (c) $B=40b$, (d) $B=50b$, (e) $B=100b$, and (f) $B=200b$, respectively.}\label{fig:case1b}
\end{figure}

In this example, we examine the continuum glide force in Eq.~\eqref{eq:case1-result5}  for different values of the ratio $B/D$ for distributions of dislocation walls with uniform active slip plane spacing. Recall that $B$ is the average inter-dislocation distance within a slip plane and $D$ is the average slip plane spacing. For these dislocation distributions, we choose the DDPFs $\psi(y)=\frac{b}{D}y$ and $\phi(x)$ determined by the following equation
 \begin{equation}\label{eq:case1-x-phi1}
  \frac{b}{B}x=\phi+b\sin\left(\frac{2\pi\phi}{40b}\right).
  \end{equation}
This a uniform dislocation wall distribution with perturbation in the $x$ direction, and the DDPF
$\phi$ can be written as $\phi(x)=\frac{b}{B}x+\tilde {\phi}(x)$, where $\tilde {\phi}(x)$ is a small perturbation, see Fig.~\ref{fig:case1b}(a). The period of this distribution is  $N=40$ dislocation walls. We fix $D=50b$ and vary the value of $B$.

%
%Five periods $(-100b, 100b)$ are chosen to simulate the collective shear stress of dislocations in the locations $\phi$ from $-20b$ to $20b$. We give two different dislocation configurations to illustrate the approximation accuracy of the formula in Eq.~\eqref{eq:case1-result5}. In the  figures for the simulation results, red points represent the results from the continuum model, and blue points represent the results from the discrete model. We  vary the different values of the distances  $D$ and $B$ to verify the results.

The values of the glide force calculated by the continuum model and comparisons with the results of the discrete dislocation model are shown in Fig.~\ref{fig:case1b}(b)-(f) for the cases of $B=15b,40b,50b,100b,200b$.
When the inter-dislocation wall distance $B$ is smaller than the slip plane spacing $D$, as shown in Fig.~\ref{fig:case1b}(b), the continuum glide force agrees excellently with the force in discrete model. In this case, the glide force is significant: around $10^{-3}\mu b$, in agreement with the strong interaction between neighboring dislocation walls. When the inter-dislocation wall distance $B$ is comparable with the slip plane spacing $D$, as shown in Fig.~\ref{fig:case1b}(c) and (d), the continuum glide force agrees well with the force in discrete model with small errors. In this case, the glide force becomes smaller: around $10^{-4}\mu b$, which is again consistent with the weak interaction between neighboring dislocation walls in this case. When the inter-dislocation wall distance $B$ is much greater than the slip plane spacing $D$,  the interaction between neighboring dislocation walls should be negligible, which is reflected by the small values of the forces  calculated by the continuum and the discrete models shown in Fig.~\ref{fig:case1b}(e) and (f) (at the  order of $\leq 10^{-6}\mu b$  and $\leq 10^{-10}\mu b$). In this sense, the continuum model still provides a good approximation to the discrete model in this case,  although the values calculated by the two models are not necessarily exactly the same. The latter differences at the negligible orders are due to the simplification of our continuum model in Eq.~\eqref{eq:case1-result5} from its exact form in Eqs.~\eqref{eq:case1-dis-03}   and  \eqref{eq:fg1} using the simplification in Eq.~\eqref{eq:truncationg1}.

{\bf Example 3}

\begin{figure}[htbp]
\centering
\subfigure[Continuum model (full)]
{\label{fig:2d-glide-D30B50-dis}\includegraphics[width=2.2in]{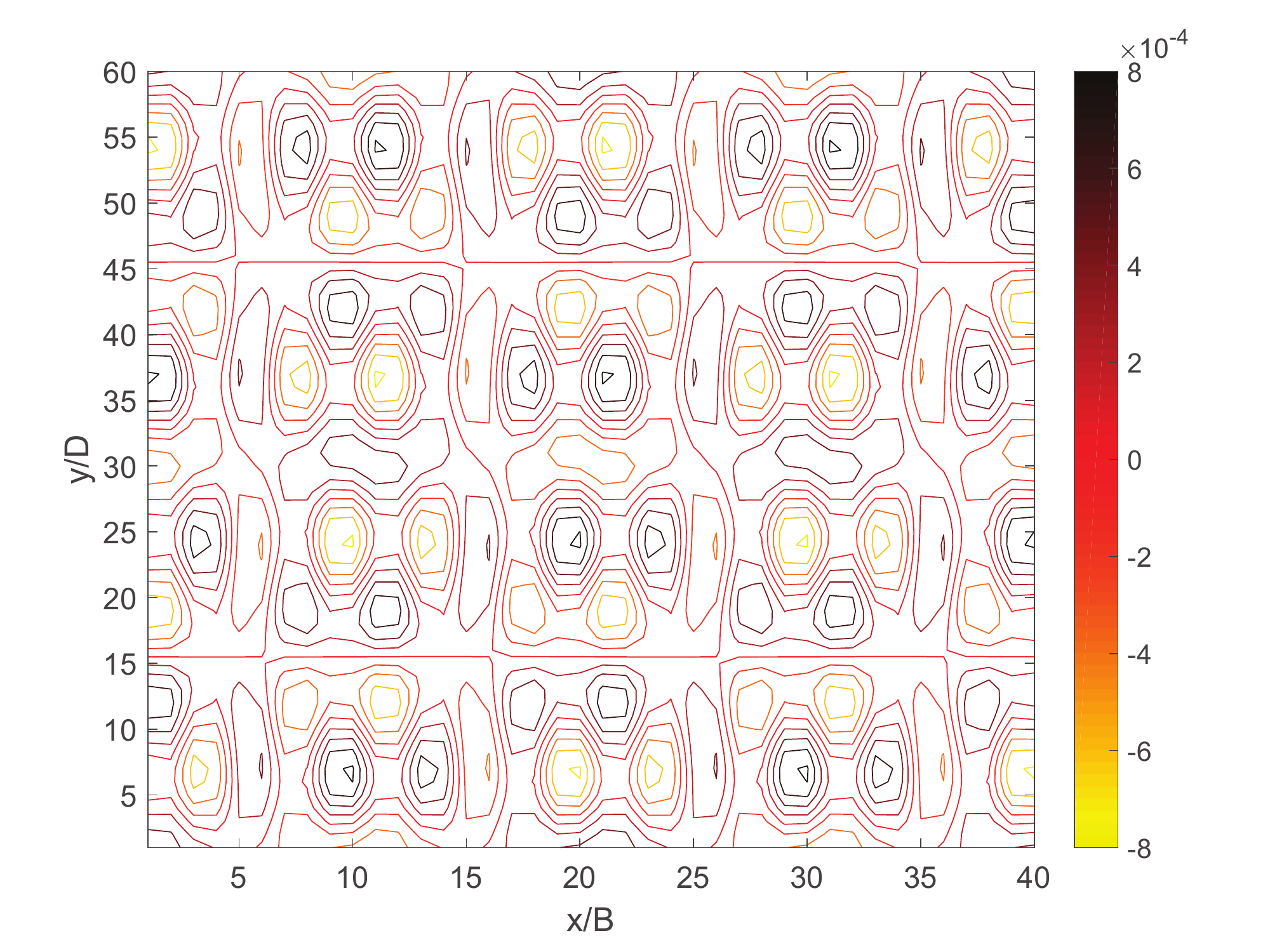}}
\subfigure[Continuum long-range force]
{\label{fig:2d-glide-D30B50-con}\includegraphics[width=2.2in]{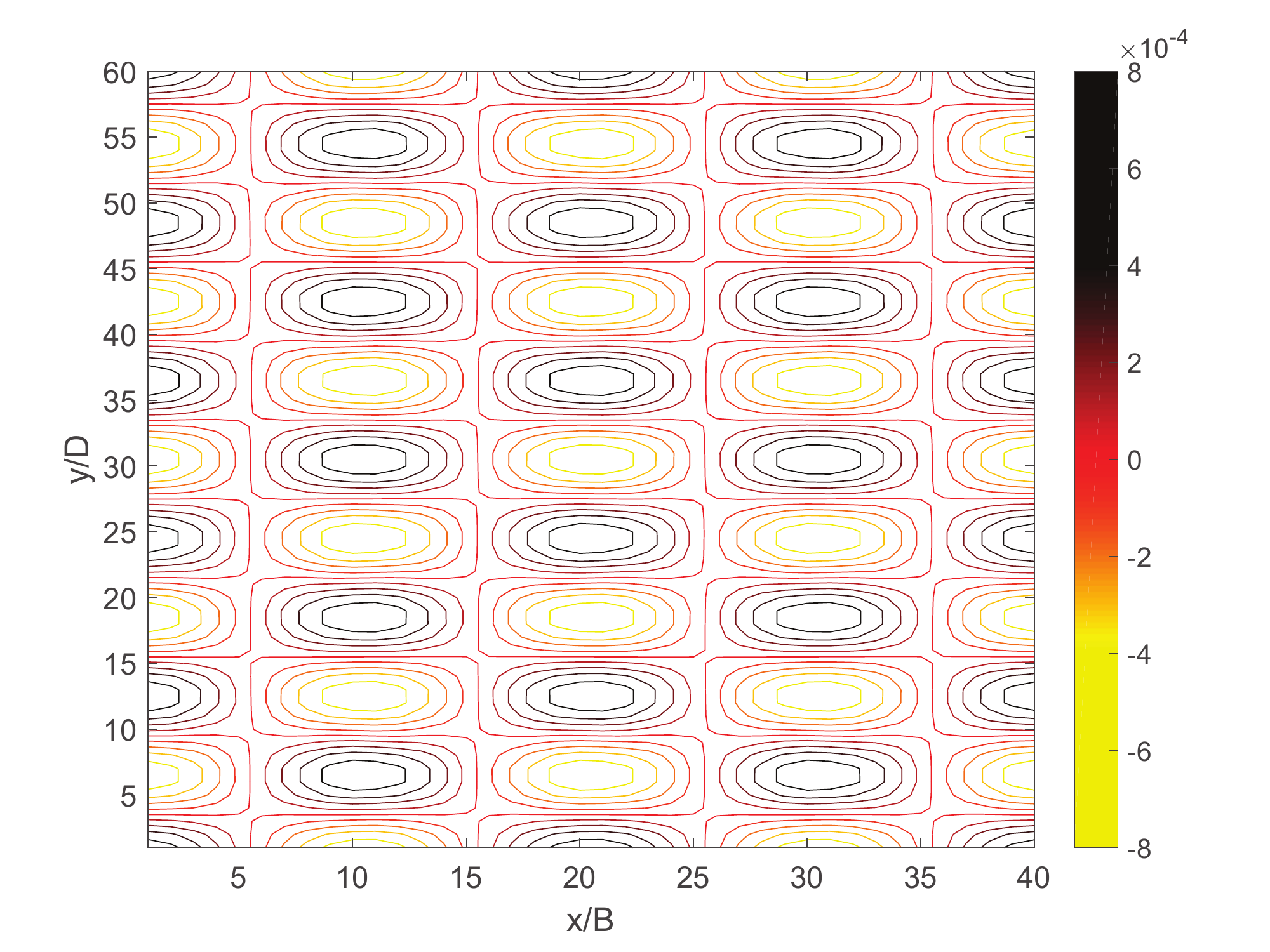}}\\
\subfigure[Discrete model]
{\label{fig:2d-glide-D30B50-con0}\includegraphics[width=2.2in]{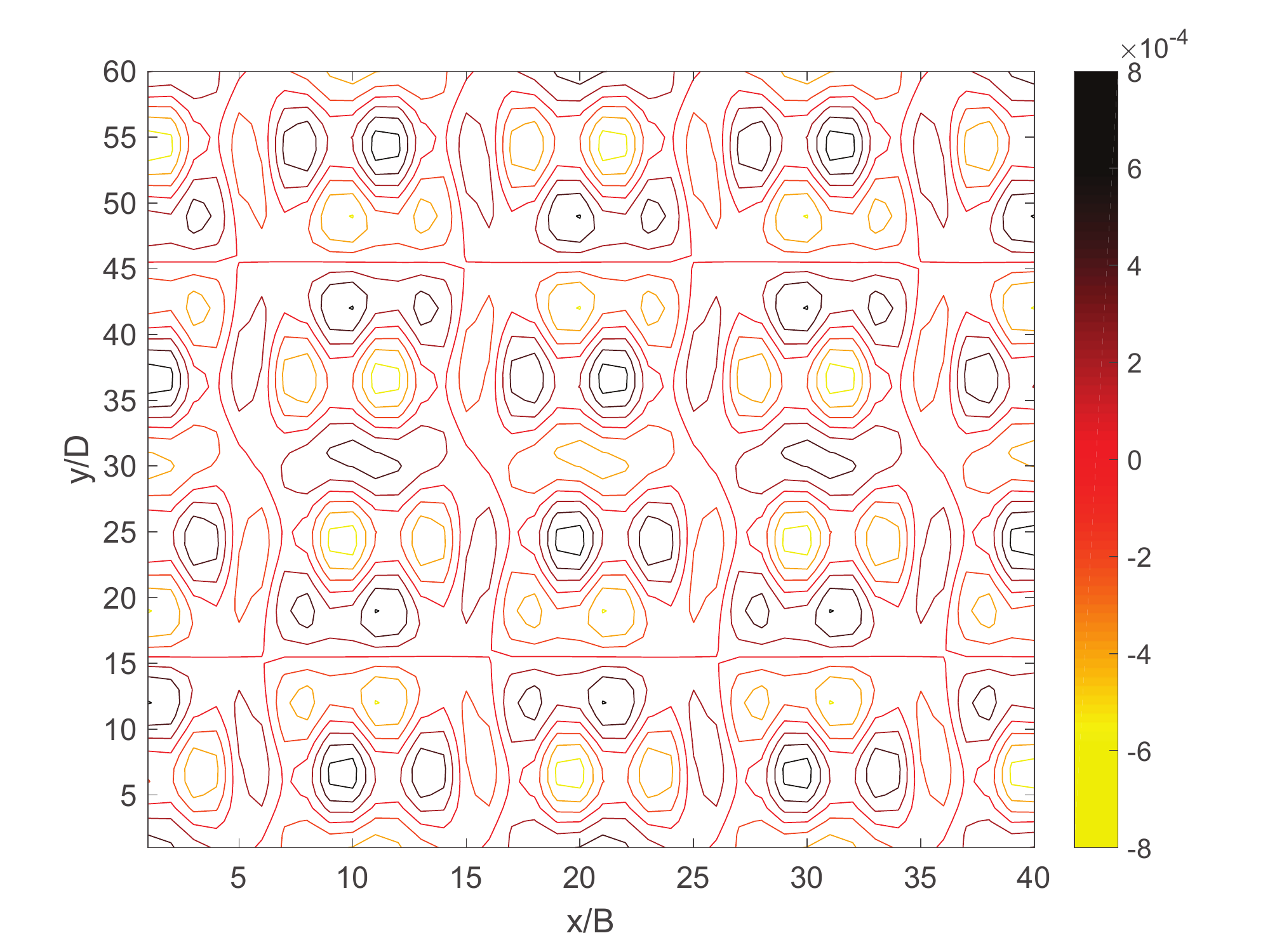}}\\
\subfigure[Error of continuum model (full)]
{\label{fig:2d-glide-D30B50-err}\includegraphics[width=2.2in]{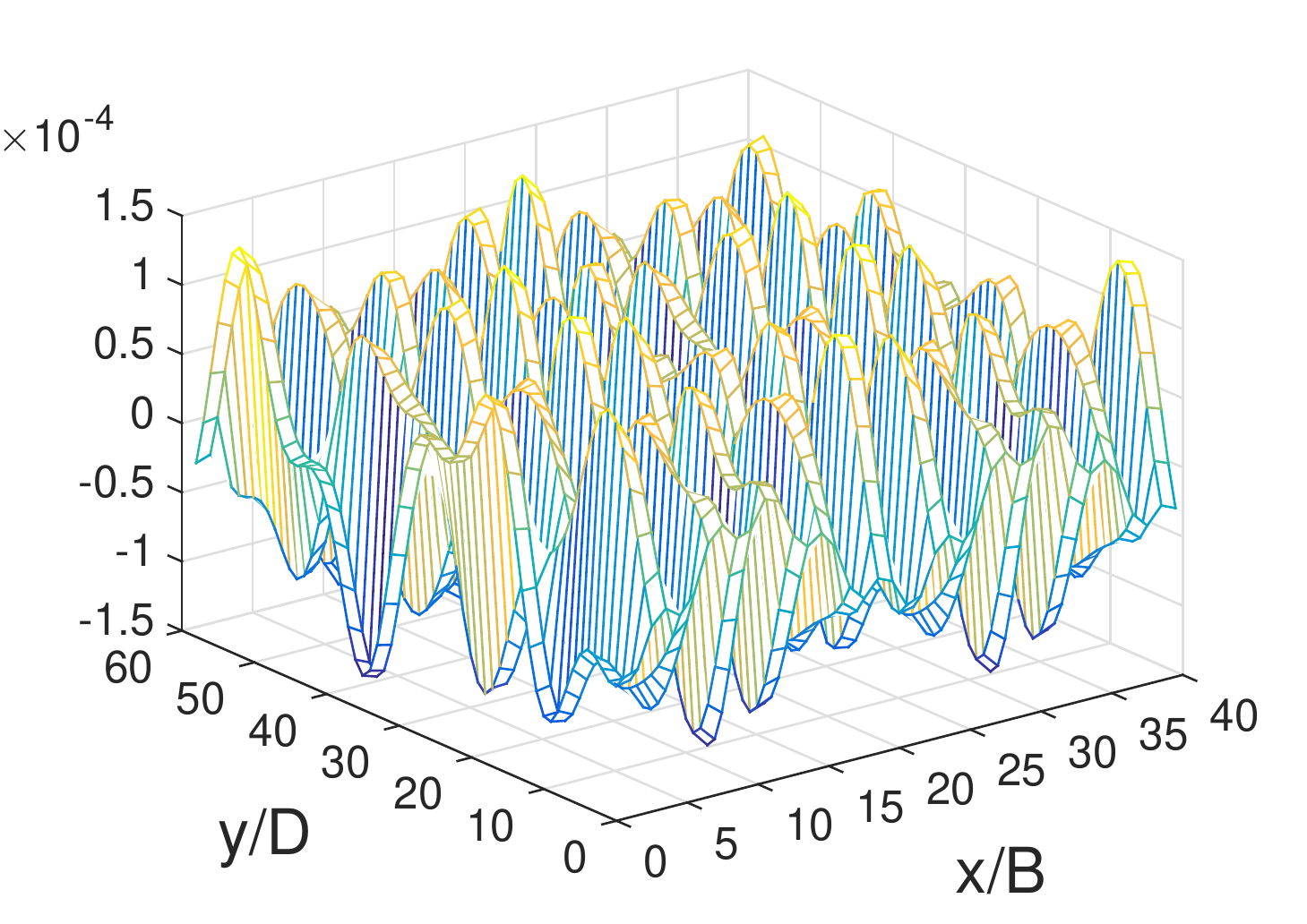}}
\subfigure[Error of continuum long-range force]
{\label{fig:2d-glide-D30B50-err0}\includegraphics[width=2.2in]{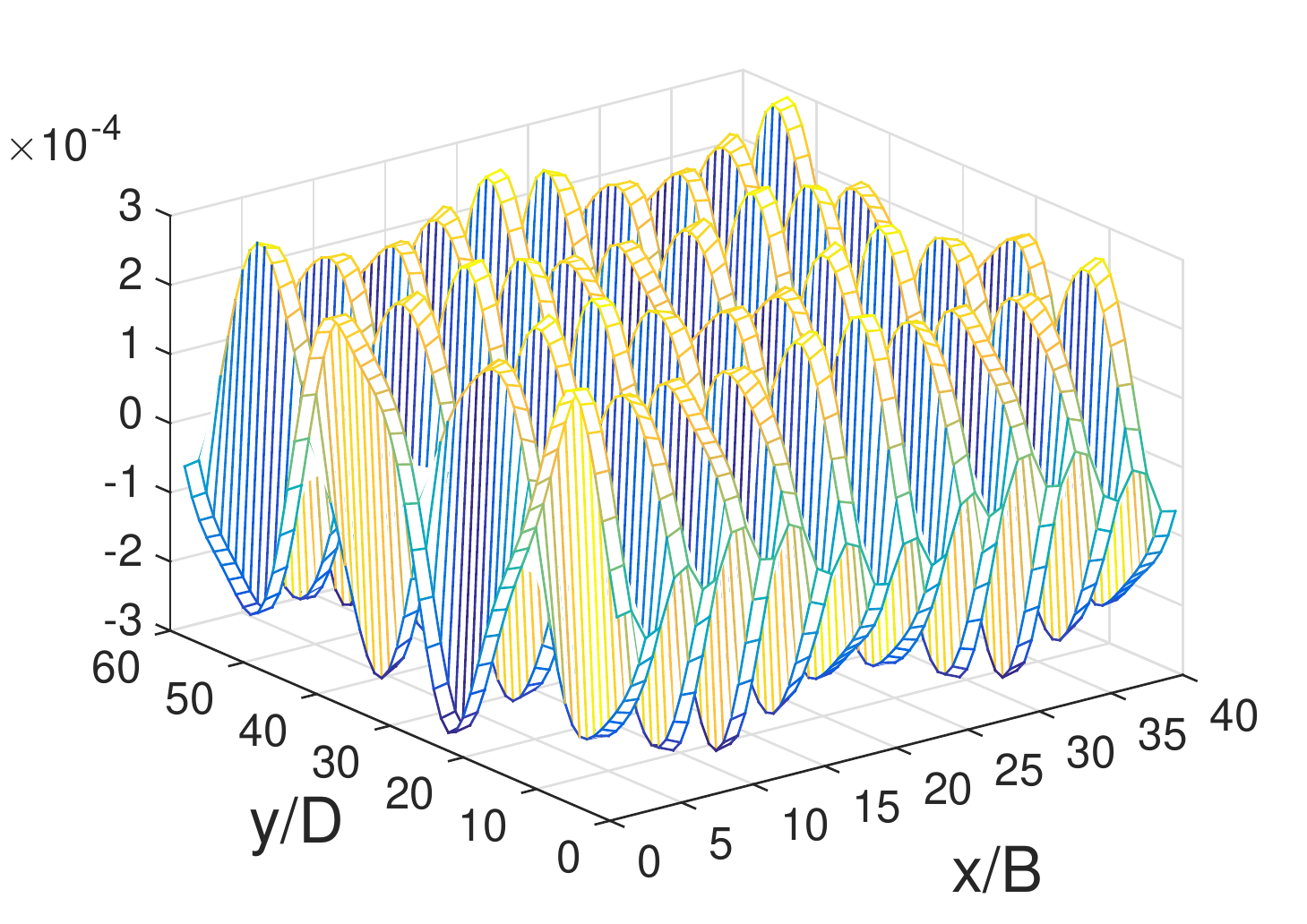}}
\caption{Example 3: Continuum glide force and comparison with that of the discrete dislocation model for a general dislocation distribution given by Eq.~\eqref{eqn:generaldistribution}. The force unit is $\mu b$.}\label{fig:fullforce}
 \end{figure}

 In this example, we examine the continuum glide force in Eq.~\eqref{eq:evn-glide-conserv} for a general dislocation distribution. The dislocation distribution is given by
 \begin{equation}\label{eqn:generaldistribution}
 \left\{
 \begin{array}{l}
\phi(x,y)=\frac{b}{B}x+
0.02 \sin \left(\frac{2\pi}{L_1}10 x\right) \sin \left( \frac{2\pi}{L_1}2y\right), \\ \psi(x,y)=\frac{b}{D}y+
 0.02\sin  \left(\frac {2\pi}{L_2}2 x \right)\sin  \left( \frac{2\pi}{L_2}5 y\right),
 \end{array}
 \right.
 \end{equation}
 where $D=50b$, $B=30b$,  $L_1=40B$ and $L_2=60D$. Here $L_1$ and $L_2$ are the periods of the perturbations in the $x$ and $y$ directions, respectively, and the wavenumbers of the perturbations in DDPFs $\phi$ and $\psi$ are $(10,2)$ and $(2,5)$, respectively.

Fig.~\ref{fig:fullforce} shows the values of the continuum glide force calculated by Eq.~\eqref{eq:evn-glide-conserv} and comparisons with the results obtained using the discrete dislocation model. It can be seen that the glide force profile calculated by the continuum model including both the long-range and short-range interactions (in Fig.~\ref{fig:fullforce}(a))  excellently keeps the overall features of the glide force distribution calculated by the discrete dislocation model (in Fig.~\ref{fig:fullforce}(c)), whereas
 the continuum long-range glide force alone (in Fig.~\ref{fig:fullforce}(b)) loses too much detailed information compared with the discrete model (in Fig.~\ref{fig:fullforce}(c)). Moreover, as shown in Fig.~\ref{fig:fullforce}(d) and (e), the full continuum force successfully reduces the maximum error of the continuum long-range force by half, although the continuum short-range terms are derived only from special distributions of dislocations.

\subsection{Dynamics simulations}

In this subsection, we present some simulation results for the dynamics of the dislocation structures and compare the results with those of the discrete model. We consider the dislocation distribution of Case 1 in Sec.~\ref{sec:continuum-short}. In this case, the continuum model is given by Eq.~\eqref{eq:case1-evn3-simulation}.
We fix the uniform active slip plane spacing $\psi_y=D=50b$.

\begin{figure}[htbp]\centering
 \label{fig-evn1-potential-trend}\includegraphics[width=3in]{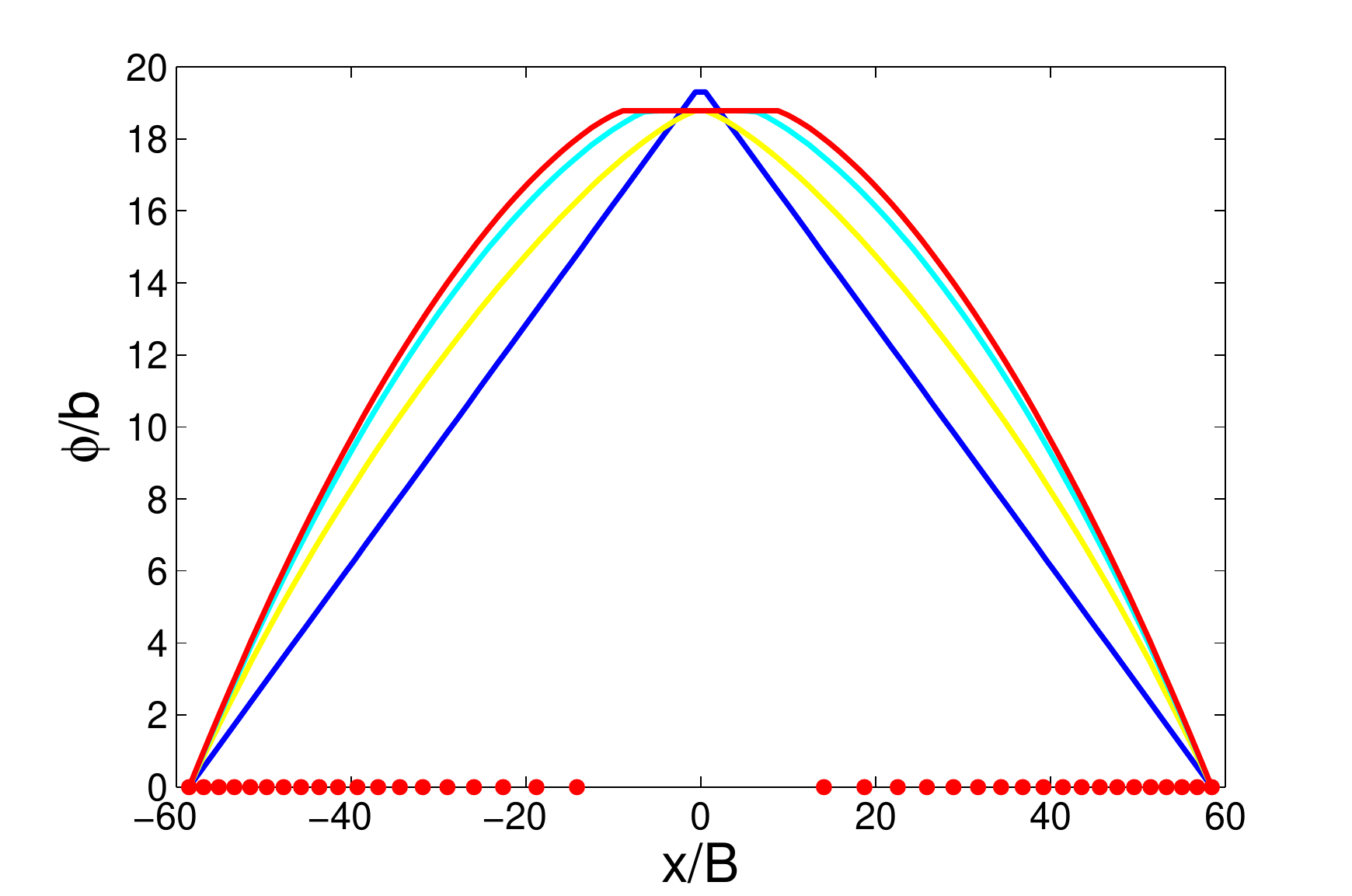}
\caption{Evolution of the dislocation walls system (represented by the evolution of the DDPF $\phi$) and equilibrium locations of dislocation walls (dots on the $x$-axis) under applied shear stress $\sigma_{xy}^0=-0.0009$. The blue curve is the initial profile of $\phi(x)$, and the red curve is the profile of $\phi(x)$ of the final, equilibrium state. }\label{fig:ddpf-evolve}
 \end{figure}

The initial state of the evolution is a dislocation wall system of $N=40$ dislocation walls with average spacing $B=30b$. The left half of these dislocation walls consist of dislocations with direction in the $+z$ direction, and the right half consist of dislocations with direction in the $-z$ direction. Initially, these dislocation walls have equal spacing. An initial profile of the DDPF $\phi$ is shown by the blue curve in Fig.~\ref{fig:ddpf-evolve}. We use periodic boundary condition in the simulations. The dislocation walls at the two ends of the simulation domain are fixed.
 We evolve the dislocation system under applied shear stress  $\sigma_{xy}^0=-0.0009\mu b$ and  $-0.09\mu b$.

\begin{figure}[htbp]\centering
\subfigure[Wall locations for $\sigma_{xy}^0=-0.0009\mu b$]
{\label{fig-evn1-equlibrium-density}\includegraphics[width=2.2in]{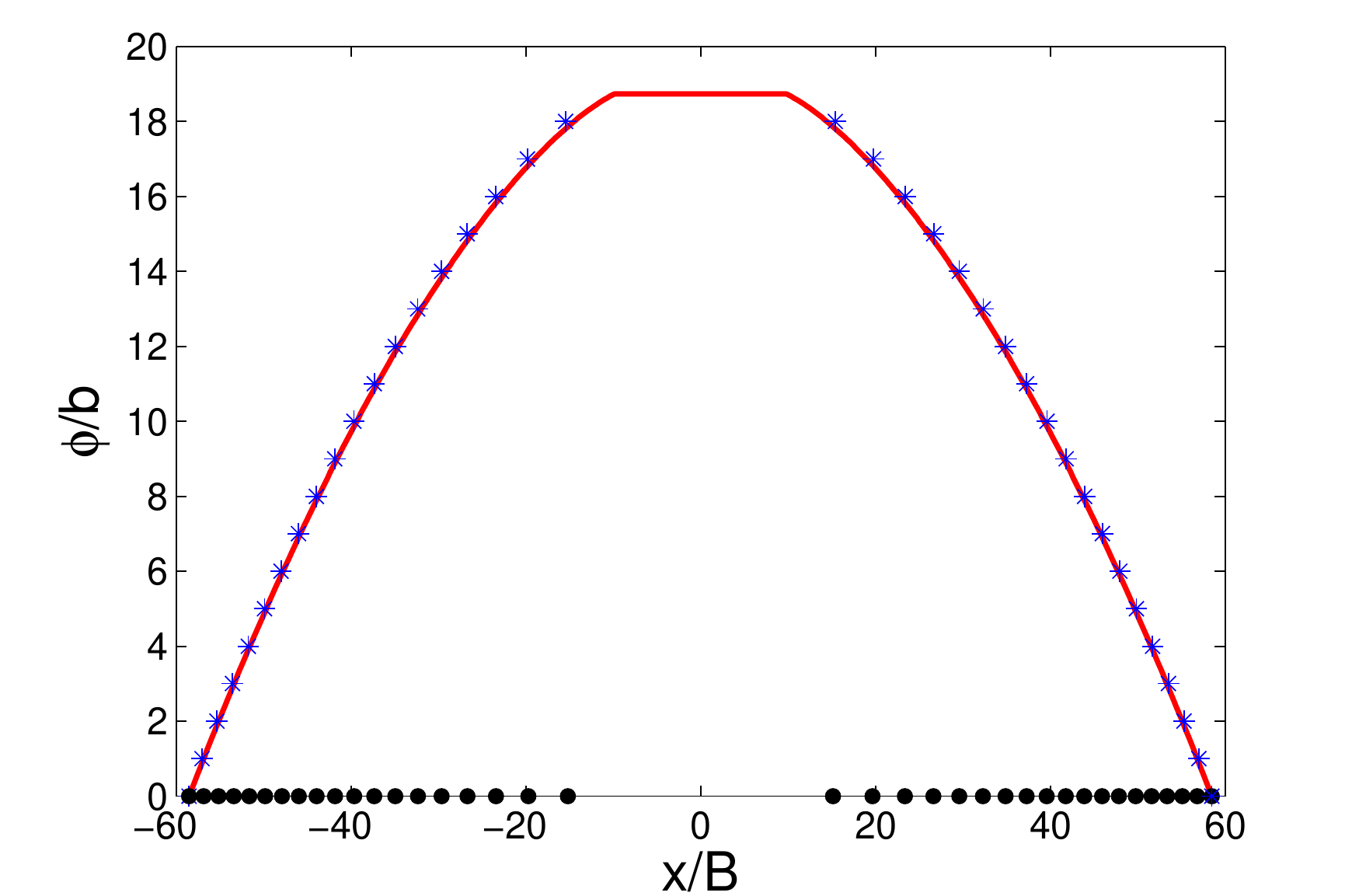}}
 \subfigure[Wall density for $\sigma_{xy}^0=-0.0009\mu b$]
 {\label{fig-evn1-equilibrium-comprison}\includegraphics[width=2.2in]{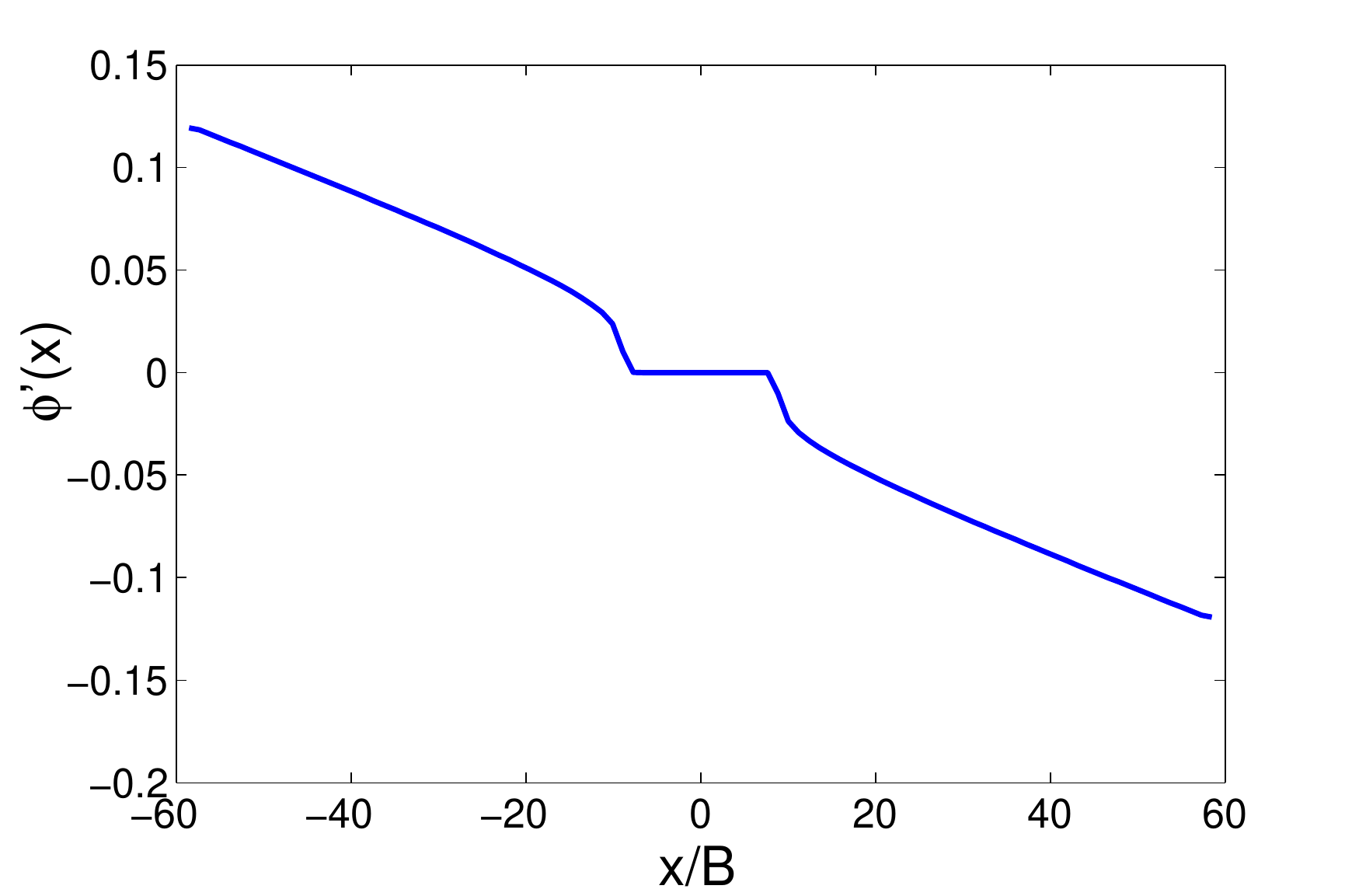}}\\
\subfigure[Wall locations for $\sigma_{xy}^0=-0.009\mu b$]
{\label{fig-evn2-equlibrium-density}\includegraphics[width=2.2in]{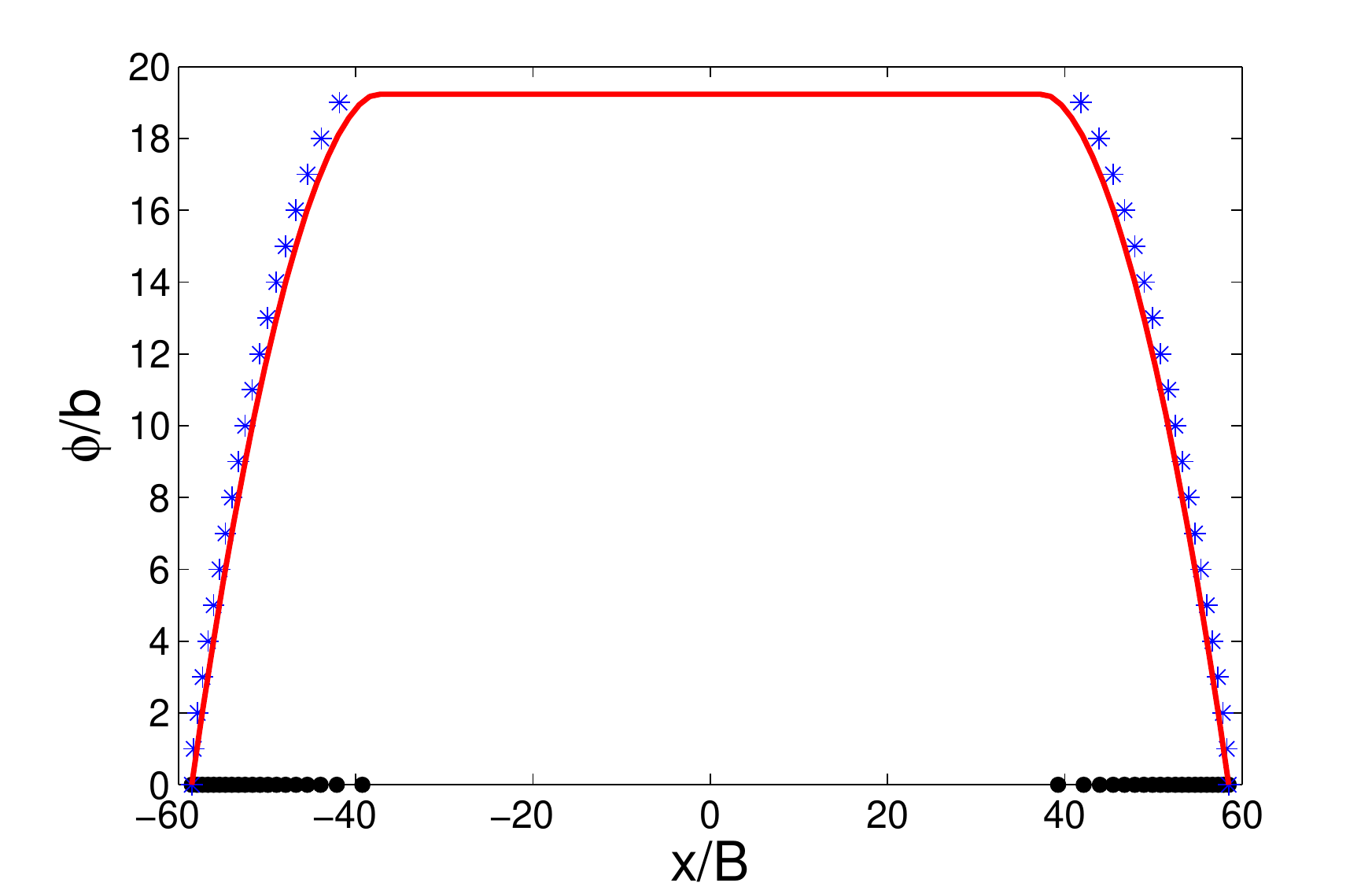}}
\subfigure[Wall density for $\sigma_{xy}^0=-0.009\mu b$]
{\label{fig-evn2-equilibrium-comprison}\includegraphics[width=2.2in]{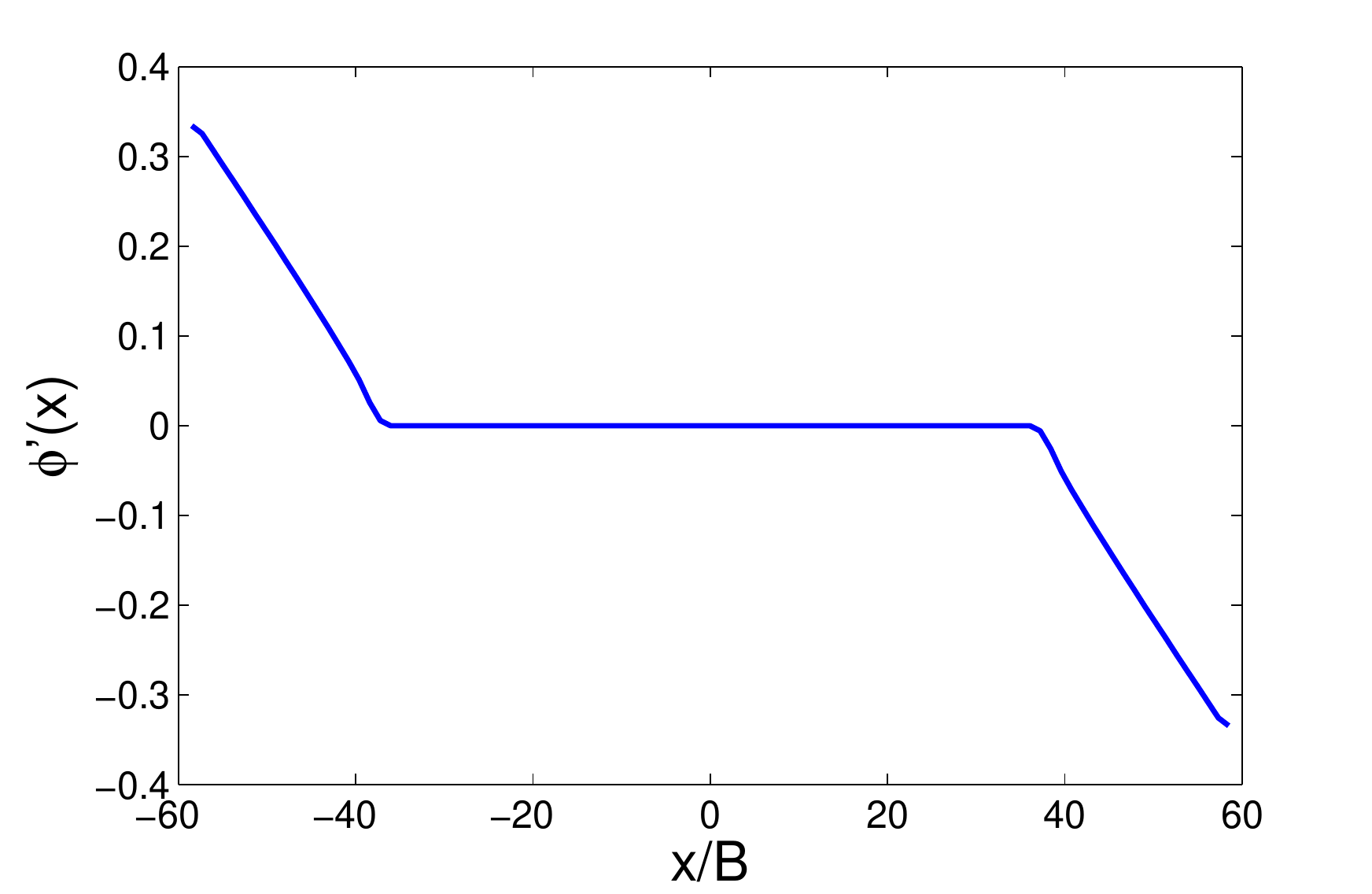}}
 \caption{Equilibrium dislocation wall pile-ups calculated using our continuum model and comparisons with the results of the discrete dislocation model under the applied shear stress  $\sigma_{xy}^0=-0.0009\mu b$ (images (a) and (b)) and  $\sigma_{xy}^0=-0.009\mu b$ (images (c) and (d)). The profiles of the DDPF $\phi$  and locations of  the dislocation walls in the equilibrium states are shown in (a) for $\sigma_{xy}^0=-0.0009\mu b$ and (c) for $\sigma_{xy}^0=-0.009\mu b$. The black dots on the $x$-axis indicate the locations of the dislocation walls, and the blue dots show the corresponding values of $\phi$ in the continuum model. The densities of the dislocation walls (given by $\phi'(x)$) in the equilibrium states are shown in (b) for $\sigma_{xy}^0=-0.0009\mu b$ and (d) for $\sigma_{xy}^0=-0.009\mu b$.}\label{fig:dynamics-equilibrium}
 \end{figure}

Evolution of the dislocation walls system represented by the DDPF $\phi$  under applied shear stress $\sigma_{xy}^0=-0.0009$ is shown in Fig.~\ref{fig:ddpf-evolve}. It can be seen that during the evolution,
   some  opposite-direction dislocation wall pairs initially in the middle annihilate, and the remaining dislocation walls are piled-up at the two ends of the domain. Finally, an equilibrium state is reached, in which the $+z$ dislocation walls are piled up at the left end of the domain and the $-z$ dislocation walls are piled up at the right end of the domain.

 The obtained equilibrium dislocation wall distributions under applied shear stress  $\sigma_{xy}^0=-0.0009\mu b$ and  $-0.09\mu b$  and comparisons with the results obtained by discrete dislocation model  are shown in Fig.~\ref{fig:dynamics-equilibrium}.
In both cases, the simulation results using the continuum model agree excellently with the results of discrete dislocation model for these pile-ups of dislocation walls, even though the dislocation wall densities are high in the pile-ups and vanishes in the middle of the domain.

\section{Conclusions}

In this study, we have considered
systems of parallel straight dislocation walls and have identified four cases of these dislocation structures where the continuum long-range glide or climb force vanishes but the corresponding Peach-Koehler force from the discrete dislocation model does not. We have developed continuum descriptions
for the short-range dislocation interactions  for these four cases by using asymptotic analysis.
The obtained continuum short-range interaction formulas are incorporated
in the continuum model for dislocation dynamics based on a pair of
dislocation density potential functions that represent continuous distributions of dislocations.
 This derived continuum model is able to describe the anisotropic dislocation interaction and motion. It has been shown that after incorporating these short-range interaction terms, the continuum model is able to provide strong stabilizing effect as does by the discrete dislocation dynamics model.
 Since these short-range interaction terms are in the form of second order partial derivatives of the DDPFs $\phi$ and $\psi$, they also serve as regularization terms in the evolution equations of $\phi$ and $\psi$.
 The derived continuum model is validated by
comparisons with the discrete dislocation dynamical simulation results.

Multiple pairs of the DDPFs can be employed in continuum model to describe the dynamics of dislocations with multiple Burgers vectors \cite{Zhu_continuum3D}. The short-range interactions between dislocations with different Burgers vectors may involve dislocation reaction and dissociation in addition to the elastic interactions \cite{Hirth,ZhuScripta2016}. Continuum formulations incorporating these interactions will be explored in the future work.

\bibliography{mybib}
\bibliographystyle{siam}
\end{document}